\documentclass[11pt]{article}
\usepackage[pctex32]{graphics}
\usepackage{amssymb}
\usepackage{latexsym}
\usepackage{epsfig}
\usepackage{multirow}  
\usepackage{pstcol} 
\usepackage{float}      
\usepackage{epstopdf}
\usepackage{mathrsfs}

\usepackage{graphicx} 
\usepackage{amsmath}
\usepackage{relsize}  

\usepackage{graphicx}
\usepackage{multicol}

\usepackage{ragged2e}

\usepackage{times}
\usepackage{subfig} 
\usepackage{tcolorbox}
\newtcolorbox[auto counter]{pabox}[2][]{
colback=white,
fonttitle=\bfseries, title=Box~\thetcbcounter: #2,#1}

\definecolor{Blue}{rgb}{0.,0.,0.6}
\definecolor{Red}{rgb}{1,0.,0.}
\definecolor{turchese}{RGB}{64, 224, 208}
\definecolor{cyan}{RGB}{0, 183, 235}
\definecolor{midnigth}{RGB}{0, 51, 102}

\usepackage{fancyhdr}
\usepackage{setspace}

\newcommand{\R}{{\mathbb R}}

\newcommand{\mSigma}{{\mathsf\Sigma}}

\newcommand{\mE}{{\mathsf E}}

\newcommand{\mC}{{\mathsf C}}

\newcommand{\mI}{{\mathsf I}}
\newcommand{\mL}{{\mathsf L}}
\newcommand{\mM}{{\mathsf M}}
\newcommand{\mP}{{\mathsf P}}

\newcommand{\mT}{{\mathsf T}}

\title{Brain activity mapping from MEG data via a hierarchical Bayesian
algorithm with automatic depth weighting: \\ sensitivity and specificity
analysis}
\author{D Calvetti$^1$ \and A Pascarella$^2$  \and F Pitolli$^3$ \and E Somersalo$^1$ \and B Vantaggi$^3$}
\date{$^1$ Case Western Reserve University \\ Department of Mathematics, Applied Mathematics and Statistics \\ 10900 Euclid Avenue, Cleveland, OH 44106 \\
$^2$ CNR - National Research Council \\ Istituto per le Applicazioni del Calcolo ``Mauro Picone'' \\ Via dei Taurini 19, 00185 Rome, Italy\\
$^3$ University of Rome ``La Sapienza''\\  Department of Basic and Applied Science for Engineering \\ Via Scarpa 16, 00161 Rome, Italy
}
\begin{document}
 
\maketitle

\newpage
\begin{abstract}
A recently proposed IAS MEG inverse solver algorithm, based on the coupling of a hierarchical Bayesian model with computationally efficient Krylov subspace linear solver, has been shown to perform well for both superficial and deep brain sources. However, a systematic study of its sensitivity and specificity as a function of the activity location is still missing. We propose novel statistical protocols to quantify the performance of MEG inverse solvers, focusing in particular on their sensitivity and specificity in identifying active brain regions. We use these protocols for a systematic study of the sensitivity and specificity of the IAS MEG inverse solver, comparing the performance with three standard inversion methods, wMNE, dSPM, and sLORETA. To avoid the bias of anecdotal tests towards a particular algorithm, the proposed protocols are Monte Carlo sampling based, generating an ensemble of activity patches in each brain region identified in a given atlas. The sensitivity is measured by how much, on average, the reconstructed activity is concentrated in the brain region of the simulated active patch. The specificity analysis is based on Bayes factors, interpreting the estimated current activity as data for testing the hypothesis that the active brain region is correctly identified, vs. the hypothesis of any erroneous attribution. The methodology allows the presence of a single or several simultaneous activity regions, without assuming the knowledge of the number of active regions. The testing protocols suggest that the IAS solver performs well in terms of sensitivity and specificity both with cortical and subcortical activity estimation.
\end{abstract}

{\bf Key words:} MEG inverse problem, Activity map, Brain region, Bayes factor, Deep activity

\newpage

\pagestyle{fancy}
\lhead{D. Calvetti}
\section{Introduction}

In the ongoing quest for understanding brain functions and connectivity, during  specific task and in resting states, with the hope of finding a functional fingerprint of autism, pathological depression or schizophrenia, the potential of MEG as a non-invasive and non-disturbing brain activity mapping modality cannot be emphasized enough. The high temporal resolution of MEG offers enormous possibilities for understanding the fine details of brain dynamics, however the reliable reconstruction of activity patterns in cortical and deep brain regions relies on the sensitivity and specificity of the underlying inverse solver. In this work we propose a comprehensive, statistically sound methodology to assess the sensitivity and specificity of different inverse solvers, and demonstrate the viability of the approach by applying it to four different algorithms.

A fundamental difficulty in the solution of the MEG inverse problem \cite{Hamalainen1, baillet2001, brette2012handbook} comes from the non-uniqueness of the solution and the high sensitivity of the problem to noise in the data. To address the non-uniqueness, it is necessary to augment the data with additional information about the solution, which entails either the introduction of regularization techniques in a deterministic setting, or the introduction of prior models in the Bayesian framework. The sensitivity of the computed solution to noise, with low signal-to-noise ratio of data, in turn, requires a good understanding of the noise sources, which in the Bayesian framework is related to a design of a reliable likelihood. These considerations highlight the importance of being able to assess the reliability of a given inversion algorithm in the multitude of tasks it may be applied to, and to quantify the uncertainty in computed solutions.

In this article, we present a systematic study of the sensitivity and
specificity of the Iterative Alternating Sequential (IAS)  MEG inversion algorithm \cite{calvetti3,IAS} which pairs hierarchical Bayesian modeling, with a computationally very efficient prior-based preconditioned iterative linear solver. 
Numerous algorithms based on hierarchical Bayesian modeling have been suggested in the literature, see, e.g., \cite{auranen,calvetti3,henson2009,henson2010,kiebel2008,lucka2012,Lopez2014,Mattout2006,nummenmaa,nummenmaa2,owen,sato,stephan2009,trujillo,wipf2, wipf3}. A common feature of these methods is the use of parameter-dependent prior models, the model parameters representing the second layer of the unknowns. A standard approach in hierarchical Bayesian modeling is to introduce a hyperprior model for the parameters, and to either marginalize, or model  average, the parameters, or to estimate them by maximizing the evidence, see, e.g.,  \cite{bernardo} for general reference.  Alternatively, an empirical Bayes procedure in which the design of the prior includes the selection of a set of hyperparameters driven by the data, under the assumption that for a given dataset, the most likely priors are those that maximize the model evidence. A more time-consuming approach to the hierarchical modeling avoiding optimization techniques is to use Markov chain Monte Carlo (MCMC) methods.

The IAS MEG algorithm, too, is based on the use of hierarchical, conditionally Gaussian model for the prior. The hyperparameters are the variances of the elementary 
sources in the current density model, for learning focality from data. To reduce the computational complexity of the problem, especially in case of high dimensionality of the MEG data,
the algorithm approximates the maximum of the posterior distribution instead of computing the posterior distribution itself. The iterative alternating sequential minimization procedure was shown to converge in \cite{IAS}, where an easily implementable procedure for the estimation of the maximizer of the posterior with respect to both the unknown of primary interest and the hyperparameters was proposed. The maximization algorithm uses a dynamically preconditioned Krylov subspace iterative solver equipped with a suitable stopping rule that allows us to control the fidelity of the solution to data as well as the speed of convergence.
The solution of the regularized subproblem does not rely on Tikhonov-regularized linear solution, but a very fast converging preconditioned Conjugate Gradient Least Squares (CGLS) algorithm, which makes the solution to depend non-linearly from the data \cite{BayesKrylov}. Moreover, the IAS MEG inverse method employs an anatomically justified prior: Based on the segmented subject-specific MRI image containing the information of the cortical surface orientation, each dipole has a direction which is favored, but not forced, by the prior model. Similar ideas with different implementation have been suggested in the literature, see. e.g. \cite{Lin2006}.

Some of the appeal of standard MEG solvers such as MNE \cite{Hamalainen2,Lin} or LORETA \cite{Pasqual} is their simplicity of use, with minimal user intervention required , e.g., for setting parameters or controlling the optimization process, which is not always the case with the hierarchical models. One of the goals in developing the IAS algorithm was to have a robust and convergent method that depends on very few user-supplied parameters with an intuitive meaning. In \cite{calvetti3}, the connection between the conditionally Gaussian hierarchical models and several sparsity-promoting methods \cite{Gorodnitsky,Nagarajan,Uutela} was pointed out. 

A novel contribution of this paper is to provide a physical interpretation of the hyperparameters and to establish a connection with sensitivity weighting.  The IAS  algorithm, as it was rigorously shown in \cite{IAS}, contains only one user-supplied parameter that controls the sparsity of the solution and, as pointed out in Section 2.2, the other hyperparameters can be semi-automatically set based on the information about the signal-to-noise ratio. This formulation provides a proper Bayesian interpretation for the sensitivity weighting that is commonly used with,  e.g., the Minimum Norm Estimate (MNE) and the Minimum Current Estimate (MCE) \cite{Uutela} algorithms to overcome the methods' tendency to favor superficial sources over deep ones,  thereby filling the gap between Bayesian modeling and traditional regularization.

The quest for MEG inverse solvers capable of identifying activities both on superficial cortex and in deeper brain regions continues to be ongoing. In \cite{attal, attal2013} the localization accuracy of subcortical generators was studied  for  three different algorithms: the weighted Minimum Norm Estimate (wMNE), dynamic Statistical Parametric Mapping (dSPM), and standardized Low Resolution brain Electromagnetic Tomography (sLORETA). The procedure relies on a realistic electrophysiological model of deep-brain activity and compares the performance of the three methods in recovering activity in the neocortex, hippocampus, amygdala and thalamus. In the present study the comparison is extended to include the IAS method using a model of the subcortical regions that considers their surface envelopes, adding basal ganglia, brainstem and cerebellum to the list of brain regions, and introducing a statistically justified metric to evaluate the performances of the different algorithms.

More specifically, our investigation addresses the following question: Given an atlas of anatomo-functional  brain regions (BR),  how well does the IAS method for the MEG inversion identify active regions, and to what extent non-active regions will be identified correctly as inactive ones? The former question is asking for the sensitivity of the MEG inverse solver to the location of the active patch in the brain, and the latter for the specificity. In line with the Bayesian paradigm that IAS is based on, we investigate these questions using statistical tools. For this purpose we introduce an Activity Level Indicator (ALI), measuring the mean activity level of each BR given a simulated patch activity in a selected BR, and compute the first and second order statistics of this indicator over a sample of random activation patches in the selected BR. To assess the relative performance of IAS, we compare these statistics with those of three other standard inversion methods, wMNE \cite{Lin}, dSPM \cite{Dale2000}, and sLORETA \cite{Pasqual}, using the implementations provided in Brainstorm \cite{Brainstorm}.
Furthermore, to test the relative specificity of the four methods, we observe that a common approach in the Bayesian methodology to
test hypotheses or models is to use Bayes factors of probabilities of the data, given the competing hypotheses or models \cite{bernardo, KassRaftery}. In the current setting, we interpret the estimated activity as data, and test the probabilities of them under the competing hypotheses of having activity in the correct BR versus having it in another randomly chosen BR.  We compare the specificity of the IAS algorithm to that of the other three standard solvers using simulated data under two different scenarios: when there is only one active patch in a cortical or subcortical region, and when there are multiple (from two to six) active patches in cortical and subcortical regions.  In this way we can test not only the sensitivity and specificity of each inverse solver with respect to a single BR, but also test their ability in recovering an activation pattern when other regions disturb the identification.  We conclude the computed experiments by comparing the reconstruction obtained by the four inverse solvers in an example starting from real data, where the underlying activation pattern is not known exactly but is inferred from the protocol used for the data collection.

\section{Materials and Methods}

We start by presenting a brief review of the {\em Iterative Alternating Sequential} (IAS) MEG inverse solver, described in detail in \cite{IAS}. Subsequently, we introduce the computational tools that will be used for testing the algorithm's performance, focusing in particular on its sensitivity and specificity in identifying active areas of interest from the MEG data and comparing it to three other mainstream MEG inverse solvers.

To discretize the problem, we approximate the cortical surface, as well as the surface of the subcortical structures, by a triangular mesh with $n$ vertices, whose coordinates are denoted by $v_j$. We associate to each vertex a unit length vector $\vec \nu_j$ pointing in the direction normal to the local cortical/subcortical surface.

The forward model is of the form
\begin{equation}\label{forward}
 b = \sum_{j=1}^n \mM_j \vec q_j + \varepsilon,
\end{equation}
where $b \in \R^m$ is the observation vector representing the measured magnetic field components, $\mM_j\in\R^{m\times 3}$ is the lead field matrix associated with the $j$th dipole located at $v_j$, $\vec q_j$ is the $j$th dipole moment and $\varepsilon\in\R^m$ is additive observation noise. The lead field matrices take into account the conductivity structure of the head model. While changing the forward model does not affect the inversion algorithm, the results are to some extent sensitive to the model used (cf. \cite{vorwerk2014}).

We model the noise term as a zero mean Gaussian random variable, $\varepsilon\sim{\mathscr N}(0,\mSigma)$, where
$\mSigma\in\R^{m\times m}$ is the covariance matrix. Therefore, the likelihood density of $b$ conditioned on $\vec q_1,\ldots,\vec q_n$ can be written as
\[
 \pi(b\mid \vec q_1,\ldots,\vec q_n) \propto  {\rm exp} \left(-\frac 12 \|b -\sum_{j=1}^n \mM_j\vec q_j\|^2_\mSigma\right),
\]
where $\|z\|_\mSigma^2 = z^\mT\mSigma^{-1}z$ is the square of the Mahalanobis norm and ``$\propto$'' stands for ``proportional up to a normalizing constant''.

\subsection{The inverse solver: an overview}\label{sec:solver}

The IAS algorithm is based on a Bayesian hierarchical prior model of the activity of a single dipole of the form
\[
 \pi_{\rm prior}(\vec q_j\mid\theta_j)\sim{\mathscr N}(0,\theta_j \mC_j) \propto
 \frac 1{\theta_j^{3/2}} {\rm exp}\left(- \frac{ \vec q_j^{\mT}\mC_j^{-1}\vec q_j}{2\theta_j}\right) = {\rm exp}\left(- \frac{\|\vec q_j\|^2_{\mC_j}}{2\theta_j} - \frac 32\log\theta_j\right);
\]
here $\mC_j\in\R^{3\times 3}$ is the local anatomical prior matrix,
\[
 \mC_j = \vec\nu_j\vec\nu_j^\mT + \delta(\vec\xi_j\vec\xi_j^\mT + \vec\zeta_j\vec\zeta_j^\mT),
\]
where $0<\delta<1$ and $(\vec \xi_j,\vec\zeta_j,\vec\nu_j)$ is a local
orthonormal frame at the  $j$-th vertex and $ \vec\nu_j$ is orthogonal to the
cortical/subcortical surface. The parameter $\theta_j>0$ scaling the local prior covariance of the dipole $\vec q_j$, is modeled further as a random variable following the gamma distribution,

\[
 \theta_j \sim{\Gamma}(\theta_j^*,\beta_j) \propto \theta_j^{\beta_j-1}{\rm
 exp}\left(-\frac{\theta_j}{\theta_j^*}\right).
\]

The  scaling parameter $\theta_j^*$ of the gamma density can be chosen to express our a priori belief about the order of magnitude of the expected value of the
variance $\theta_j$, while the value of the shape parameter $\beta_j$ controls the sparsity of the solution. In all our computed experiments the latter is held constant for all dipoles, that is, $\beta_j = \beta$, while
$\theta_j^*$ is chosen by an empirical Bayes argument to correspond to a sensitivity scaling. The details and justification for the choice of the hyperparameters can be found in 
the Appendix (see also the original article \cite{IAS}).
By Bayes' theorem the posterior model can be written as
\begin{eqnarray}\label{posterior}
 && \pi(\vec q_1,\ldots,\vec q_n,\theta_1,\ldots,\theta_n\mid b) \propto \prod_{j=1}^n\pi_{\rm prior}(\vec q_j\mid\theta_j)\pi (\theta_j\mid\theta_j^*,\beta)\pi(b\mid \vec q_1,\ldots,\vec q_n) \nonumber\\
 &&\phantom{XX}\propto{\rm exp}\left(  - \frac 12\sum_{j=1}^n\frac{\|\vec q_j\|^2_{\mC_j}}{\theta_j} +
 \sum_{j=1}^n\left[(\beta -  \frac 52)\log\theta_j
 -\frac{\theta_j}{\theta_j^*}\right] -\frac 12 \|b - \sum_{j=1}^n\mM_j\vec q_j\|_{\mSigma}^2 \right)
 \end{eqnarray}

The IAS algorithm for computing an estimate of the maximum a posteriori (MAP)
solution of (\ref{posterior}), is based on the following alternating iterative
minimization scheme:
\begin{enumerate}
\item Initialize $\theta_j = \theta_j^*$, $1\leq j\leq n$, and set $k=0$;
\item Until the convergence criterion is met:
\begin{itemize}
\item[(i)] Update $\vec q_1,\ldots,\vec q_n$, setting
\[
 (\vec q_1^{\,k+1},\ldots, \vec q_n^{\,k+1}) = {\rm argmax} \{\pi(\vec q_1,\ldots,\vec q_n,\theta_1^k,\ldots,\theta_n^k\mid b)\};
\]
\item[(ii)] Update $\theta_1,\ldots,\theta_n$, setting
\[
 (\theta_1^{k+1},\ldots,\theta_n^{k+1}) = {\rm argmax}\{\pi(\vec q_1^{\,k+1},\ldots,\vec q_n^{\,k+1},\theta_1,\ldots,\theta_n\mid b)\};
\]
\item[(iii)] Increase $k\rightarrow k+1$.
\end{itemize}
\end{enumerate}

It was proved in \cite{IAS} that the above optimization algorithm converges to a unique local maximum. Moreover, the iterations can be performed effectively by noticing that when minimizing the energy function defined as,
\begin{equation}\label{energy}
{\mathscr E}(\vec q_1,\ldots,\vec q_n;\theta_1,\ldots,\theta_n) =
\lefteqn{ \overbrace{\phantom{\|b - \sum_{j=1}^n\mM_j\vec q_j\|_{\mSigma}^2 + \sum_{j=1}^n\frac{\|\vec q_j\|^2_{\mC_j}}{\theta_j} }}^{(a)}}
\|b - \sum_{j=1}^n\mM_j\vec q_j\|_{\mSigma}^2+
\underbrace{ \sum_{j=1}^n\frac{\|\vec q_j\|^2_{\mC_j}}{\theta_j}
-2
 \sum_{j=1}^n\left[(\beta -  \frac 52)\log\theta_j -\frac{\theta_j}{\theta_j^*}\right]}_{(b)}
\end{equation}
the optimization with respect to $\vec q$, affecting only part (a) of the above expression, reduces to a quadratic minimization problem, while the minimization with respect to the hyperparameters $\theta_j$, depending only on part (b) of the energy function, can be done independently of other components, and admits a solution in closed form.
Finally, we point out that an approximate solution of the quadratic minimization problem can be found very efficiently using a priorconditioned CGLS algorithm oulined below.

\centerline{
\includegraphics[width=7.5in]{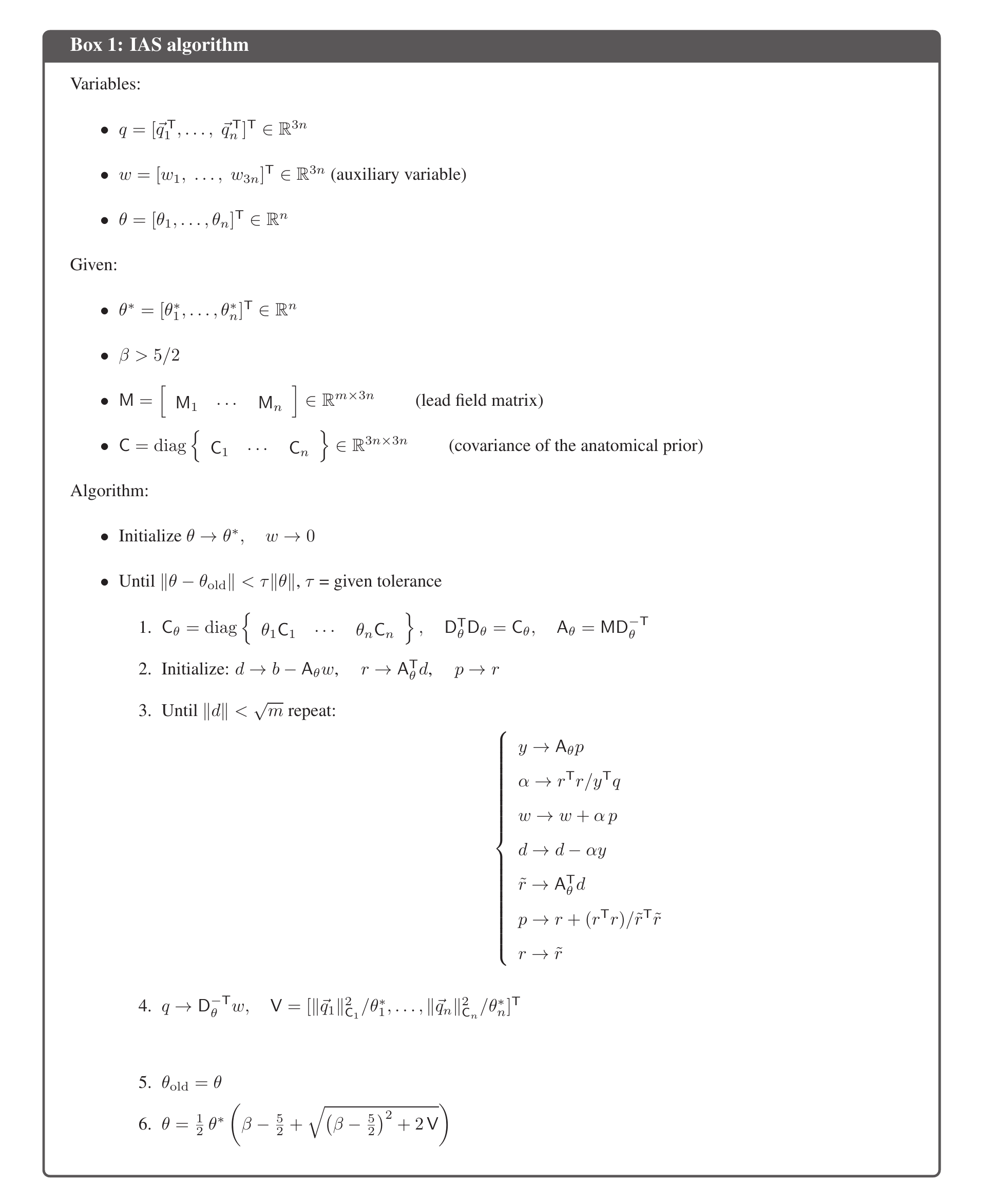}}

For the omitted details about the algorithm and a discussion about the meaning of $w$, we refer to \cite{IAS}.

\subsection{Initializing the IAS algorithm}

To initialize the IAS algorithm we have to choose the scaling parameters $\theta_j^*$, $1 \le j \le n$. This can be done observing that $\theta_j^*$ can be 
viewed as a sensitivity weight that takes into account the distribution of the active sources and the signal-to-noise ratio SNR.
If we assume  
\[
\varepsilon \sim {\cal N}(0,{\mathsf \Sigma}),
\]
and, furthermore, that the expectation of the number of active dipoles is $\bar k$, we get
\[
\theta_{j} ^*= \sum_{k=1}^n\frac {p_k}k \,\frac {{\rm SNR}\times{\rm trace}({\mathsf\Sigma})}{\beta \|\mM_j{\mathsf C}_j^{1/2}\|^2_{\rm F}}, \quad 1\leq j\leq n, \quad p_k\sim{\rm Poisson}(\overline k),
\]
where the subscript ${\rm F}$ denotes the Frobenius norm (see the Appendix for details). 

We point out that the expectation of the variance of a single dipole cannot exceed a physiologically meaningful upper bound $\theta_{\rm max}$, which in practice means that we need to choose $\theta_j^*$ as
\[
 \theta_j^* = \min\left\{\theta_{\rm max},\frac{C}{ \|\mM_j{\mathsf C}_j^{1/2}\|^2_{\rm F}}\right\}.
\]
Thus, the analysis leaves only three parameters to be selected by the user: $(\beta,\overline k, \theta_{\rm max})$, each of which has a clear physiological interpretation.

\subsection{Test protocols for sensitivity and specificity}

The protocols for the validation of the IAS algorithm focus on the sensitivity of the method, i.e., how well the method is able to identify an active area, and on its specificity, i.e., how often the algorithm misidentifies a non-active area as active. The validation methodology discussed here is general and can be used for any MEG inversion algorithm, therefore providing a flexible platform for comparing the performance of different methods. 

In the sequel, we assume that the source space representing the brain is divided into anatomo-functional brain regions (BR) following a given atlas. We denote by $L$ the number of BRs in the atlas, and assume that every vertex $v_j$ can be uniquely attributed to a single BR.

\subsubsection{Patch activity generation}\label{sec:patches}

We study the sensitivity and specificity of the method in the case of synthetic
data produced by the activity of a patch ${\mathcal{P}}$ in a given
anatomo-functional BR. The patch activity is generated with a Monte Carlo
algorithm that allows us to test the effect of random source-dependent
variations in the data on the reconstructions. The activity patch
${\mathcal{P}}$ consists of a preselected number $N_{\mathcal{P}}$ of vertices
and corresponding dipole moments. (see the Appendix for details on the
construction of ${\mathcal{P}}$).

Let ${\mathcal{P}}=\{(v_1,\vec q_1),\ldots,(v_{N_{\mathcal{P}}}, \vec q_{N_{\mathcal{P}}}) \}$ denote the coordinates of the vertices and the corresponding dipole moments on the patch ${\mathcal{P}}$. For later reference, we define the barycenter of the activity patch by the formula
\begin{equation}\label{barycenter}
 m_{\mathcal{P}} = \frac 1{\sum_{v_j\in
 \mathcal{P}} \|\vec q_j\|} \sum_{v_j\in
 \mathcal{P}} v_j\|\vec q_j\|,
\end{equation}
where the summation is over all vertices in the patch. Observe that the barycenter of an activity patch need not coincide with any of the nodes; in fact, there is no guarantee that the barycenter is inside the BR of interest due to the lack of convexity of the brain regions.

When testing the performance of the algorithms, we consider two simulation protocols: A single active BR, or several active BRs.
For both protocols, we generate a sample ${\mathscr B}=\{b^1,b^2,\ldots,b^K\}$ of data vectors as follows. In the single active BR case, we select one of the regions from the atlas, and generate $K$ independent random activity patches in the selected BR. For each activity, we then compute the magnetic field components at the sensor locations with the model (\ref{forward}) then add noise following the random dipole brain noise model  \cite{DeMunck, Huizenga}. In the second protocol, we consider activity occurring simultaneously in $N$ different BRs. For each of the $N$ active regions, we generate independently an activity patch, repeating the process independently $K$ times, and compute the magnetic field data at the sensors, corrupting the data with the additive simulated brain noise as in the single region case.

\subsubsection{ Sensitivity and BRs: Activity level indicator vectors}
An important criterion for assessing the performance of an MEG inverse solver for analyzing the brain activity is how reliably it maps the observed data to the correct region of interest. Because it is unreasonable to expect a perfect performance due to the severe ill-posedness of the problem, we propose an evaluation tool to quantify how well the BR where the simulated activity takes place can be identified, and what are the most likely confounding regions. The correct identification of regions further away from the sensors, e.g., deep brain structures, is expected to be most challenging. In general, however, for a reliable solver,
it is reasonable to expect that the confounding regions should be anatomically close to the one where the activity occurs. The procedure for quantifying the sensitivity of a MEG inverse solver to the BR of the activity is outlined below.

Let
\[
F:\R^m\to\R^{3n}, \quad b\mapsto  \left[\begin{array}{c} \widehat{\vec q}_{1} \\ \vdots \\ \widehat{\vec q}_{n}\end{array}\right],
\]
denote the map defined by the MEG inversion algorithm, where $\widehat{\vec q}_{j}$, $j=1,\ldots,n$ are the estimated dipole moments at the $n$ vertices. We introduce the estimated brain activity vector $a\in\R^n$, whose $j$th component  is the norm of the $j$th current dipole found by the MEG inverse solver, that is, $a_{j} = \|\widehat{\vec q}_{j}\|$. Since each entry of  the vector $a$ can be associated to one of the BRs in which the corresponding vertex is located, 
we may define the estimated BR Activity Level Indicator (BR-ALI) vector $ \alpha\in\R^L$, corresponding to the brain activity estimate $a$ as
\[
 \alpha = \left[\begin{array}{c}  \alpha_{1} \\ \vdots \\  \alpha_{L}\end{array}\right],\quad  \alpha_{\ell} =
{\rm mean} \ \{ a_{j}, j\in  {\cal R}_\ell\},
\]
where ${\cal R}_\ell$, $\ell = 1,\ldots,L$, is the set of the vertex indices corresponding to the $\ell$-th brain region.

Assuming that the data $b$ are generated by an active patch in the $\tilde \ell$th BR alone, if $F$ were a {\em perfect} MEG inversion algorithm for identifying active regions, the estimated ALI vector would have  only the $\tilde\ell$th component different from
zero, while  $\alpha_{\ell} = 0$ for all $\ell \ne \tilde \ell$.
Based on this observation, we propose to measure the sensitivity of an MEG inversion algorithm to activity in a given BR by how many components of $ \alpha$ corresponding to other BRs are different from zero and how large they are.

In our study of the sensitivity and specificity of the IAS  MEG inverse solver and of the other three algorithms considered, we assess the quality of the associated $ \alpha$ using the Monte Carlo simulations of $K$ independent activity patches as described in Section~\ref{sec:patches}. More specifically, for each BR, to evaluate  the MEG inverse solver, we 
\begin{enumerate}
\item[i)] apply the MEG inverse solver with the synthetic data sample
${\mathscr B} = \{ b^1,\ldots,b^K\}$, to estimate the corresponding activity vectors,
$\{a^1,\ldots, a^K\}$, $a^k\in\R^n$,
\item[ii)] compute the corresponding BR-ALI vectors $\{ \alpha^1,\ldots,\alpha^K\}$, $\alpha^k\in\R^L$,
 \item[iii)] compute the sample mean and the sample covariance of the $K$ BR-ALI vectors,
 \[
 \overline{\alpha} = \frac 1K\sum_{k=1}^K \alpha^k,\quad  \Gamma_\alpha = \frac {1} {K-1}\sum_{k=1}^K(\alpha^k -
 \overline\alpha)( \alpha^k - \overline\alpha)^\mT.
\]
\end{enumerate}

The display of the mean  BR-ALI vector $\overline\alpha$ in the form of a
histogram provides a visual clue of the method's performance; the more tightly
the histogram is concentrated on the active patch, the more reliable the
estimate is. The diagonal entries of  $\Gamma_\alpha$, which are the marginal
variances of the estimated activity levels in various BRs and measure the
consistency of the mean BR-ALI vector retrieval, are indicated in the
histograms displayed in Figures ~\ref{fig:LatOc_Baffo}, ~\ref{fig:Cereb_Baffo}
as variance whiskers.

The methodology proposed for the sensitivity and specificity analysis generalizes naturally to several active regions of interest. 
In the case where there are N simultaneously active BRs, the statistics of the estimated BR-ALI vectors can be investigated as in the case of a single patch.

\subsubsection{Specificity and Bayes factors}

To set up a protocol to assess the specificity of an MEG inverse solver,
consider the following simple elementary test that constitutes one of the basic
building blocks of Bayes factor analysis, see e.g. \cite{KassRaftery}.

Given activity in a patch ${\mathcal{P}}$ with barycenter $m_{\mathcal{P}}$ defined by (\ref{barycenter}), consider two spheres with the same fixed radius $R$, $B_0 = B(m_{\mathcal{P}},R)$ centered at the barycenter of the activity, and $B_1 = B(v_j,R)$, centered at a randomly selected vertex $v_j$ in the brain, respectively.
Let $ a \in\R^n$ be the activity vector estimated from the data $b$ generated by the activity in the patch ${\mathcal P}$, and formulate the following pair of hypotheses:
\begin{eqnarray*}
(H_0) &\quad& \mbox{The activity is in the sphere $B_0$,} \\
(H_1) &\quad& \mbox{The activity is in the sphere $B_1$.}
\end{eqnarray*}
We consider the estimated activity vector $a$ as data, and associate to the hypotheses $H_0$  and $H_1$ the probabilities 
\[
 P(a\mid H_0) = \frac 1{|  a |} \sum_{v_j\in B_0}  a_{j} \quad  \mbox{and} \quad P( a \mid H_1) = \frac 1{| a|} \sum_{v_j\in B_1}  a_{j},
\]
where $| a | = \sum_{j=1}^n  a_{j}$ is a normalizing factor. We quantify the strength of the hypothesis $(H_0)$ against the hypothesis $(H_1)$ in terms of the Bayes factor,
\[
{\rm BF}(H_0,H_1) = \frac{P( a\mid H_0)}{P( a\mid H_1)},
\]
in the following manner. If ${\rm BF}(H_0,H_1)<1$ we argue that the data support the hypothesis $H_1$, while if ${\rm BF}(H_0,H_1)>1$, the evidence is in favor of the hypothesis $H_0$, and the support for $H_0$ is stronger the larger the value of the ratio. Following \cite{KassRaftery}, we classify the relative strength of the two hypotheses according to the following scale:
\begin{eqnarray}
 {\rm BF}(H_0,H_1)&\in (0,1)  \quad &\mbox{The evidence is against $(H_0)$,} \nonumber\\
 {\rm BF}(H_0,H_1)&\in [1,3)  \quad &\mbox{The evidence is weakly in favor of $(H_0)$,} \label{evidence strength}\\
 {\rm BF}(H_0,H_1)& ~\in [3,10) \quad &\mbox{The evidence is strongly in favor of $(H_0)$,} \nonumber\\
 {\rm BF}(H_0,H_1)&\geq10 \quad &\mbox{The evidence is overwhelmingly in favor of $(H_0)$.} \nonumber
\end{eqnarray}

Clearly, the hypothesis $(H_1)$ depends on the choice of the point $v_j$ defining the sphere $B$. Because we are, in fact, interested in assessing whether the MEG inversion method used to compute $\widehat a$ is favoring the correct activation area over {\em any} other area, we enrich the Bayes factor test by generating not one, but a family of $M$ competing hypotheses,
\[
 (H_m) \quad \mbox{The activity is in the sphere $B_m = B(v_{j_m},R), \quad 1\leq m\leq M$},
\]
where the center points $v_{j_m}$ are drawn from the set of vertices with uniform probability. This leads to a set of $M$ Bayes factors,  
${\rm BF}(H_0,H_m)$, $1\leq m\leq M$, and we can compute the frequencies of
occurrences in the scale defined above. A good solver should have most of the
Bayes factors in the intervals $[3,10)$ or above 10. Numerous accurrences of
Bayes factors below 1 indicate that the solver suggests often false positive
activities.

It may be argued that the Bayes factor analysis as described above implicitly uses the information that the MEG data was generated by a single active patch, while in reality the number of active brain regions is unknown. To address this issue, we test the algorithm also with several simultaneous patch activities. More specifically, in the case where the activity occurs simultaneously in $N$ different BRs, we select one of these active regions at a time and set the activity spheres $B_0$ centered at the barycenter of the selected patch, while the center of the competing activity spheres $B_m$ are drawn randomly from any other BR.  Subsequently we compute the Bayes factors supporting the hypothesis
\[
(H_0):\quad  \mbox{The activity is in the sphere $B_0= B(m_{\mathcal P},R)$}.
\]
versus the family of competing hypotheses,
\[
 (H_m) \quad \mbox{The activity is in the sphere  $B_m = B(v_{j_m},R), \quad 1 \leq m \leq M$,}
\]
where the centers  $v_{j_m}$ are again drawn from the uniform distribution. Observe that if  $B_m$ overlaps with one of the active BRs other than the one defining the hypothesis $(H_0)$, the Bayes factor may be low; thus, the other active regions can be seen as confounding factors in this case. However, this simulation is closer to the situation with the real data in which the number of active regions of interest is unknown.

\subsection{Model settings}

\subsubsection{Source space, brain regions, and simulated data}

\begin{figure}[tbh]
\centerline{
\includegraphics[width=14cm]{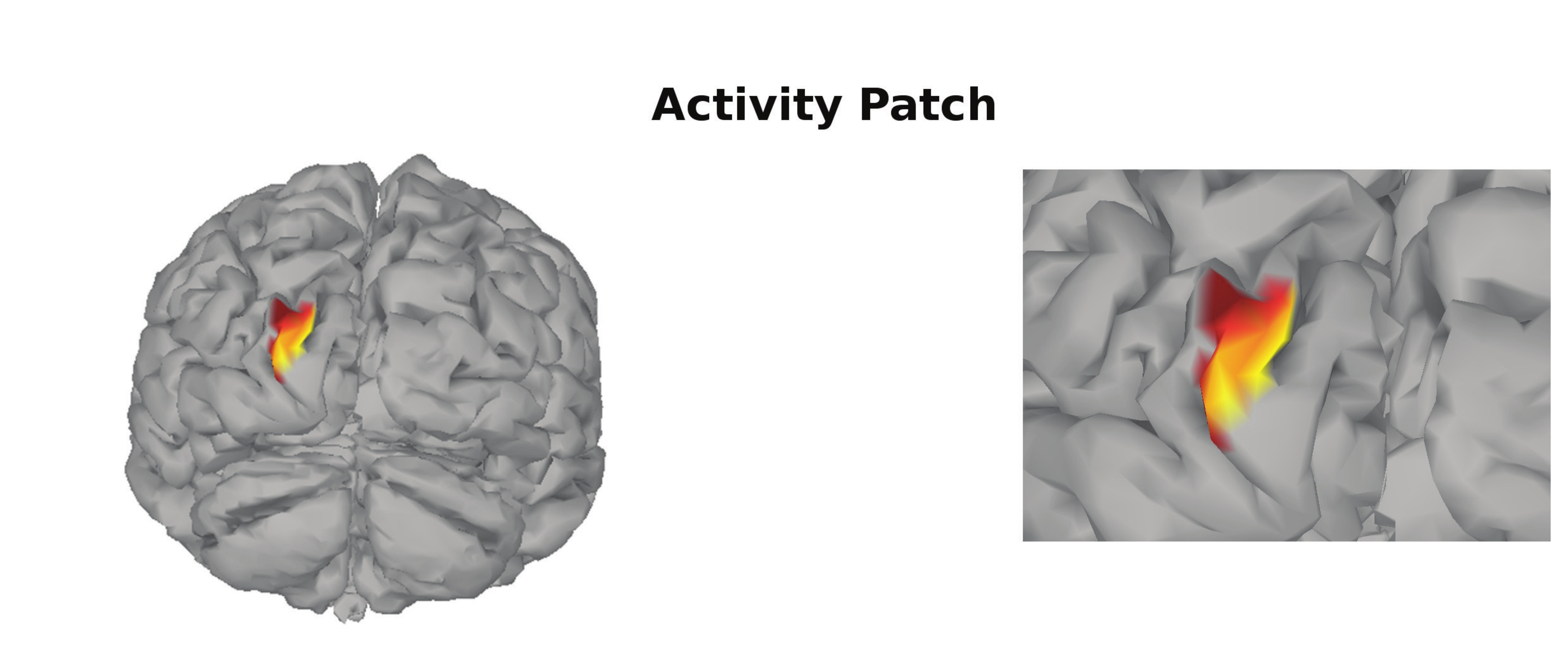}}
\caption{\label{fig:patch_LO_activity} A randomly generated patch of activity in the left lateral
occipital cortex region of the brain.}
\end{figure}

\begin{figure}[tbh]
\centerline{
\includegraphics[width=14cm]{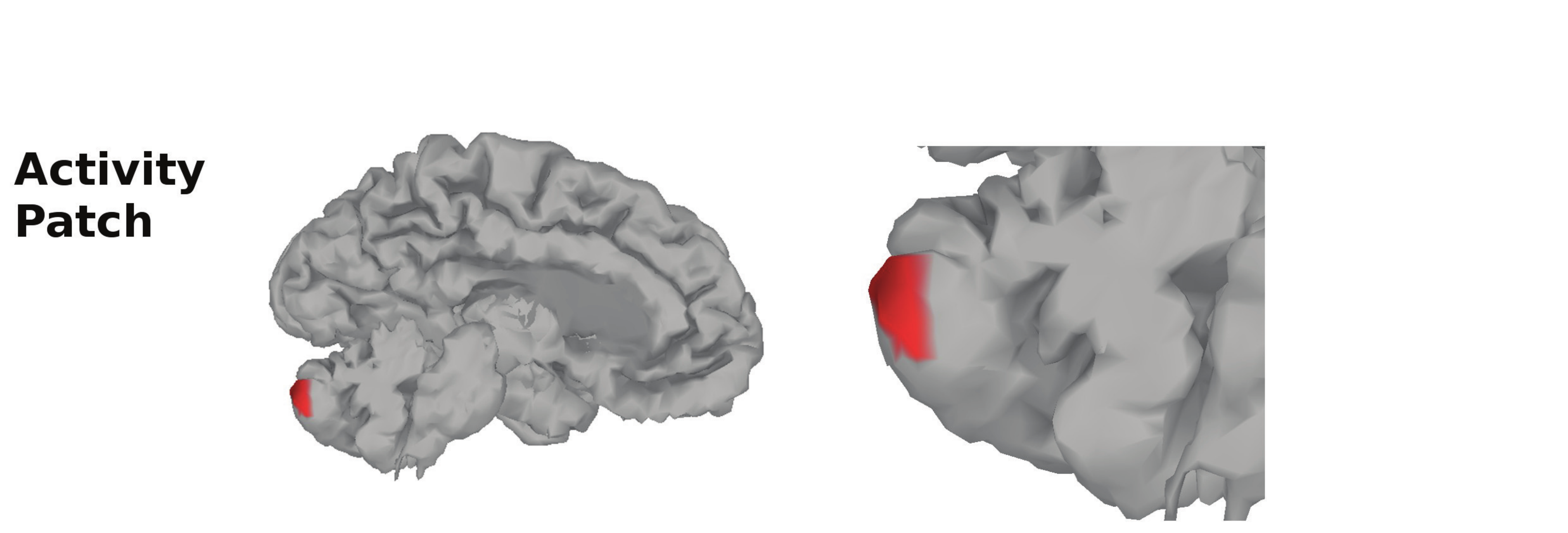}}
\caption{\label{fig:patch_cerebellum_activity}  A randomly generated patch of activity in right cerebellum.}
\end{figure}

We generate the synthetic data using the anatomical and geometric information provided by the MEG-SIM web portal \cite{megsim}. The MRI data provided in the portal were segmented with Freesurfer \cite{Dale1999} and imported in Brainstorm \cite{Brainstorm} to generate a source space including both the cortical surface and the substructure regions. 
The source space obtained in this manner consists of $27\,000$ nodes
identifying possible dipole locations in the gray matter. A randomly generated
patch of activity in the left lateral occipital cortex is shown in
Figure~\ref{fig:patch_LO_activity} and a randomly generated patch of activity
in the right cerebellum is shown in 
Figure~\ref{fig:patch_cerebellum_activity}. To avoid the overly optimistic
results that typically result from testing inversion algorithms using the same model for data generation and inverse solver, we solved the inverse problem of a sparser mesh of $n = 20\,000$ nodes obtained through independent sampling of the  cortical and subcortical structure.

The labeling of the brain regions in the cortical surface follows the Desikan-Killiany Cortical Atlas \cite{Desikan} consisting of $34$ cortical anatomical regions in each hemisphere, to which we added $8$ bilateral structures (accumbens, amygdala, caudate, hippocampus, pallidum, putamen, thalamus, cerebellum) and $1$ central structure (brainstem). Thus the atlas we used in the simulations consists of a total of $85$ brain regions. \\
\\
The data acquisition geometry corresponds to the 306-channel Elekta Neuromag device and the lead field matrix is computed using the single layer model implemented in the OpenMEEG \cite{OpenMEEG, OpenMEEG_2010} software provided in the Brainstorm package.

\subsubsection{Real and realistic data}\label{sec:real data}
\begin{figure}[tbh]
\centerline{
\includegraphics[width=14cm]{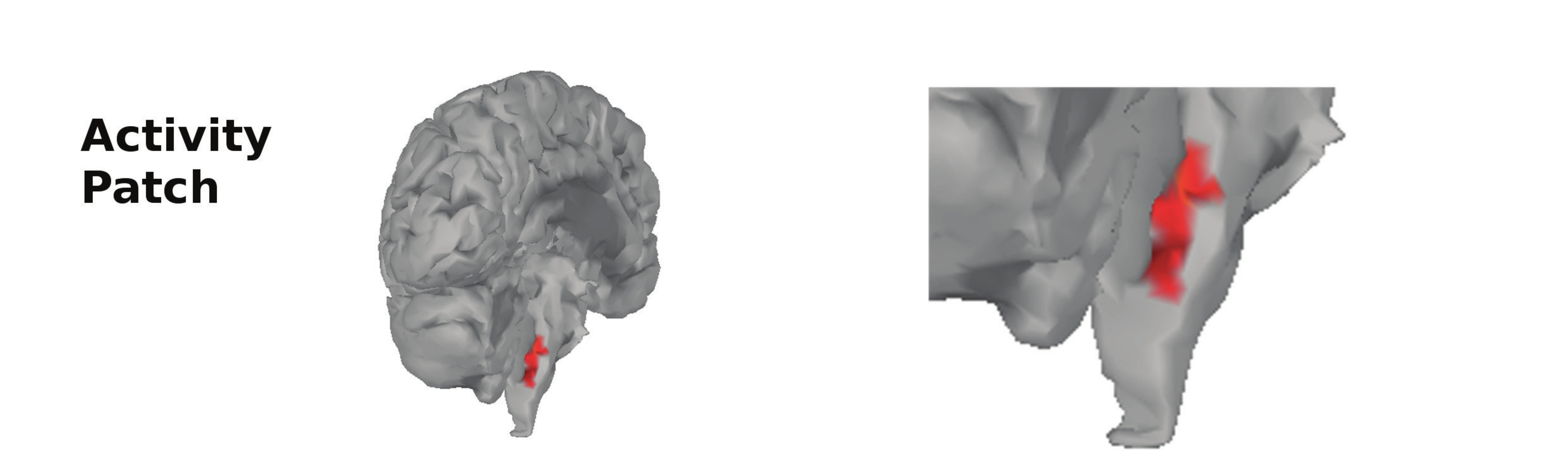}}
\caption{\label{fig:patch_brainstem_activity}The randomly generated patch of activity in the brainstem.}
\end{figure}

The real data set used for our computed examples is the MEG sample data acquired with the Neuromag Vectorview system at MGH/HMS/MIT Athinoula A. Martinos Center Biomedical Imaging and made available, together with the MRI reconstructions created with FreeSurfer, in the MNE software package
\cite{MNE}. As part of the protocol for the data collection, checkerboard patterns were presented into the left and right visual field, interspersed by
tones to the left or right ear. The interval between the stimuli was 750 ms. Occasionally a smiley face was presented at the center of the visual field. The
subject was asked to press a key with the right index finger as soon as possible
after the appearance of the face \cite{MNE}.  In our computed experiments we
only consider the trials corresponding to the left-ear auditory stimulus and
perform the averaging on these trials. To test how the inverse methods
accurately recover activity in the deep regions, we added to the real data set
described above the magnetic field evoked by a patch of $20$ active dipoles in
the brainstem, shown in Figure~\ref{fig:patch_brainstem_activity} at time
$T=5$ms, whose time series follow a Gaussian distribution peaked at $T=5$ms with standard deviation $\sigma=2$; see \cite{Lauri2009} for details.
The source space containing both cortical regions and subcortical structures
after discretization comprises $22019$ vertices.

\subsection{Setting of IAS parameters}

As shown in Sections 2.1-2.2, in the IAS algorithm the user have to select the three parameters $\beta$, $\theta_{\rm max}$, $\overline k$. 
Moreover, one has to set the value of $\delta$ in the local anatomical prior matrix ${\mathsf C}_j$. 
The choice of all the four parameters can be driven by their physiological interpretation. 

The hyperparameter $\beta > \frac 52$ control the focality of the reconstructed sources; when $\beta$ approaches the limit value $\frac 52$, the reconstructed brain activity 
becomes more and more focal. Let us define $\eta = \beta-\frac52$.
Several numerical simulations we performed, show that reasonable values of $\eta$ are between 0.1 and 0.001 (cf. \cite{IAS}). 
In the following tests we assume focal sources and set $\eta = 0.001$.

The parameter $\theta_{\rm max}$ is used to avoid unrealistic high values of the dipole variances. Even if reliable maximum value can be found in the literature, 
see, e.g., \cite{mosher1993}, an heuristic criterion is to truncate the values of the variance retaining just those values below the threshold 
$\theta_{\rm max} = 0.9 \max_j (\theta_j)$. The goodness of this choice has extensively tested in several simulations (cf. \cite{IAS}).

The parameter $\overline k$ is related to the number of sources we expect to be active. A direct computation shows that 
$\xi = \sum_{j=1}^n p_k/k$ assumes values between 0.5 and 0.2 when $\overline k$ varies from 1 to 10, so that a value of $\xi$ in this interval
is a good choice if we do not have any prior information on the number of active sources.
If we assume that only one source is active, we can set $\xi=1$ (see the Appendix).
Even if in the case of synthetic data we know exactly the number of active sources, we set $\xi=1$ in all the tests we performed to show that
this choice is a good initial guess for both single and multiple source distributions. 

The last parameter we have to set is $\delta$ that takes into account the inaccuracy of the segmentation of 
the cortical/subcortical surface normals. If the directions of the normals would be exact, $\delta$ would be set equal to zero; in the realistic case the anatomical data extracted from
the MRI are not sufficiently accurate so that the normal can be considered a
preferred direction.  To allow the dipoles to follow quasi-normal orientation, we choose a small value of  $\delta$. Several numerical tests show that a good choice is $\delta=0.05$   (cf. \cite{IAS}).

The tests in the following section show the effectiveness of our settings.

\section{Results}

\subsection{Synthetic data tests}
In this section, we systematically test the sensitivity and specificity of the IAS MEG inverse solver algorithm using the validation tools described in the previous section with synthetic data sets. For the sake of comparison, we run the same tests with other three standard MEG solvers, the {weighted Minimum Norm Estimate} (wMNE) \cite{Lin,MNE}, the dynamic Statistical Parametric Mapping (dSPM) \cite{Dale2000} and sLORETA \cite{Pasqual}.

\subsection{Sensitivity measured via BR-ALI maps: synthetic data, single activity patch}

\begin{figure}[tbh]
\centerline{
\includegraphics[width=14cm]{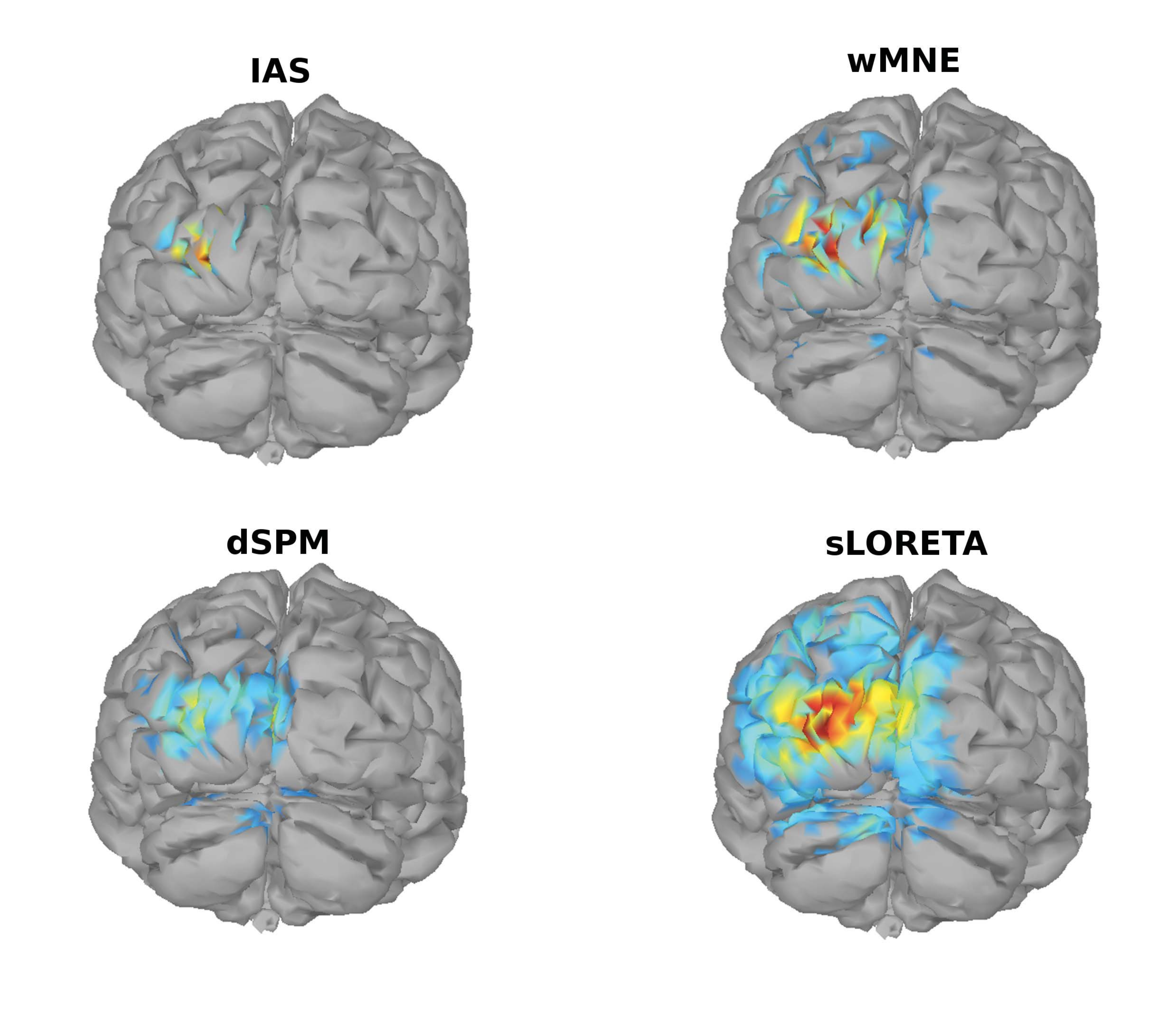}}
\caption{\label{fig:patch_LO} Reconstructions obtained with, clockwise from the top left, IAS,
	wMNE, dSPM and sLORETA, respectively, from the signal generated by the
activated patch in the left lateral occipital region shown in Figure~\ref{fig:patch_LO_activity}.}
\end{figure}

\begin{figure}[tbh]
\centerline{
\includegraphics[width=14cm]{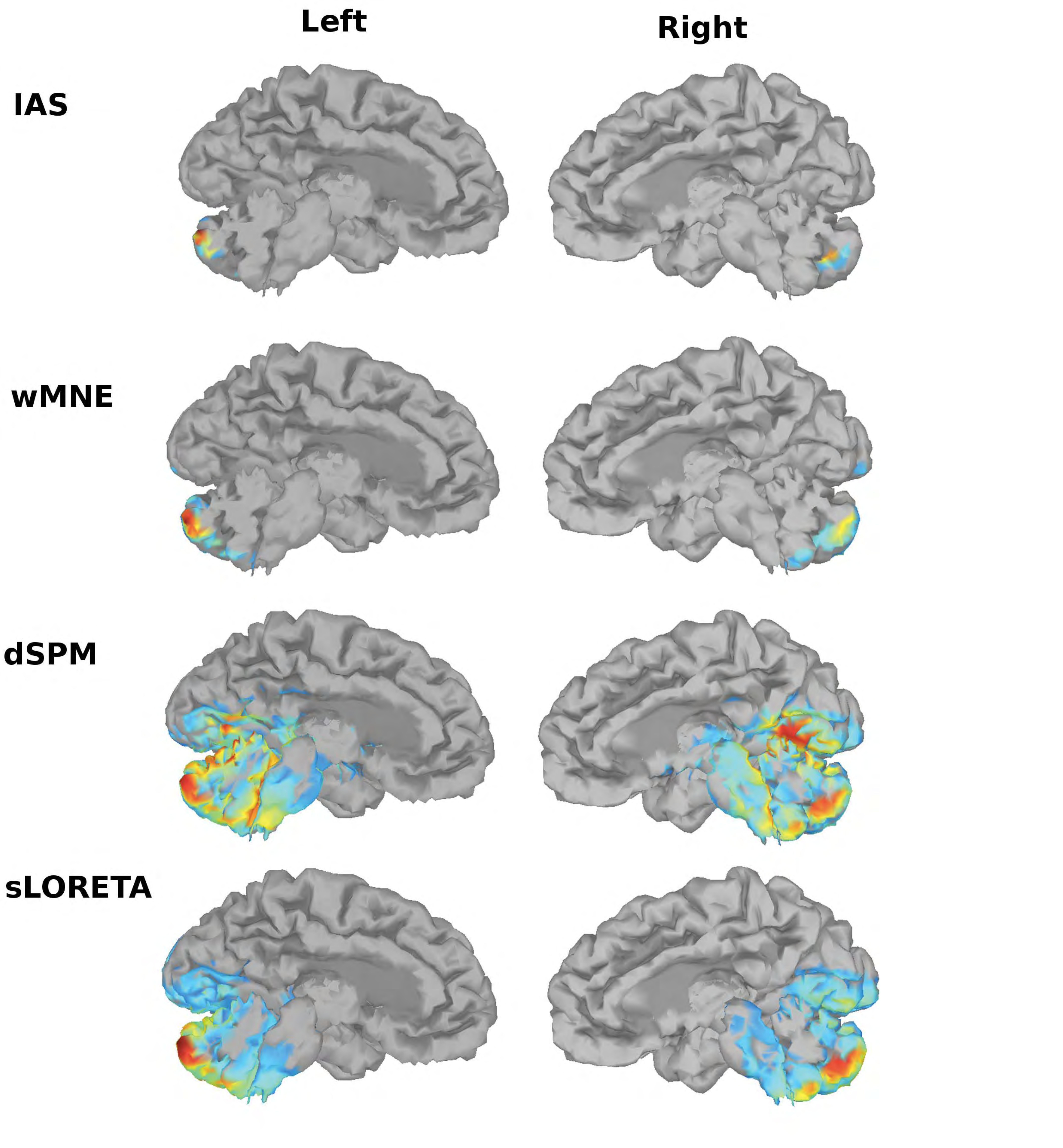}}
\caption{\label{fig:patch_Cerebellum}Reconstructions obtained with, clockwise from the top left, IAS,
wMNE, dSPM and sLORETA, respectively, from the signal generated by the
activated patch in the right cerebellum shown in Figure~\ref{fig:patch_cerebellum_activity}.}
\end{figure}

The first suite of computed experiments is designed to assess whether and how the location of the BR where the activity occurs affects the quality of the IAS reconstruction. In order to provide a measure that is robust over patches with different anatomical characteristics, for each one of the 85 BRs specified by the selected atlas, we generate a sample of $K = 100$ patch activities by the random process described in Section~\ref{sec:patches}, calculate the magnetic field measured by the sensors, and add to the latter a realization of low amplitude noise to achieve a signal to noise ratio ${\rm SNR}$ approximately equal to $15$. For each sample problem, we reconstruct the activity $\widehat a$ with the IAS algorithm described in Section~\ref{sec:solver}, as well as with the wMNE, DSPM and sLORETA algorithms as implemented in Brainstorm. The results of the reconstructions are summarized in the average BR-ALI vectors and visualized in the form of histograms with whiskers indicating the marginal standard deviation around the mean.
An example of activity reconstruction provided by the different methods for
the patch activities shown in Figures~\ref{fig:patch_LO_activity},~\ref{fig:patch_cerebellum_activity} could be find in
Figures~\ref{fig:patch_LO},~\ref{fig:patch_Cerebellum}.

\begin{figure}[tbh]
\centerline{
\subfloat[\textbf{IAS}]{
\includegraphics[width=9.5cm,trim = 0.5in 0.1in 0.5in 0.3in, clip,
scale=0.5]{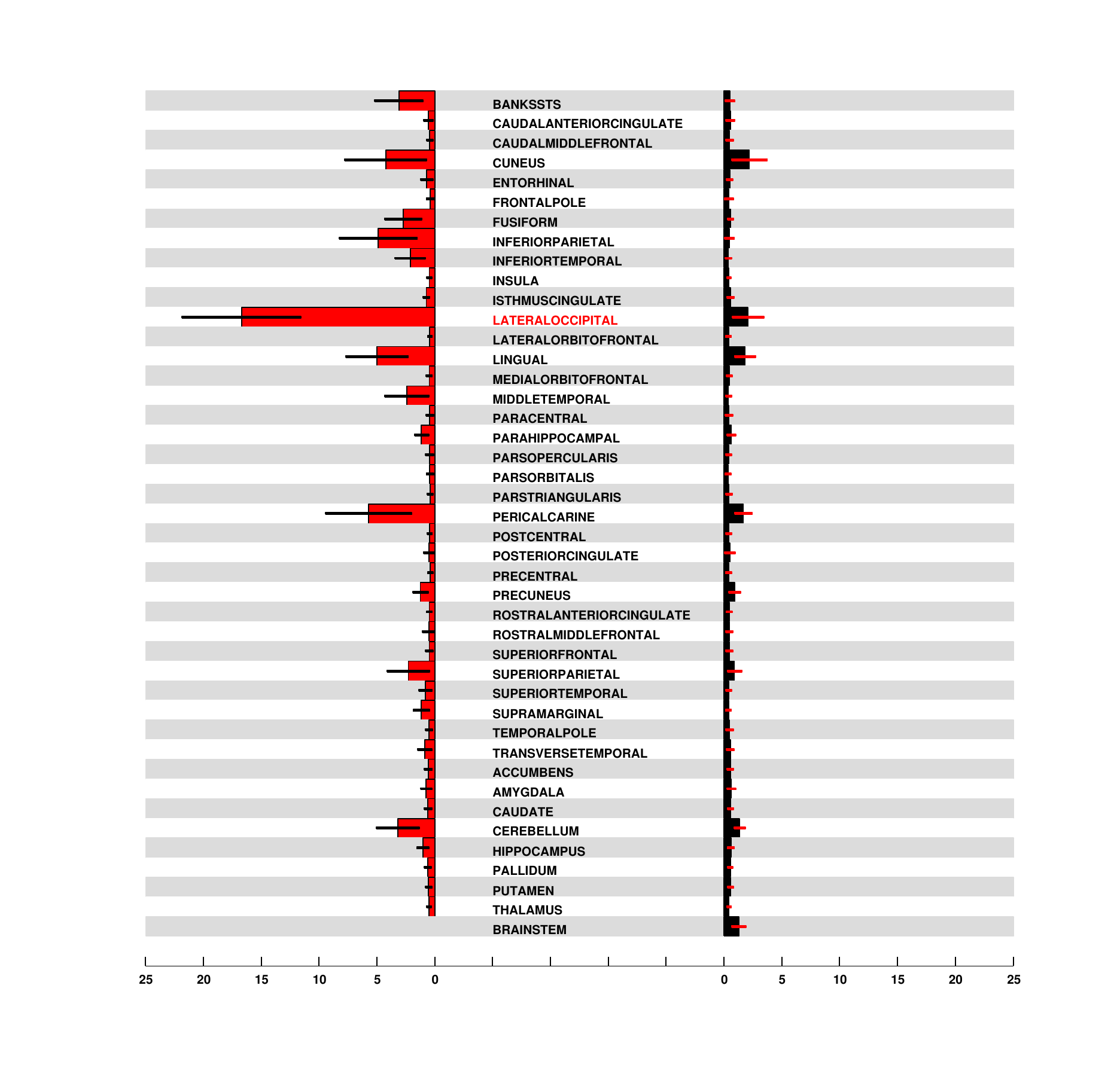}}
\subfloat[\textbf{wMNE}]{
\includegraphics[width=9.5cm,trim = 0.5in 0.1in 0.5in 0.3in, clip,
scale=0.5]{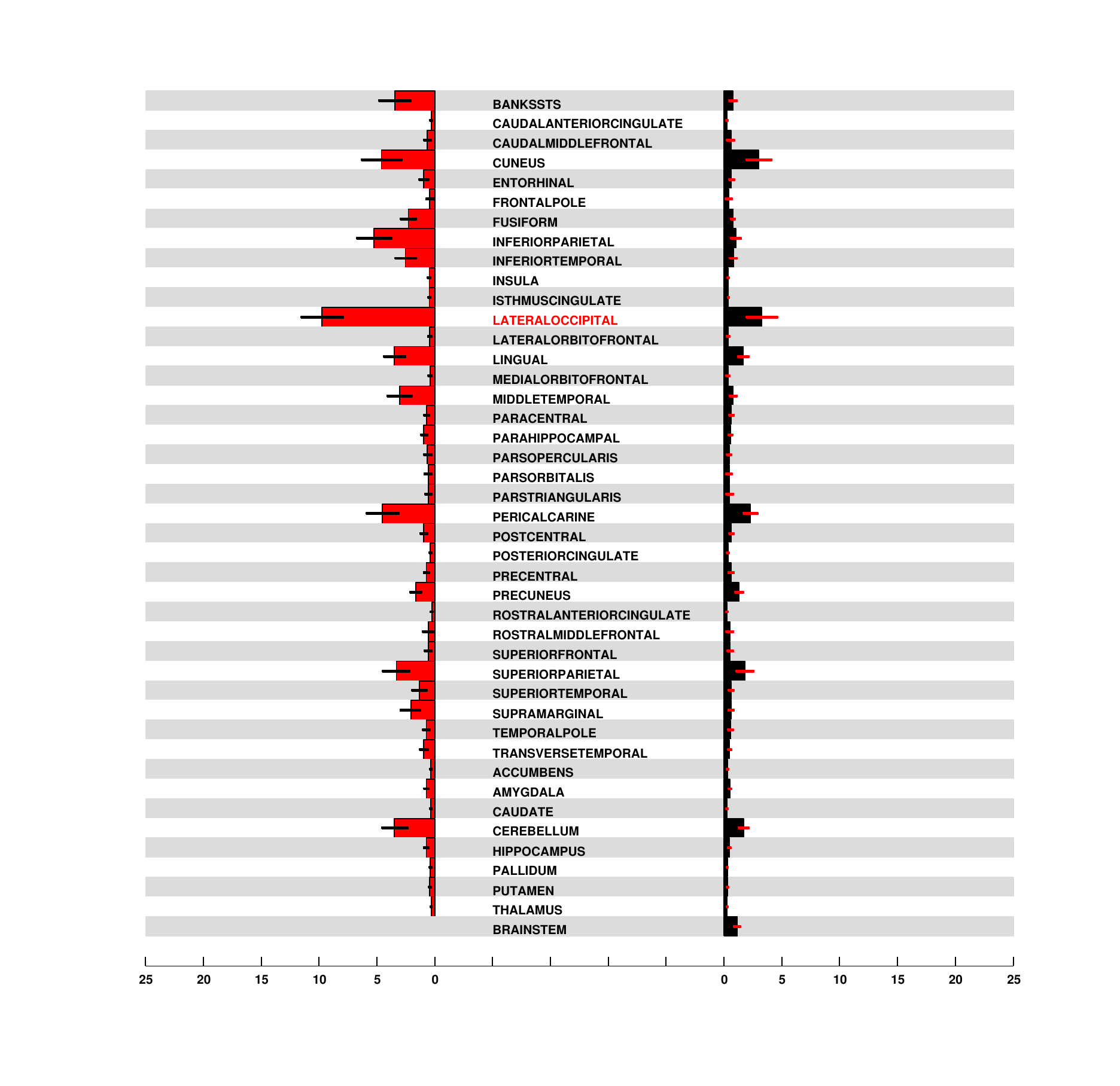}} }
\centerline{
\subfloat[\textbf{dSPM}]{
\includegraphics[width=9.5cm,trim = 0.5in 0.1in 0.5in 0.3in, clip,
scale=0.5]{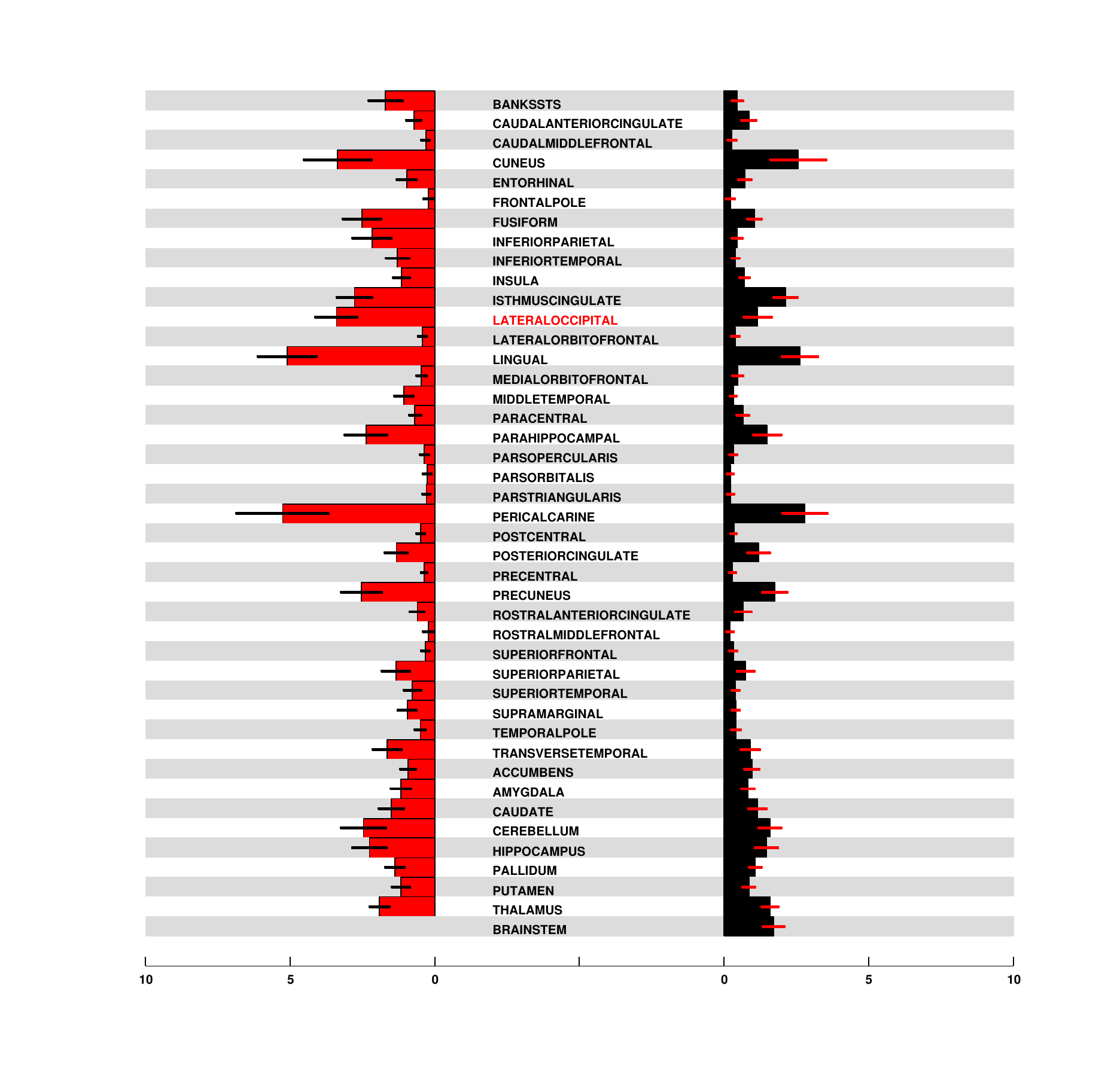}}
\subfloat[\textbf{sLORETA}]{
\includegraphics[width=9.5cm,trim = 0.5in 0.1in 0.5in 0.3in, clip,
scale=0.5]{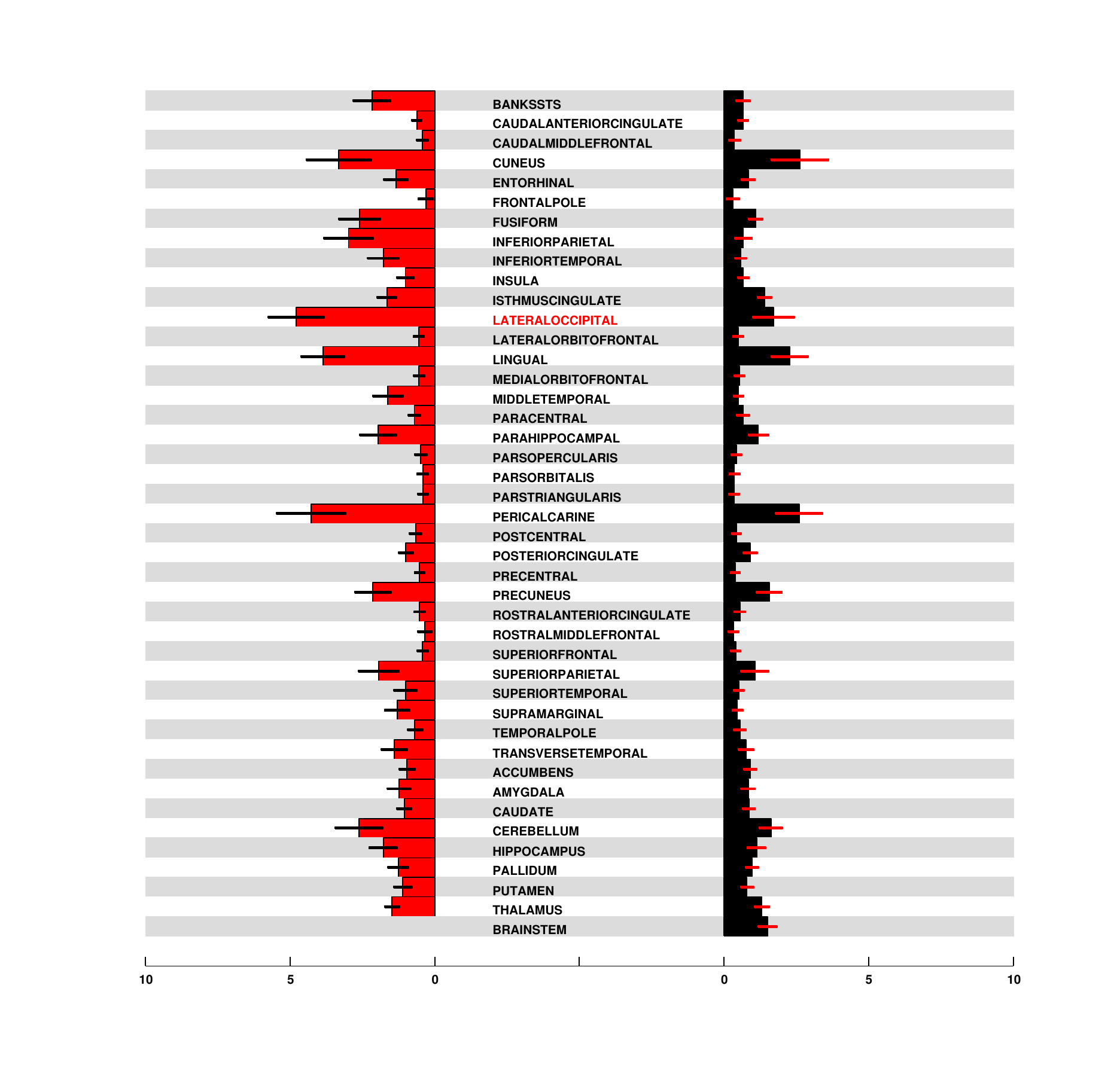}} }
\caption{\label{fig:LatOc_Baffo} Mapping of the brain activity to 85 different BRs  over 100 simulations using synthetic data corresponding to randomly generated activity patches in the {\em left lateral occipital cortex}, indicated in red in the list of the BRs reconstructed with, respectively,  IAS (a), wMNE (b),  dSPM (c) and sLORETA (d). The histograms bin the average activity in each BR: in red the BRs of the left hemisphere and in black the ones of the right hemisphere.}
\end{figure}

Figure~\ref{fig:LatOc_Baffo} shows the average BR-ALI vectors when the active patch is in the left lateral occipital cortex and the MEG inverse problem is solved with the IAS algorithm (top left), wMNE (top right), dSPM (bottom left) and sLORETA (bottom right), respectively. 
The results for the  BRs in the left hemisphere are shown in red, those for the BRs in the right hemisphere in black. 
Figure~\ref{fig:FP_Baffo} displays the histograms  relative to the case where the active patch is in the frontal pole cortical region in the right hemisphere.

\captionsetup[subfloat]{captionskip=-20pt}

\clearpage

\captionsetup[subfloat]{captionskip=-20pt}
\begin{figure}[tbh]
\centerline{
\subfloat[\textbf{IAS}]{
\includegraphics[width=9.5cm,trim = 0.5in 0.1in 0.5in 0.3in, clip,
scale=0.5]{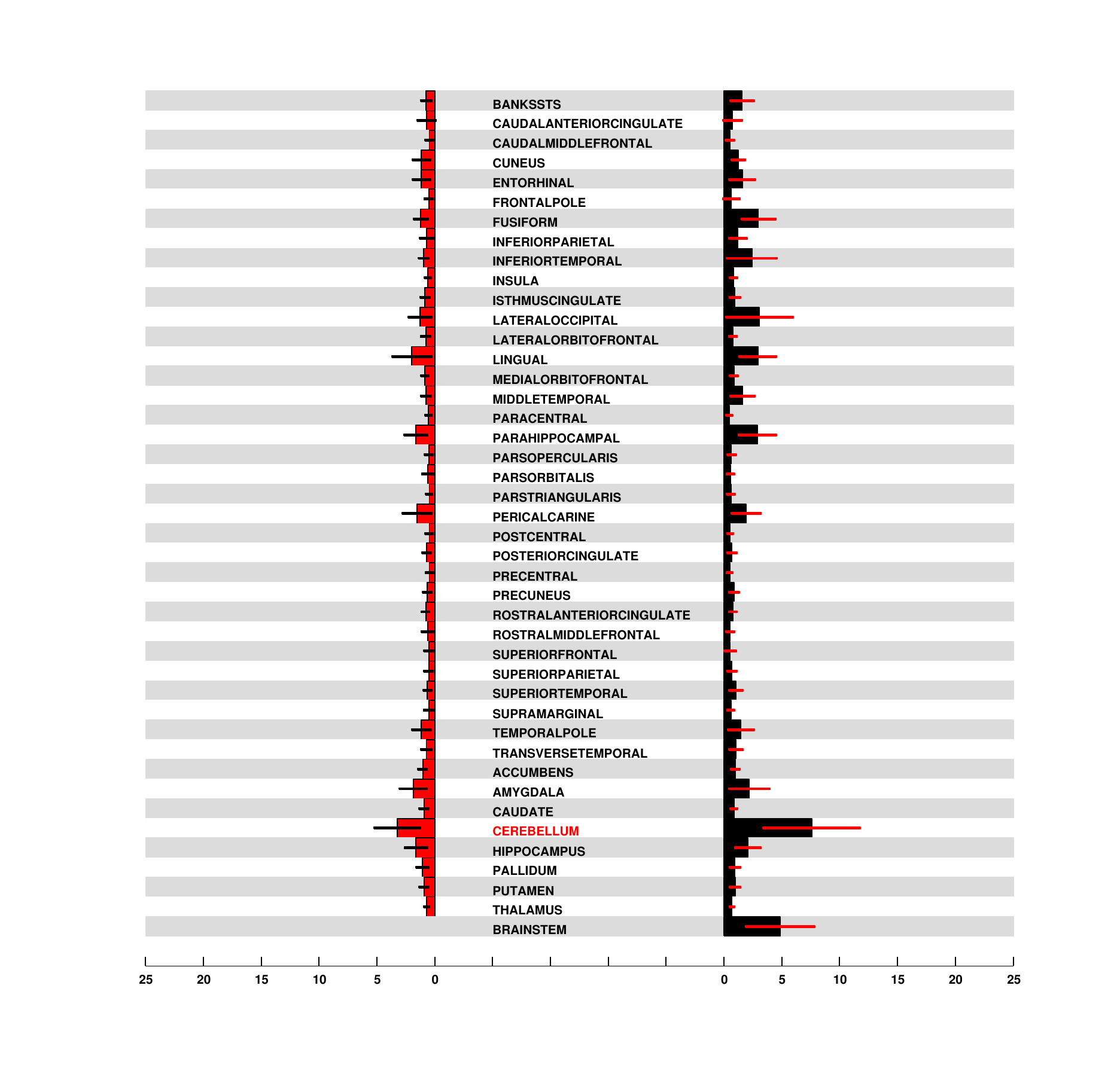}}
\subfloat[\textbf{wMNE}]{
\includegraphics[width=9.5cm,trim = 0.5in 0.1in 0.5in 0.3in, clip,
scale=0.5]{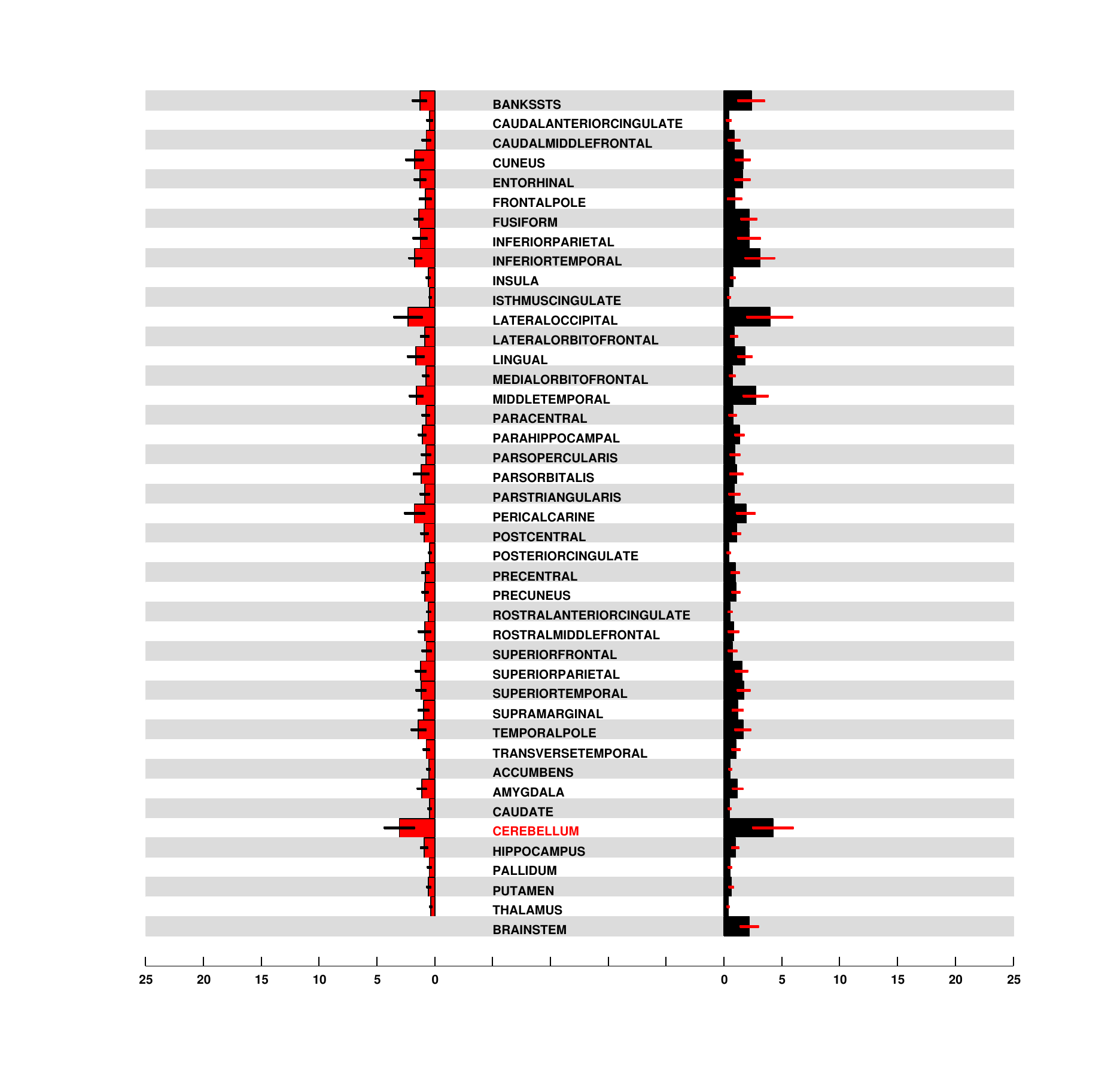}} }
\centerline{
\subfloat[\textbf{dSPM}]{
\includegraphics[width=9.5cm,trim = 0.5in 0.1in 0.5in 0.3in, clip,
scale=0.5]{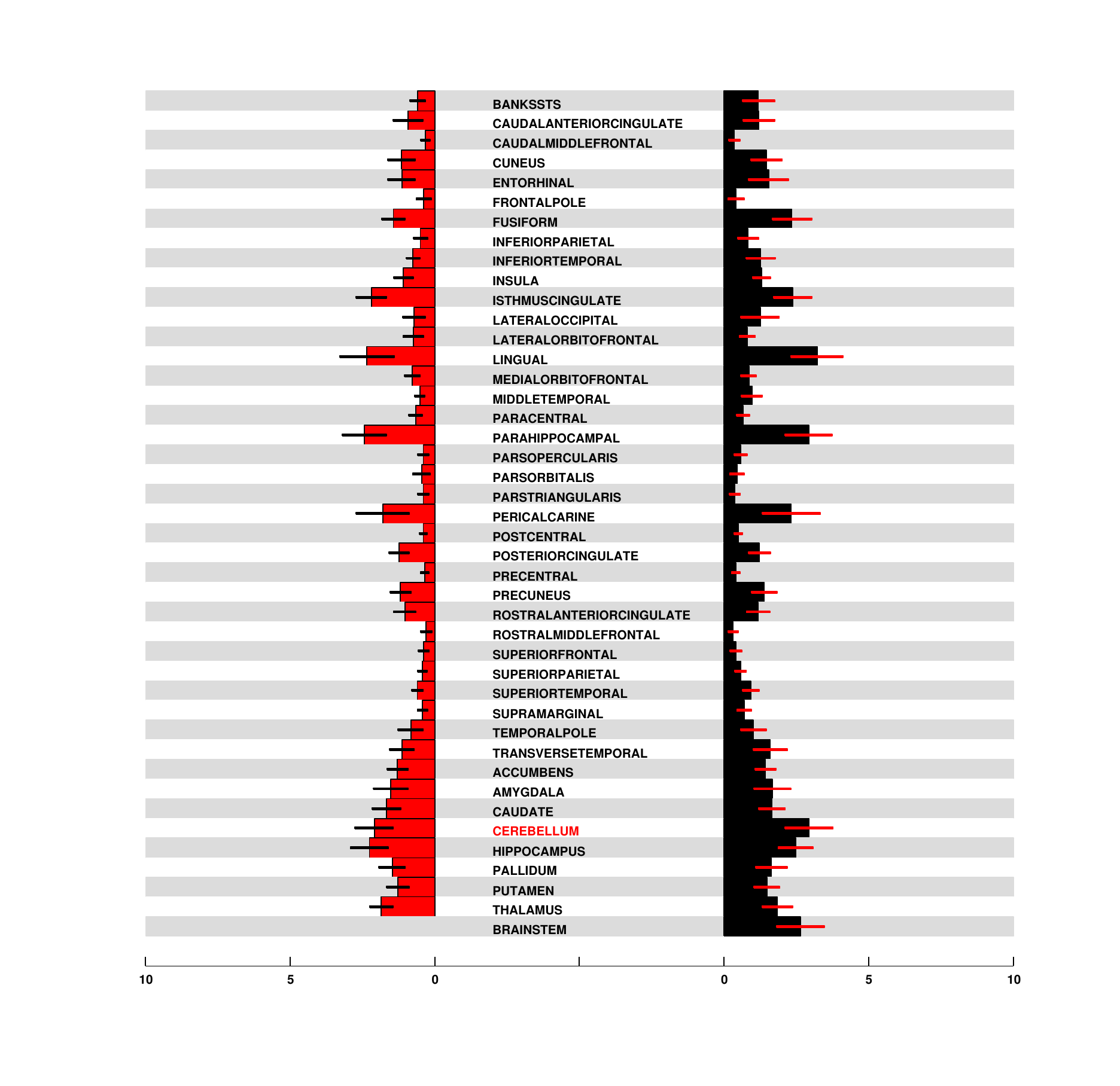}}
\subfloat[\textbf{sLORETA}]{
\includegraphics[width=9.5cm,trim = 0.5in 0.1in 0.5in 0.3in, clip,
scale=0.5]{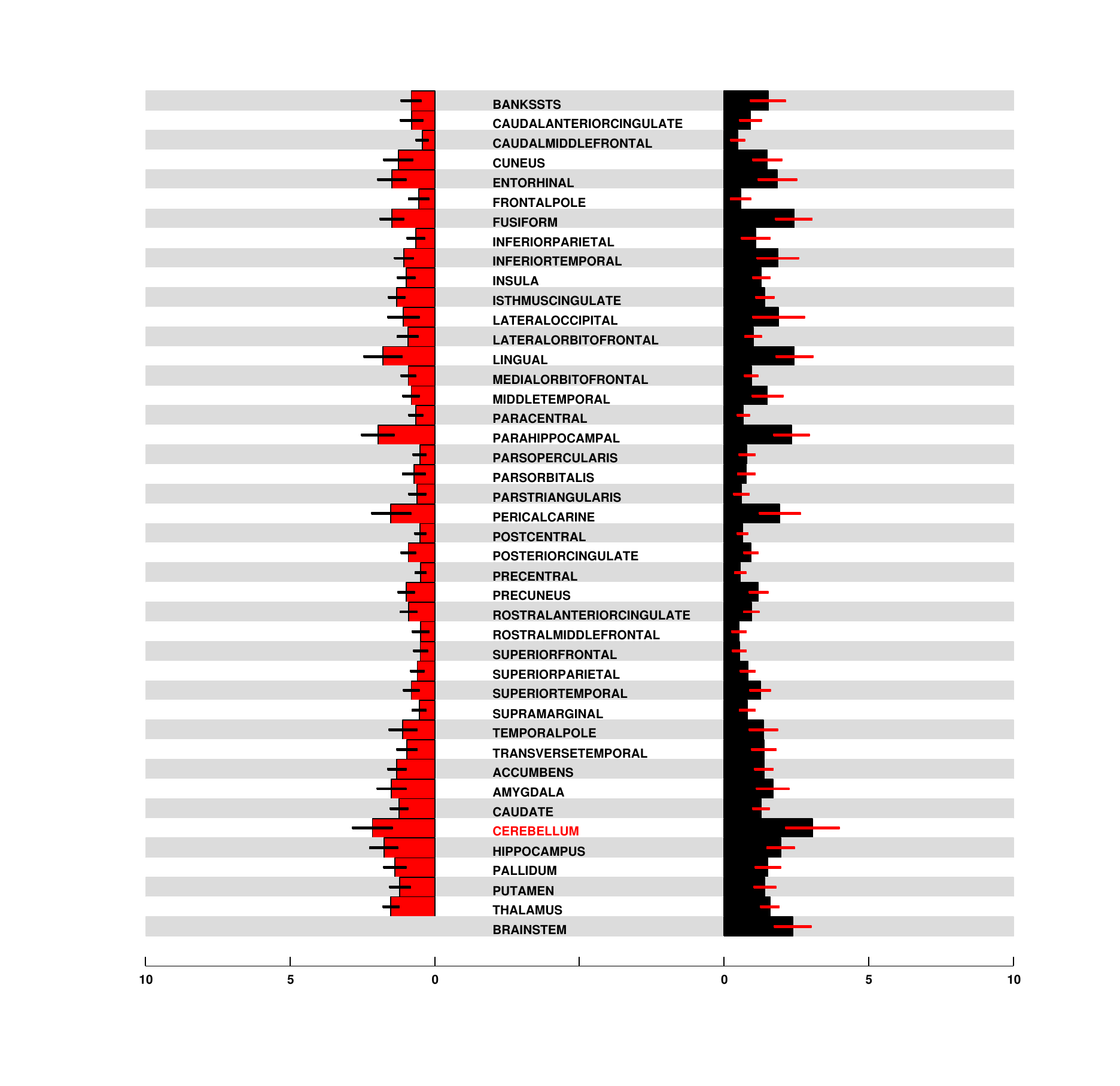}} }
\caption{\label{fig:Cereb_Baffo} Mapping of the brain activity to 85 different BRs  over 100 simulations using synthetic data corresponding to randomly generated activity patches in the {\em right cerebellum}, indicated in red in the list of the BRs reconstructed with, respectively,  IAS (a), wMNE (b),  dSPM (c) and sLORETA (d). The histograms bin the average activity in each BR: in red the BRs of the left hemisphere and in black the ones of the right hemisphere.}
\end{figure}

The deeper into the brain the active patch is, the more difficult we expect the mapping from the MEG data to the current dipoles to be. Figure~\ref{fig:Cereb_Baffo} displays the histograms relative to 100 simulations with the active patch located in the right portion of the cerebellum. Compared to the previous tests, the real challenge here arises from the distance of the activity region from the receivers.

The amygdala, a subcortical structure which is part of the limbic system, is believed to be involved in attentional and emotional processes and in the formation of memories. Its remote location and small dimensions add to the challenge of localizing activity confined into this region from MEG data. Figure~\ref{fig:Amyg_L_Baffo} displays the histograms for the four inverse methods relative to the case where the active patch is confined to the left amygdala.

The four reported results are a representative subset of the performance of the four different inverse solvers in cortical and subcortical brain regions.



\subsubsection{Specificity measure via Bayes factor analysis: synthetic data, single activity patch}

To assess the specificity of the IAS algorithm and to compare it to that of the
other three MEG inverse solvers considered in this study, we begin by performing a Bayes factor analysis with MEG data generated according to the procedure described in Section~\ref{sec:patches} with a single activity patch. After generating an ensemble of $K = 20$ randomly generated patches of activity restricted to a given BR, we compute the corresponding low noise synthetic data set, estimate the activity pattern with each inversion algorithm and compute the corresponding Bayes factors, thus testing the evidence supporting the correct identification of the active area in the reconstruction. For each of the 20 active patches in the sample, we draw $100$ random spheres as competing alternative to the hypothesis that the activity is in the  BR where it actually occurs, for a total of $2\,000$  Bayes factors per  BR.

\begin{figure}[H]
\centerline{
	\subfloat[\textbf{IAS (Left Hemisphere)}]{
	{\includegraphics[width=9cm]{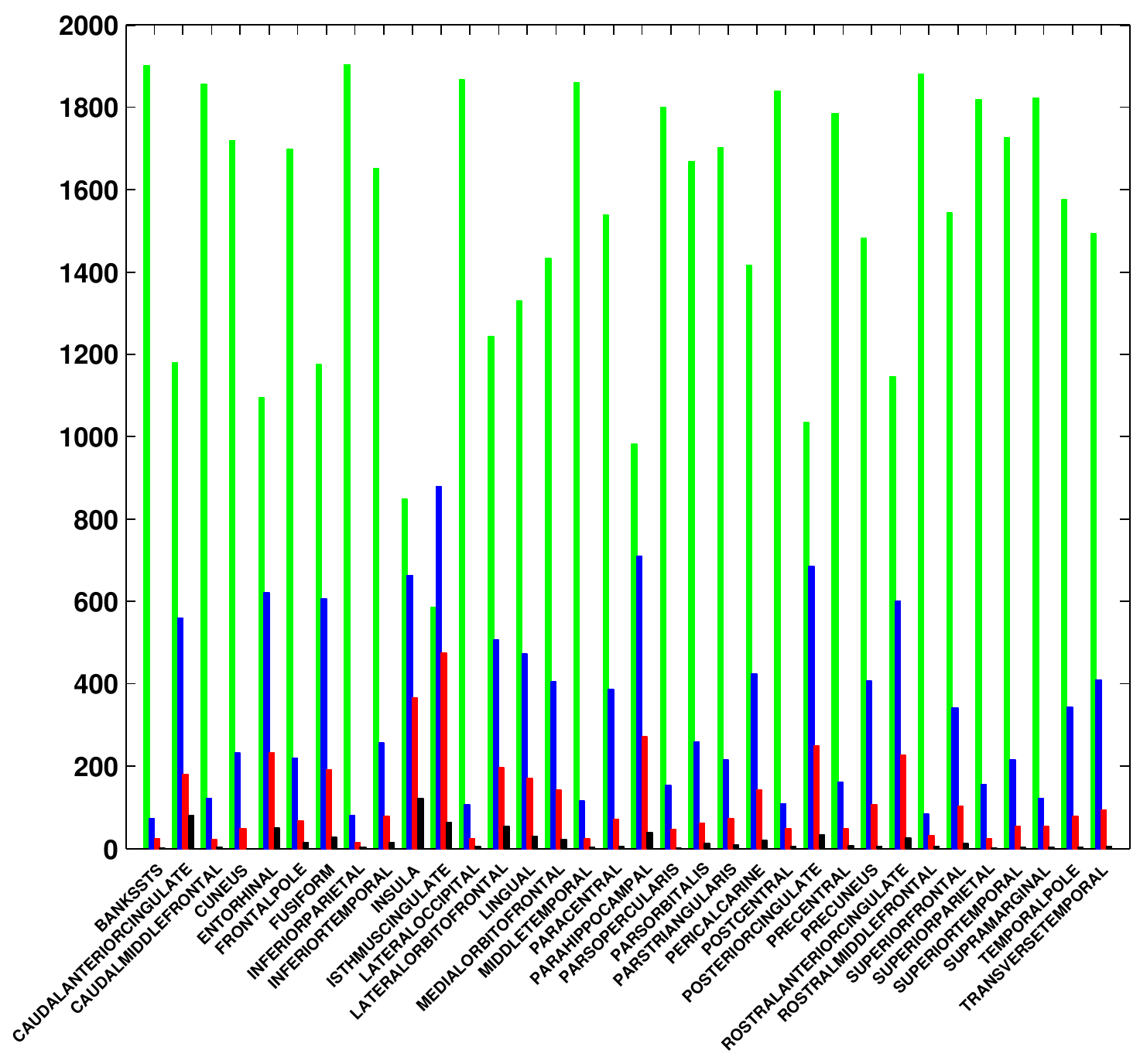}}}
	\subfloat[\textbf{wMNE (Left Hemisphere}]{
	\includegraphics[width=9cm]{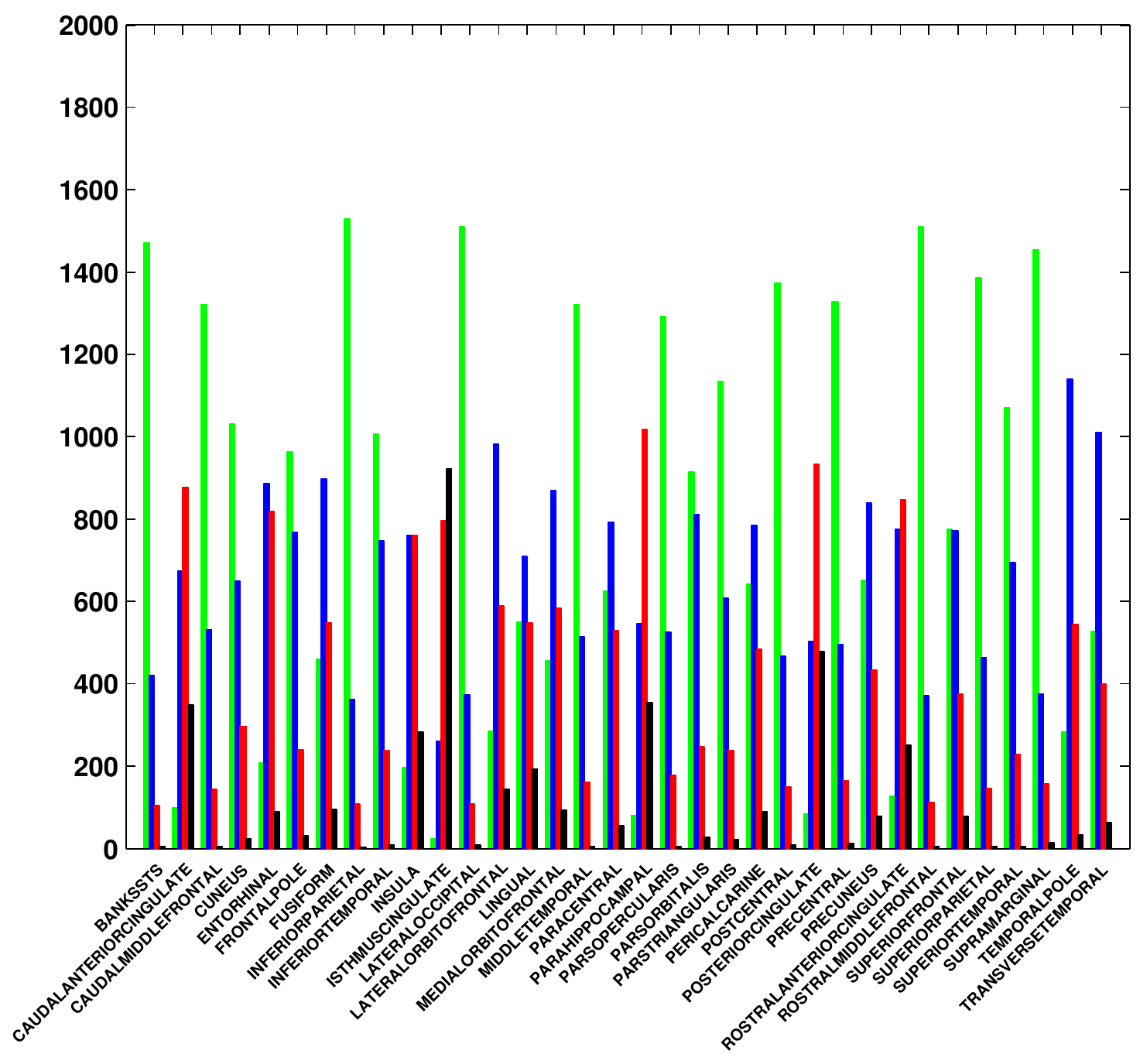}}}
\centerline{
	\hspace{.25cm}
	\includegraphics[width=8cm]{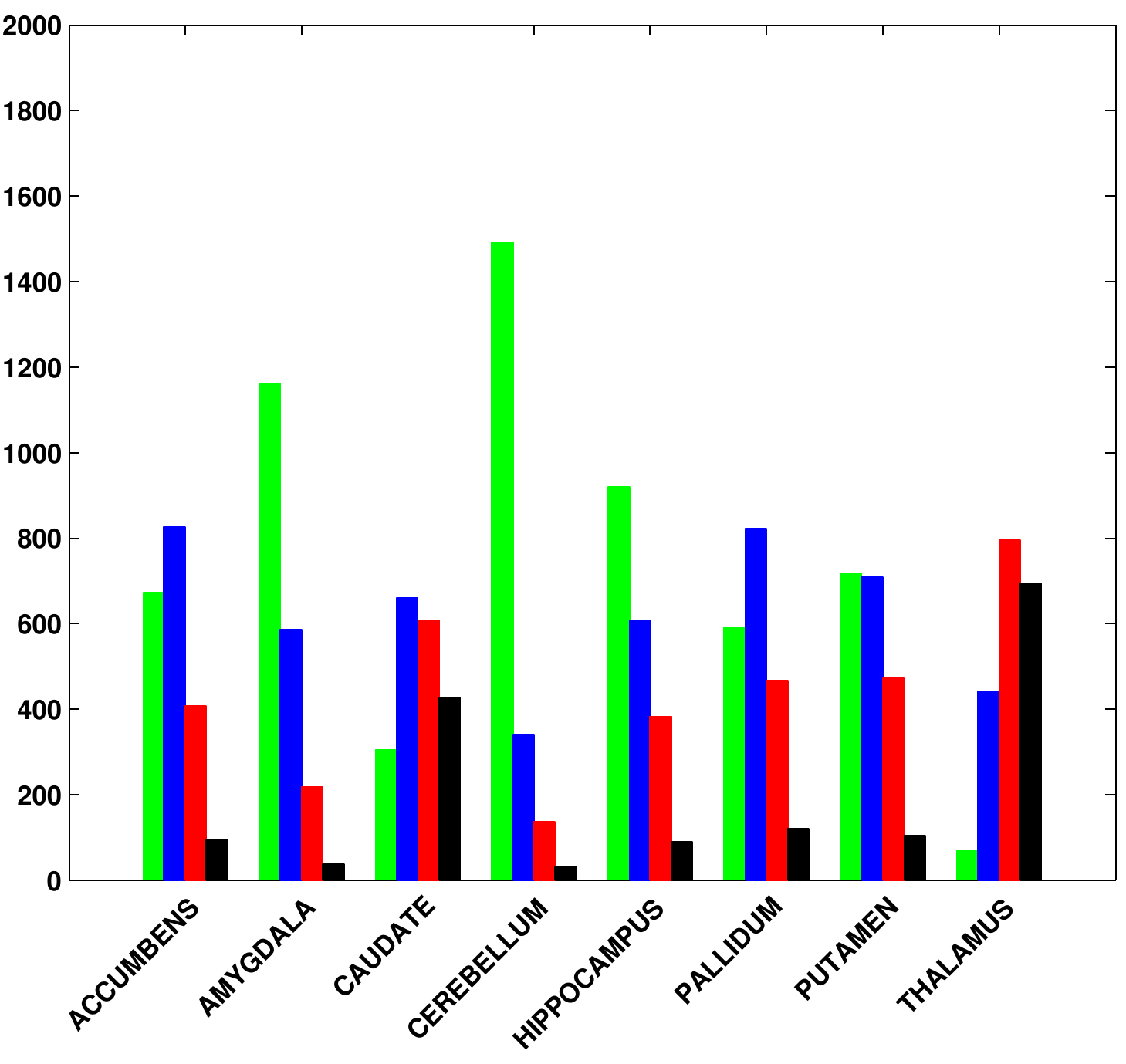}\hspace{1.25cm}
	\includegraphics[width=8cm]{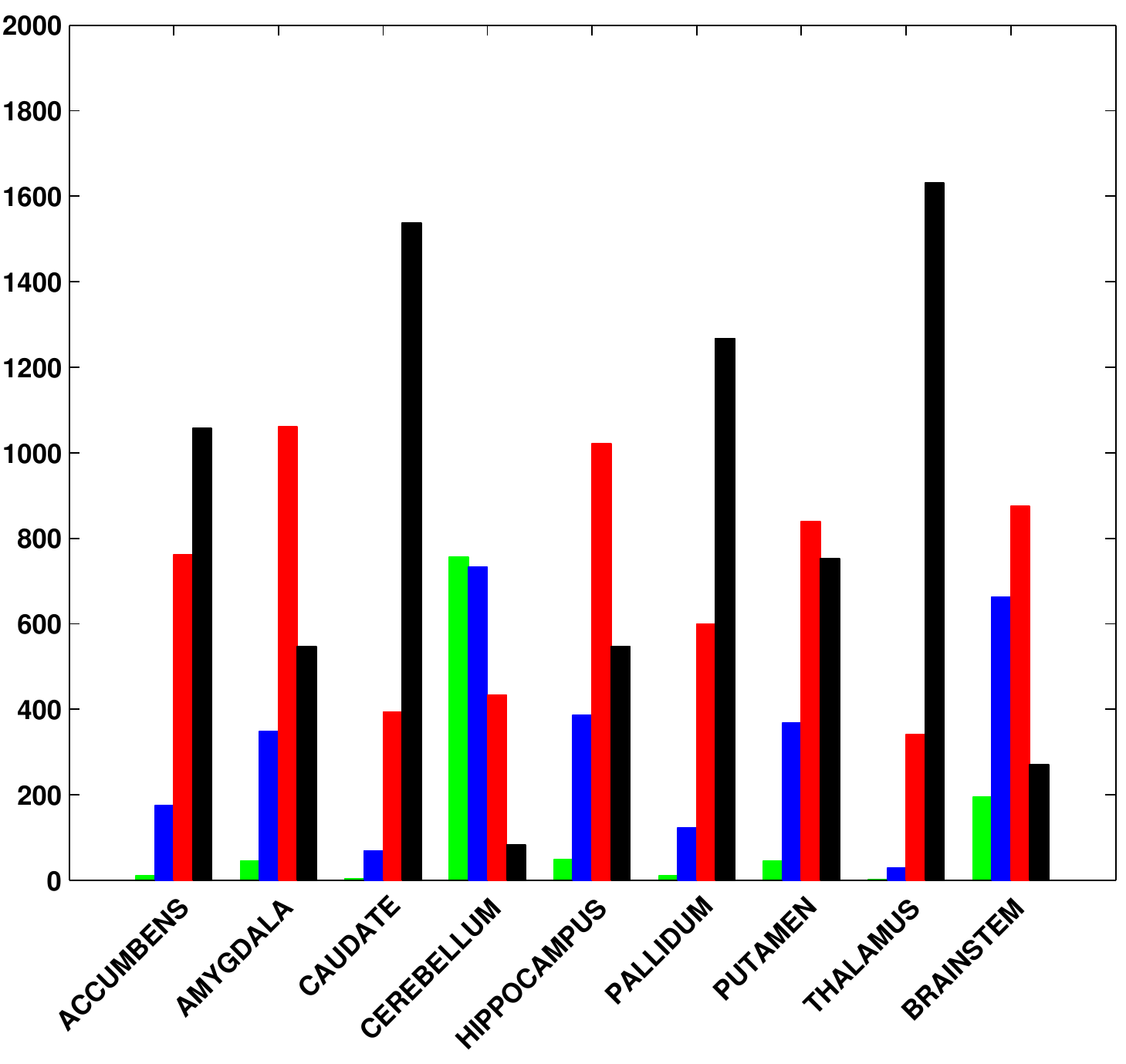}}
\caption{\label{fig:BF IAS}Histograms of the  Bayes factors for the cortical (top) and subcortical (bottom)  BRs when using the IAS algorithm (left panel) and the wMNE algorithm (right panel)
to map the MEG data to brain activity in the left hemisphere. Each histogram for each BR is the summary of 2000 Bayes factor computations color coded to indicate the different level of support of the hypothesis: green for overwhelming support, blue for strong support, red for weak support, black no support.}
\end{figure}

\begin{figure}[tbh]
\centerline{
	\subfloat[\textbf{dSPM (Left Hemisphere)}]{
	\includegraphics[width=9cm]{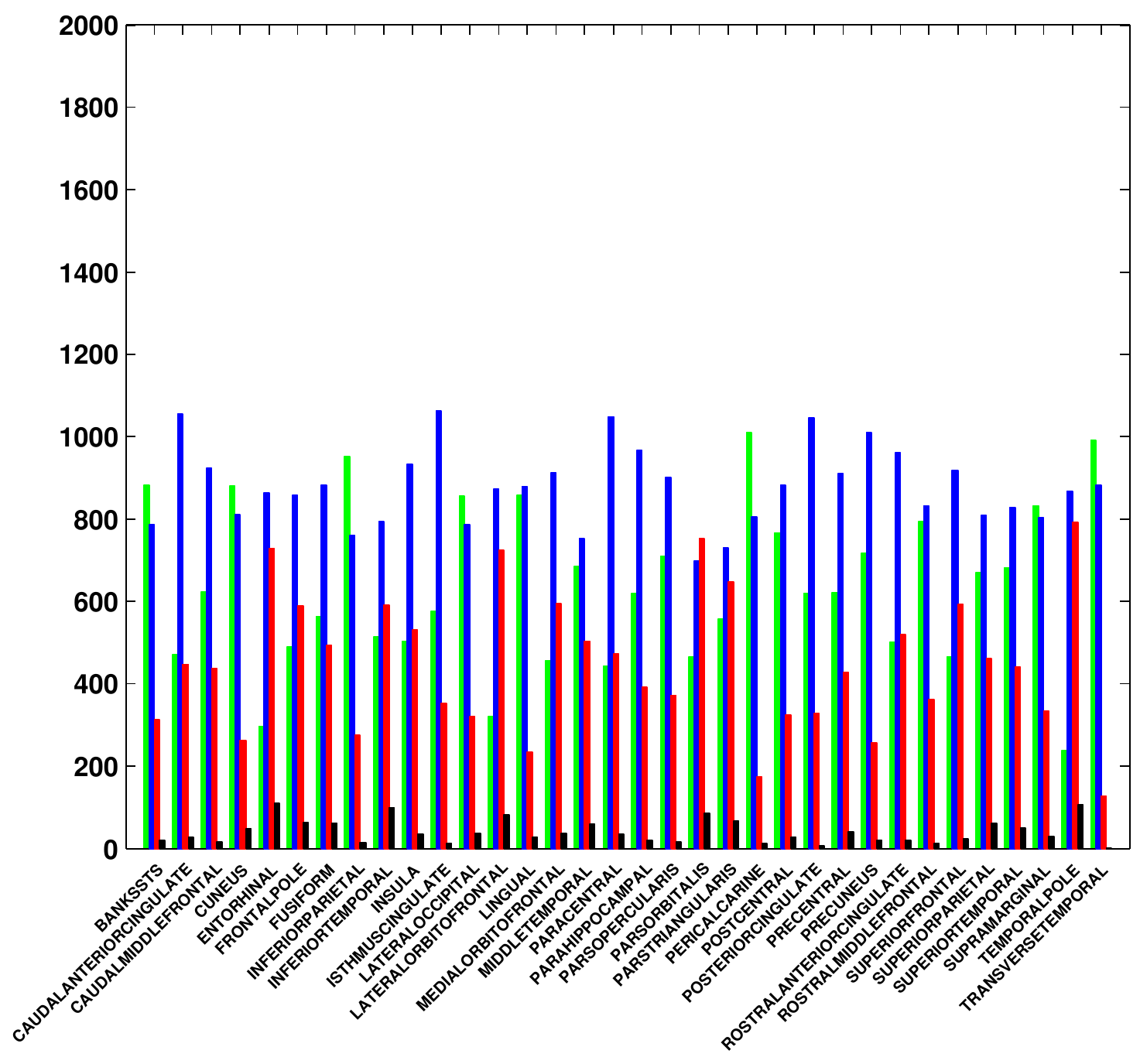}}
	\subfloat[\textbf{sLORETA (Left Hemisphere)}]{
	\includegraphics[width=9cm]{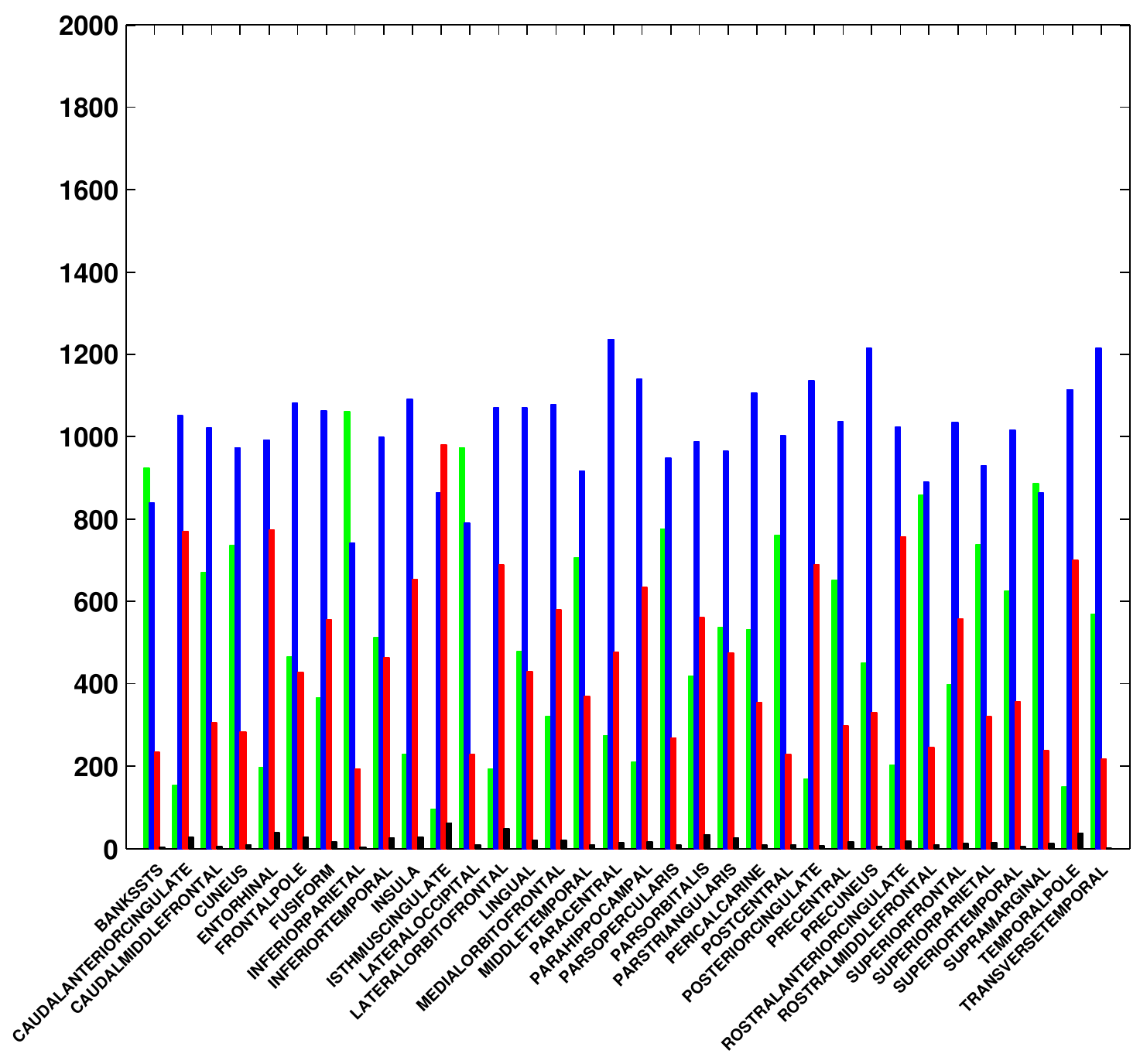}}
	}
\centerline{
	\hspace{.25cm}
	\includegraphics[width=8cm]{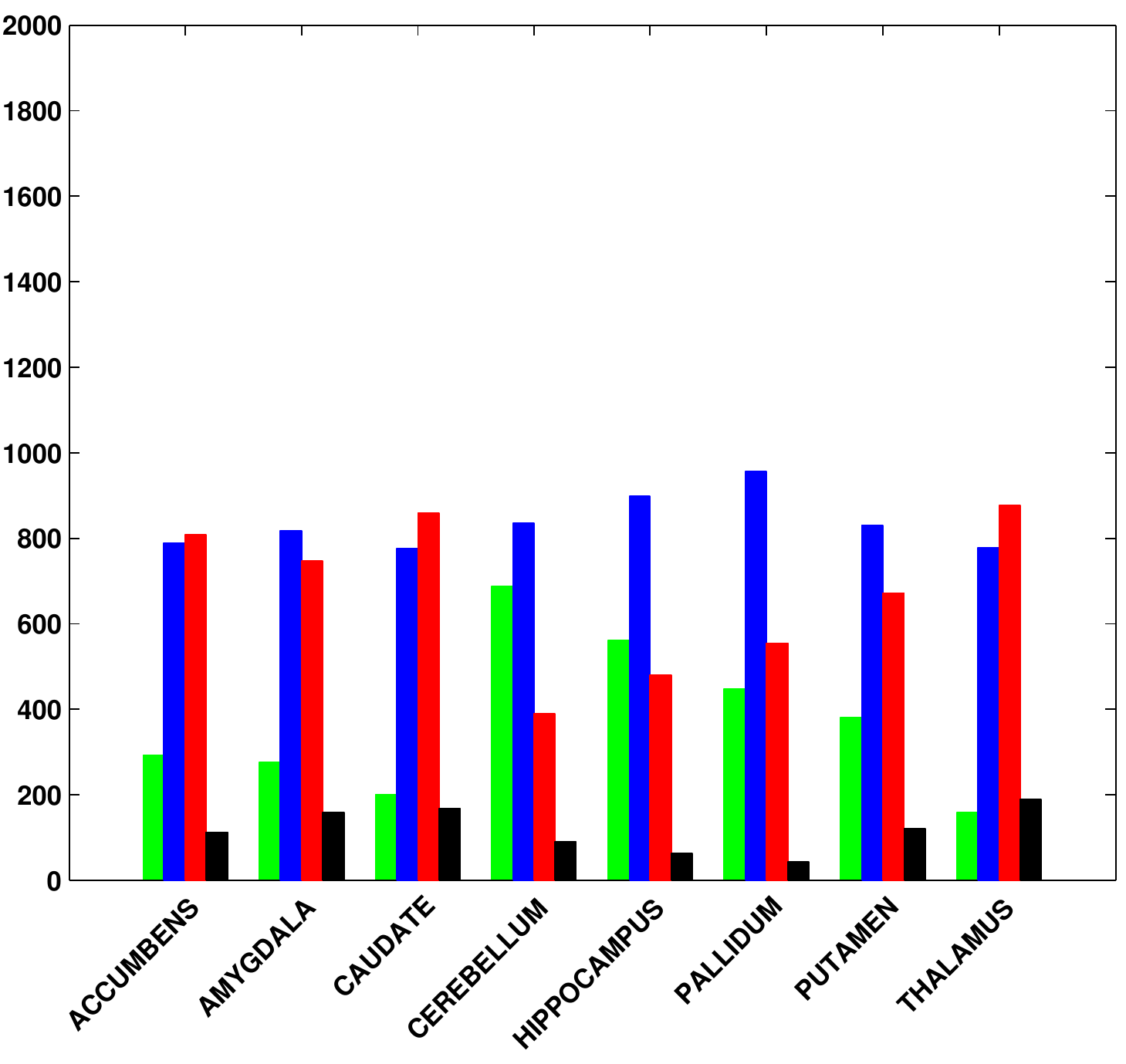}\hspace{1.25cm}
	\includegraphics[width=8cm]{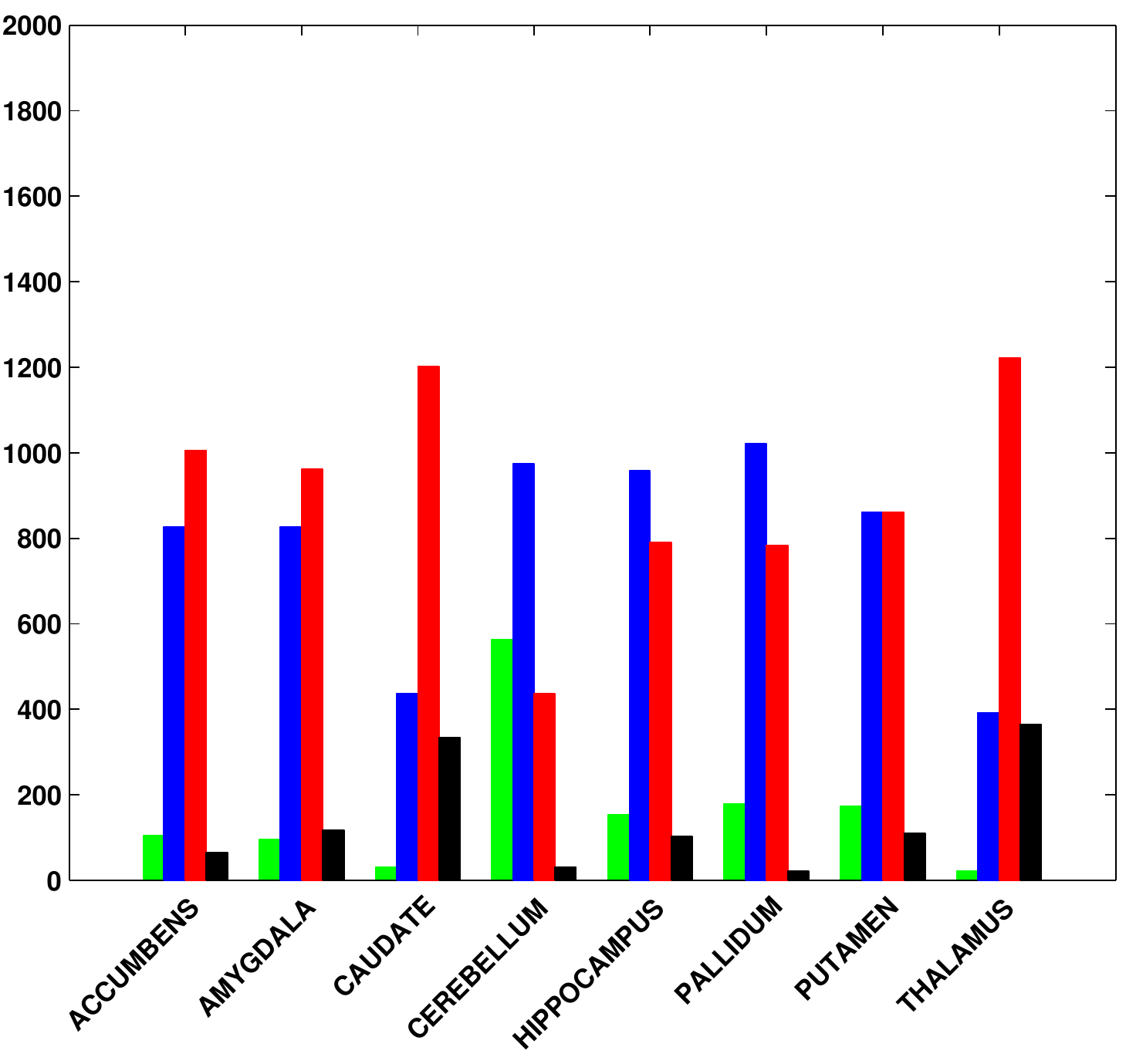}
}
\caption{\label{fig:BF DSPM} Histograms of the  Bayes factors for the cortical (top) and subcortical (bottom) BRs when using the dSPM algorithm (left panel) and the sLORETA algorithm (right panel)to map the MEG data to brain activity in the left hemisphere. Each histogram for each BR is the summary of 2000 Bayes factor computations and the color coding is as explained in Figure~\ref{fig:BF IAS}. }
\end{figure}

The summary of the Bayes factor analysis for each active BR can be represented
graphically in the form of a histogram with four bins, recording the number of
occurrence of the four levels of evidence (\ref{evidence strength}) supporting
the correct identification of the activity. The more occurrences there are in
the two top categories, the more reliable the algorithm is at correctly
identifying the area of activity.  For a easier visual assessment, we color
coded the bars indicating the numbers of times a Bayes factor falls into one of
the four categories, using green for Bayes factors greater than 10
(overwhelming support of the hypothesis that the active patch is indeed in that
BR),  blue for Bayes factors between 3 and 10 (strong support), red for Bayes
factors between 1 and 3 (weak support) and black for Bayes factor smaller than
1, in which case the support is for the hypothesis that the active patch is not
in that BR.  A prevalence of green and blue indicates a strong support of the
hypothesis that the activity has been detected in the correct region, while a
dominance of black and red suggests poor identification of active areas.
Figure~\ref{fig:BF IAS}, ~\ref{fig:BF DSPM} show the results for all cortical
and subcortical BRs included in the atlas when performing the inverse mapping with the 
four inverse solvers.

\captionsetup[subfigure]{position=top}
\captionsetup[subfloat]{captionskip=0pt}

\subsubsection{Specificity measure via Bayes factor analysis: multiple activity patches}

We extend the Bayes factor analysis to the case where several patches are active simultaneously in different BRs by performing a suite of four different tests. In the first two tests, we generate two randomly determined activity patches: In the first one, the activity patches are in the left cerebellum and the left prefrontal cortex, and in the second one, in the left amygdala and the left prefrontal cortex. In both cases one of the patches is cortical and the other one deep in the brain. In the third set of tests, the simulated data arises from three activated patches, located in the left precuneous, right precuneous and left inferior parietal cortex. In the last protocol we generate activity patches in the six different regions comprising the default mode network (DMN), namely in the left precuneus, right precuneus, left inferior parietal cortex, right inferior parietal cortex, left caudal anterior cingulate and right caudal anterior cingulate regions.

As in the case of a single active BR, we generate $K=20$ independent activity patterns in the selected BRs, compute the corresponding low noise data, solve the inverse problem with the four different algorithms and test the support of the hypothesis $H_0$ for each of the activated regions individually with $M=100$ independently drawn competing spheres $B_m$.

\begin{figure}[tbh]
\centerline{
\includegraphics[width=5cm]{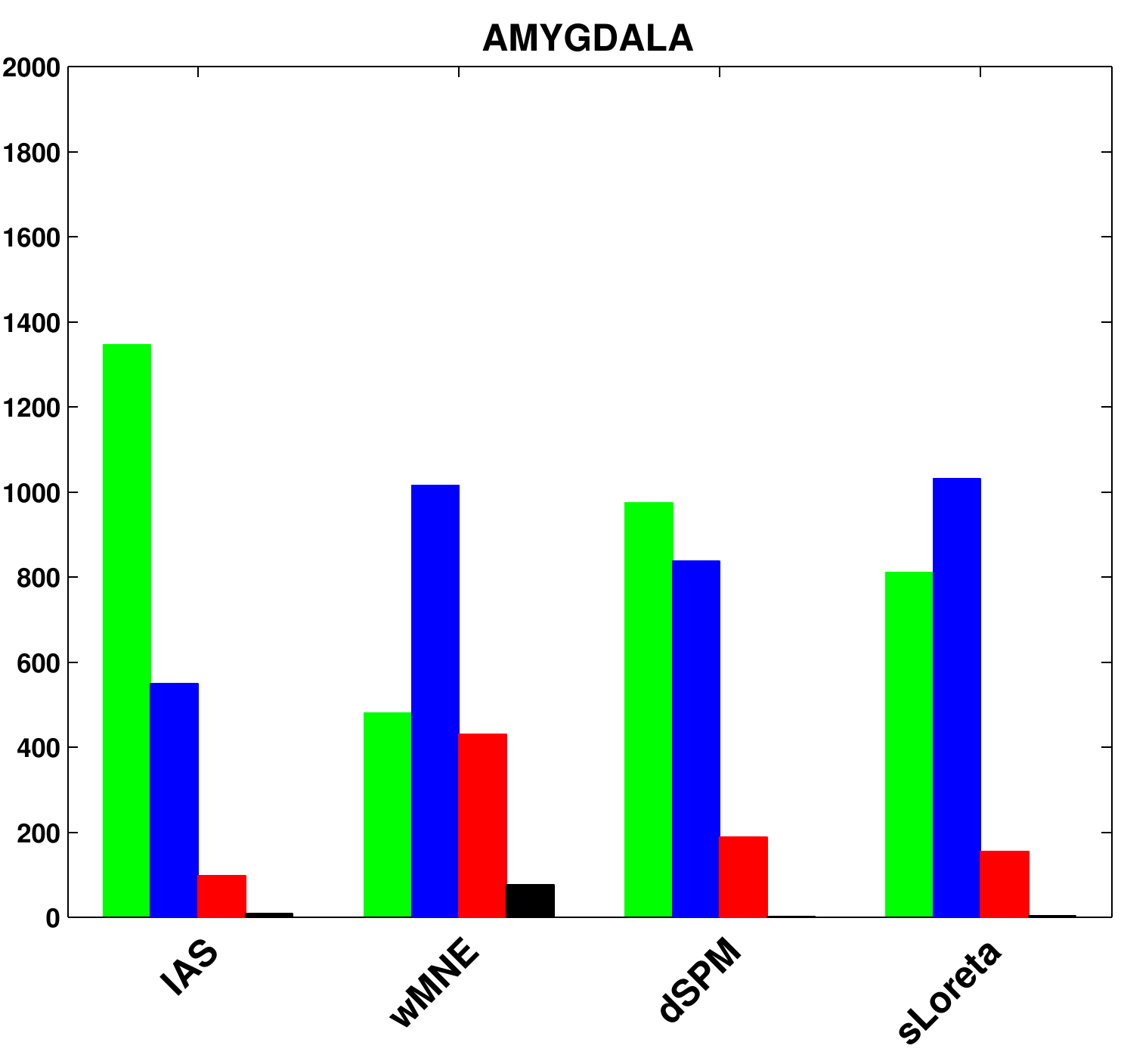}\hspace{2cm}
\includegraphics[width=5cm]{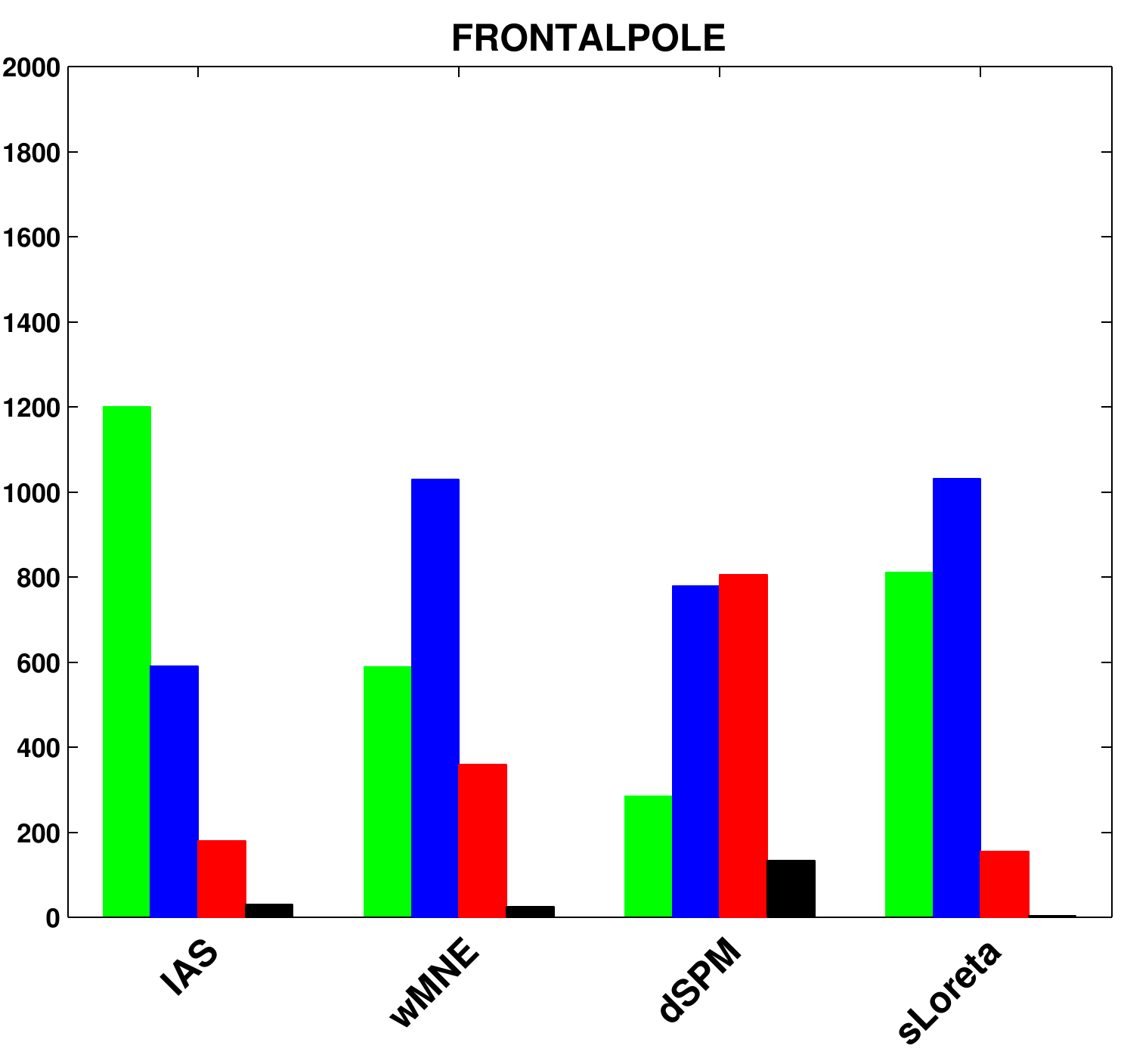}
}
\caption{\label{fig:BF Amyd_FP}Bayes factors in the case of two active patches. Left panel: Bayes factors in support of the hypothesis that the activity is in the {\em left amygdala} when the data comes from two active patches, one in the left amygdala and the other in the left frontal pole, and the MEG inverse problem is solved with the IAS, wMNE, dSPM and sLORETA algorithms, respectively. The histograms where computed from 2000 realizations and the color coding is as explained in Figure~\ref{fig:BF IAS}.  Right panel: Bayes factors in support of activity in the {\em left frontal pole} computed from the same 2000 realizations.}
\end{figure}

\begin{figure}[tbh]
\centerline{
\includegraphics[width=5cm]{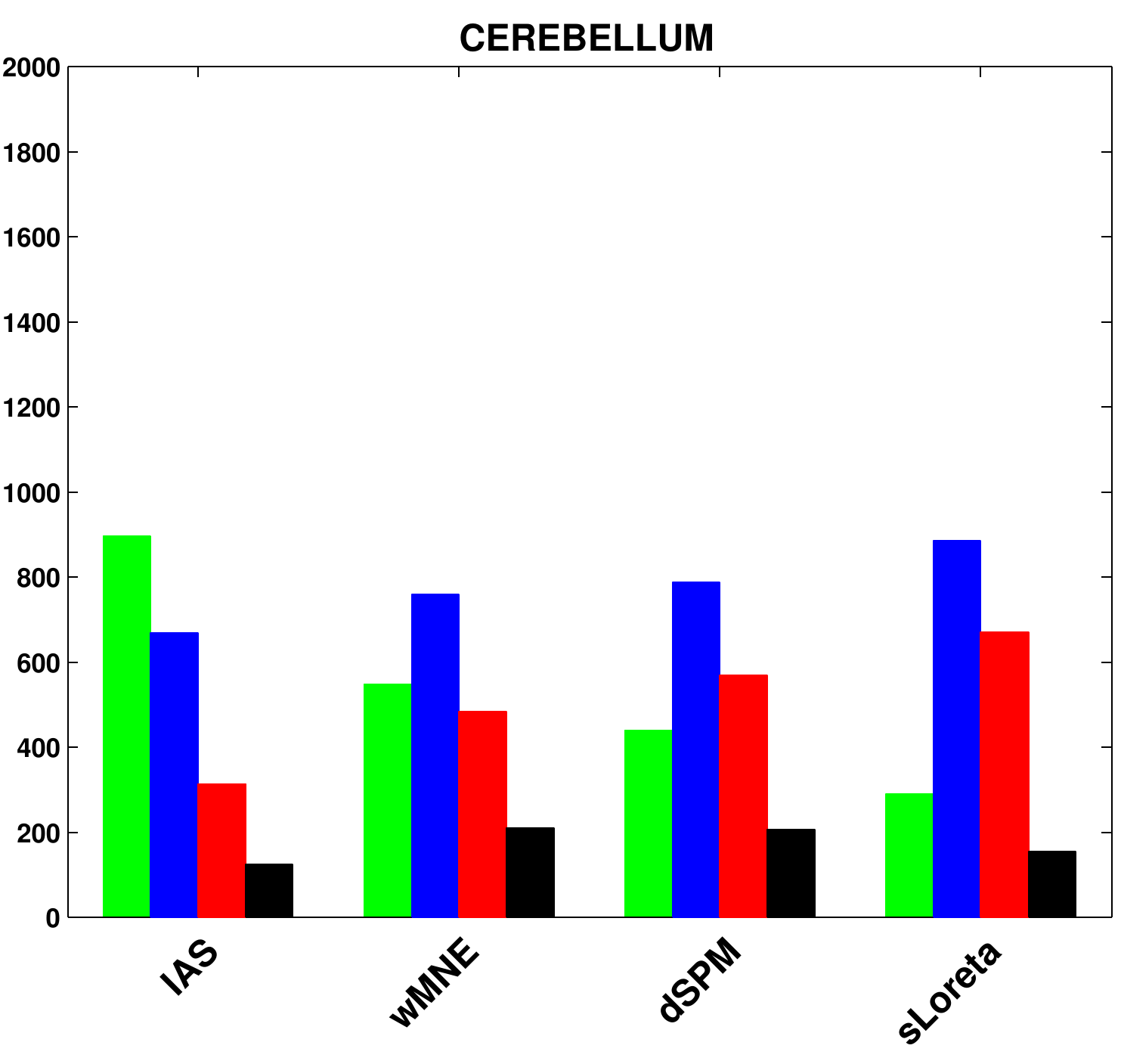}\hspace{2cm}
\includegraphics[width=5cm]{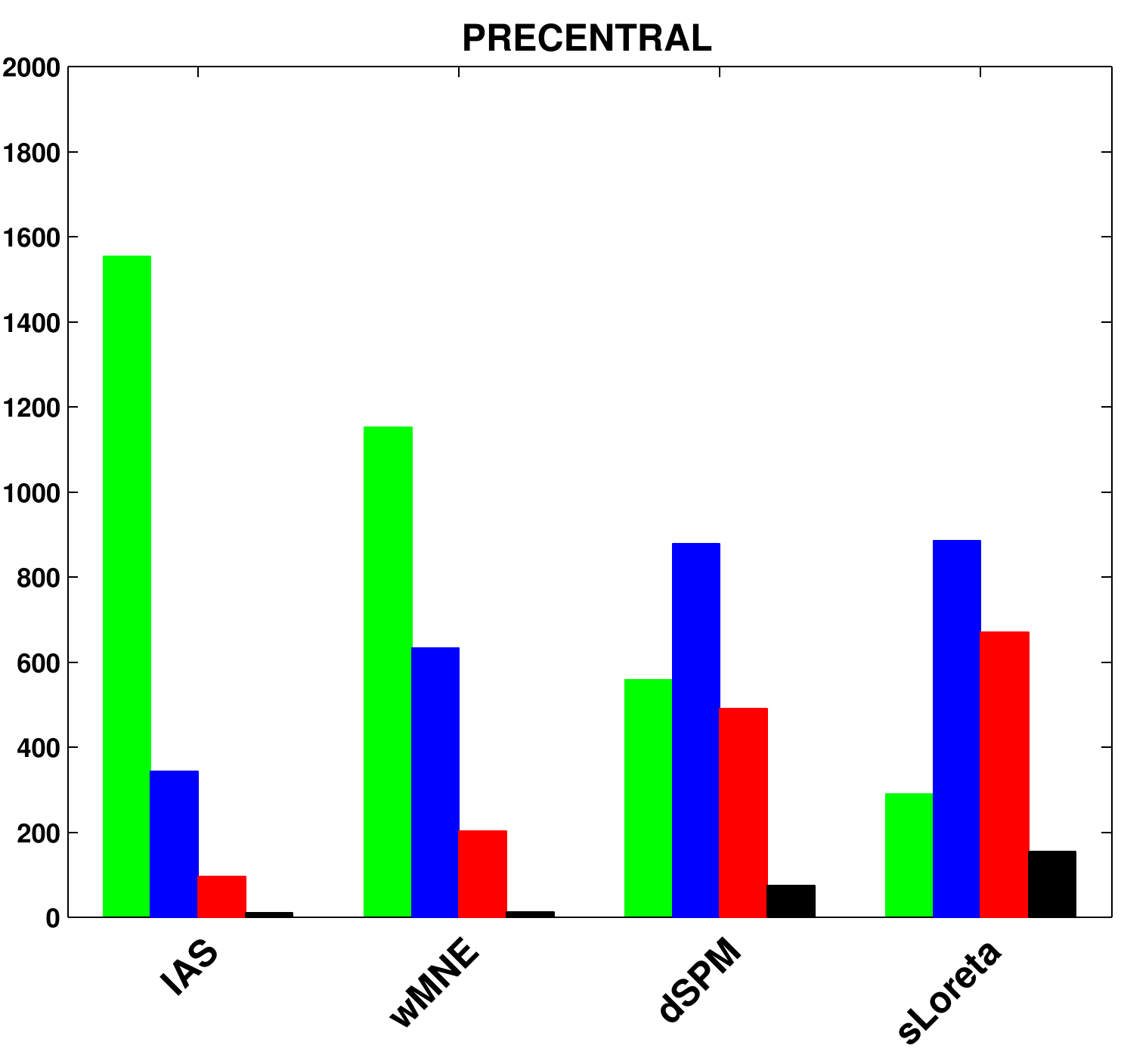}
}
\caption{\label{fig:BF_Cereb_PC}Bayes factors in the case of two active patches. Left panel: Bayes factors in support of the hypothesis that the activity is in the {\em left cerebellum} when the data comes from two active patches, one in the left cerebellum and the other in the left precentral cortex, and the MEG inverse problem is solved with the IAS, wMNE, dSPM and sLORETA algorithms, respectively. The histograms were computed from 2000 realizations and the color coding is as explained in Figure~\ref{fig:BF IAS}.  Right panel: Bayes factors in support of activity in the {\em left precentral cortex} computed from the same 2000 realizations. }
\end{figure}

Figure~\ref{fig:BF Amyd_FP} shows the histograms of the Bayes factors, binned and color coded according to the strength of evidence (\ref{evidence strength}), corresponding to the four different inverse solvers in the case when the active patches are in the left amygdala and in the left frontal pole.
Figure~\ref{fig:BF_Cereb_PC} shows the histograms of the Bayes factors for the four different inverse solvers in the case when the active patches are in the left cerebellum and in the left precentral cortex.

\begin{figure}[tbh]
\centerline{
\includegraphics[width=5cm]{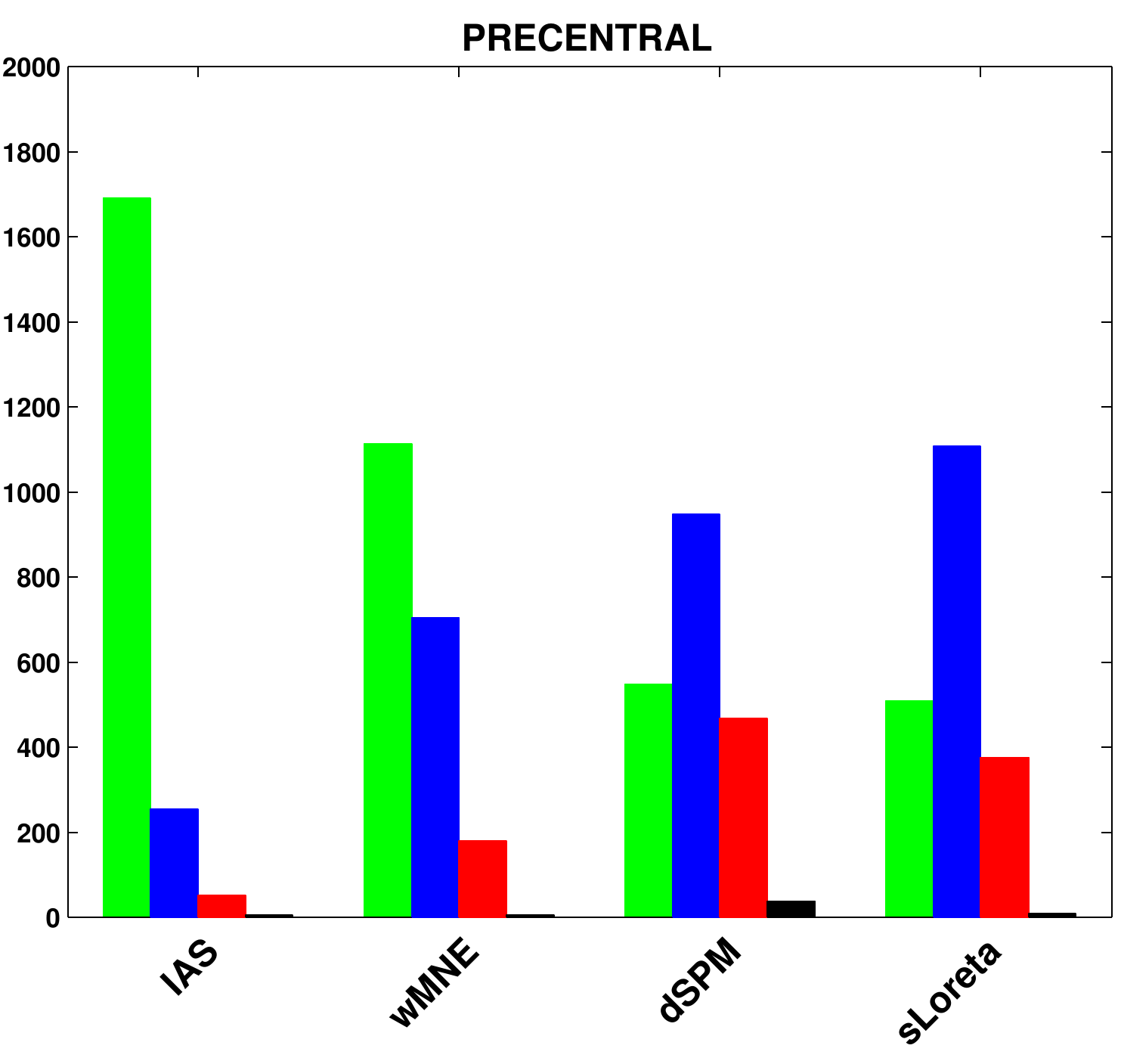}\hspace{1cm}
\includegraphics[width=5cm]{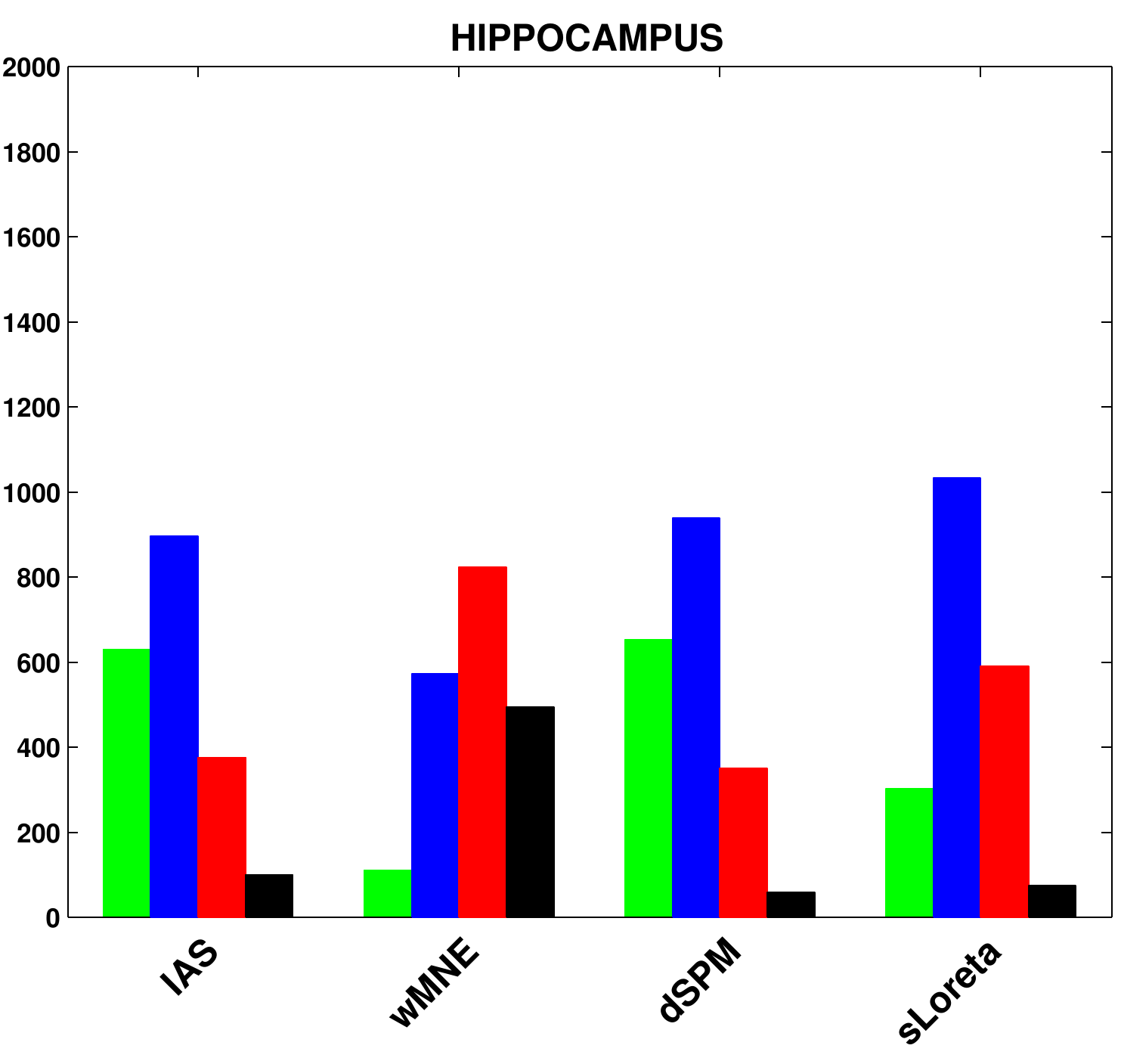}\hspace{1cm}
\includegraphics[width=5cm]{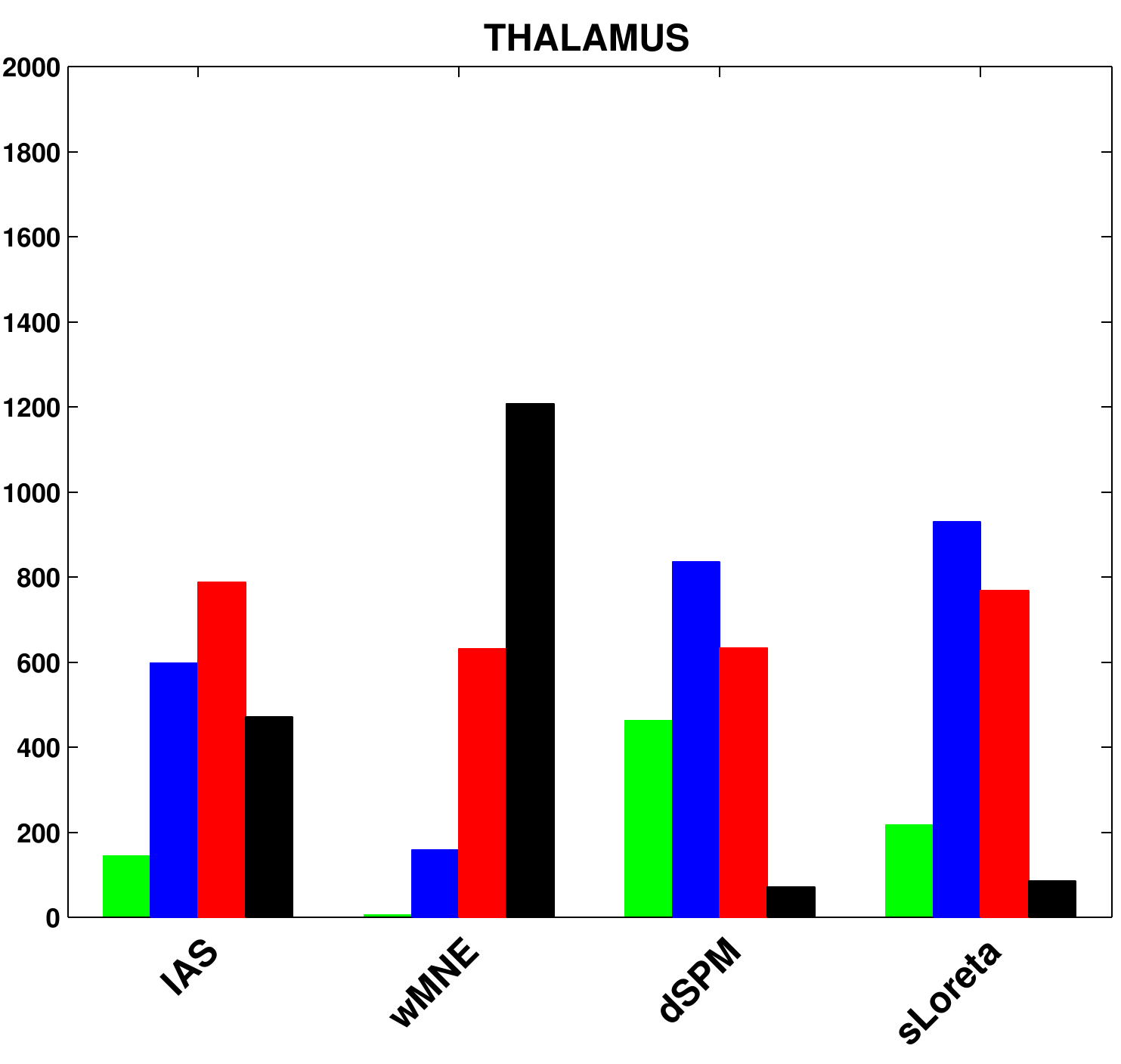}
}
\caption{\label{fig:BF_Thal_Hip_PC}  Bayes factors in the case of active patches in three different BRs, the right precentral gyrus (left), and the left hippocampus (center) and the right thalamus (right), and the MEG inverse problem is solved with the IAS, wMNE, dSPM and sLORETA algorithms, respectively. The histograms were computed from 2000 realizations and the color coding is as explained in Figure~\ref{fig:BF IAS}. }
\end{figure}

\begin{figure}[tbh]
\centerline{
\includegraphics[width=5cm]{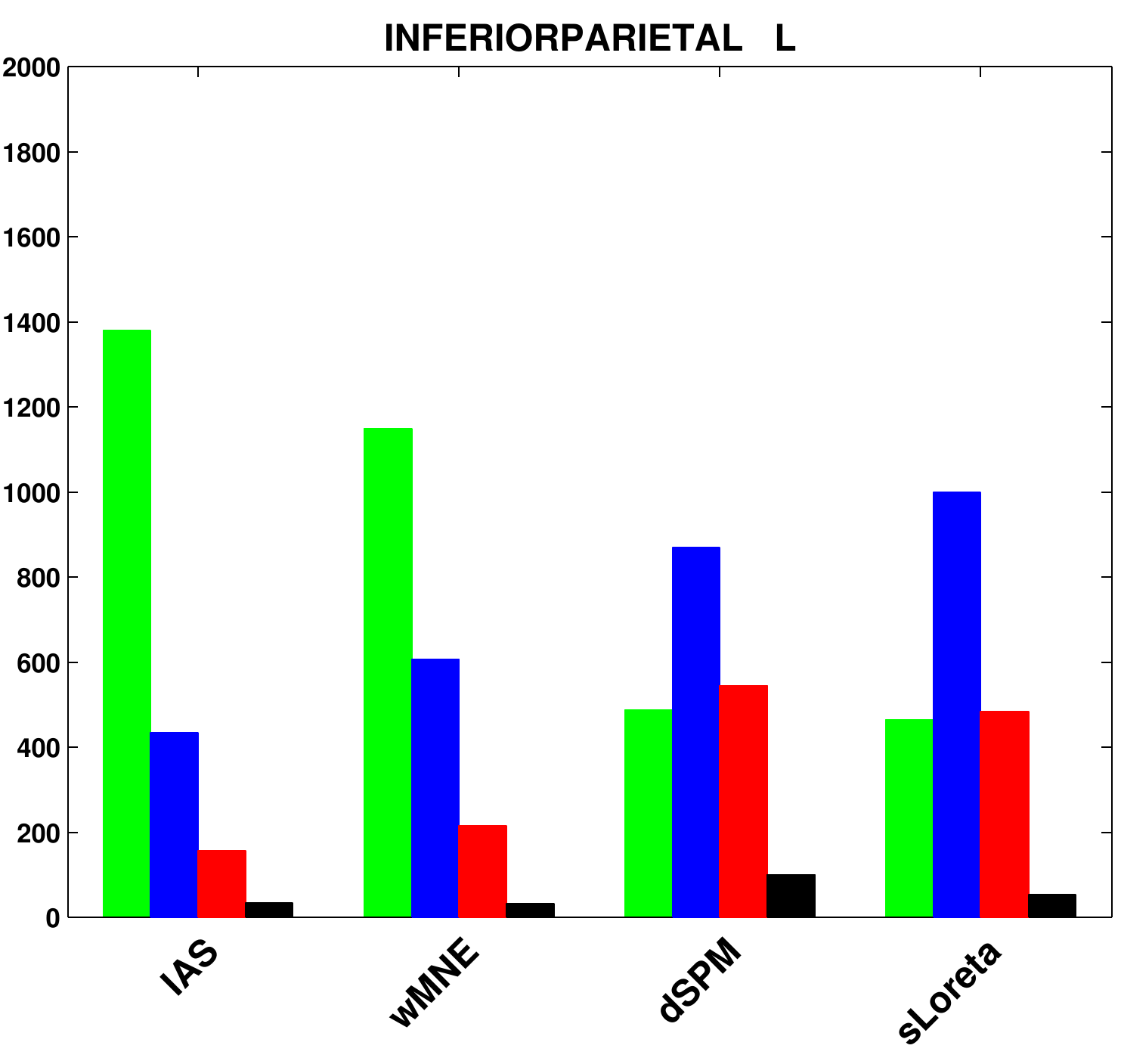}\hspace{1cm}
\includegraphics[width=5cm]{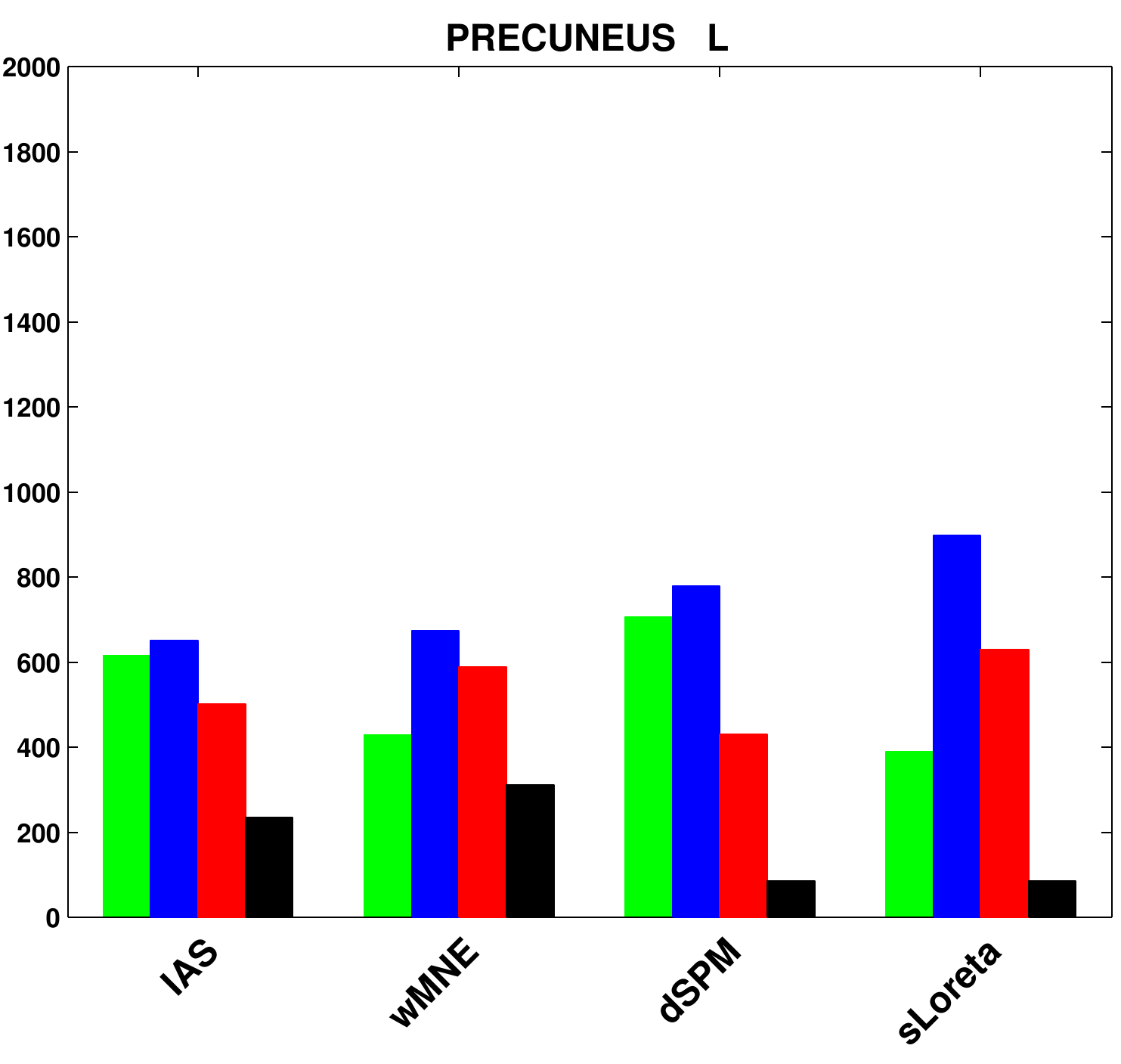}\hspace{1cm}
\includegraphics[width=5cm]{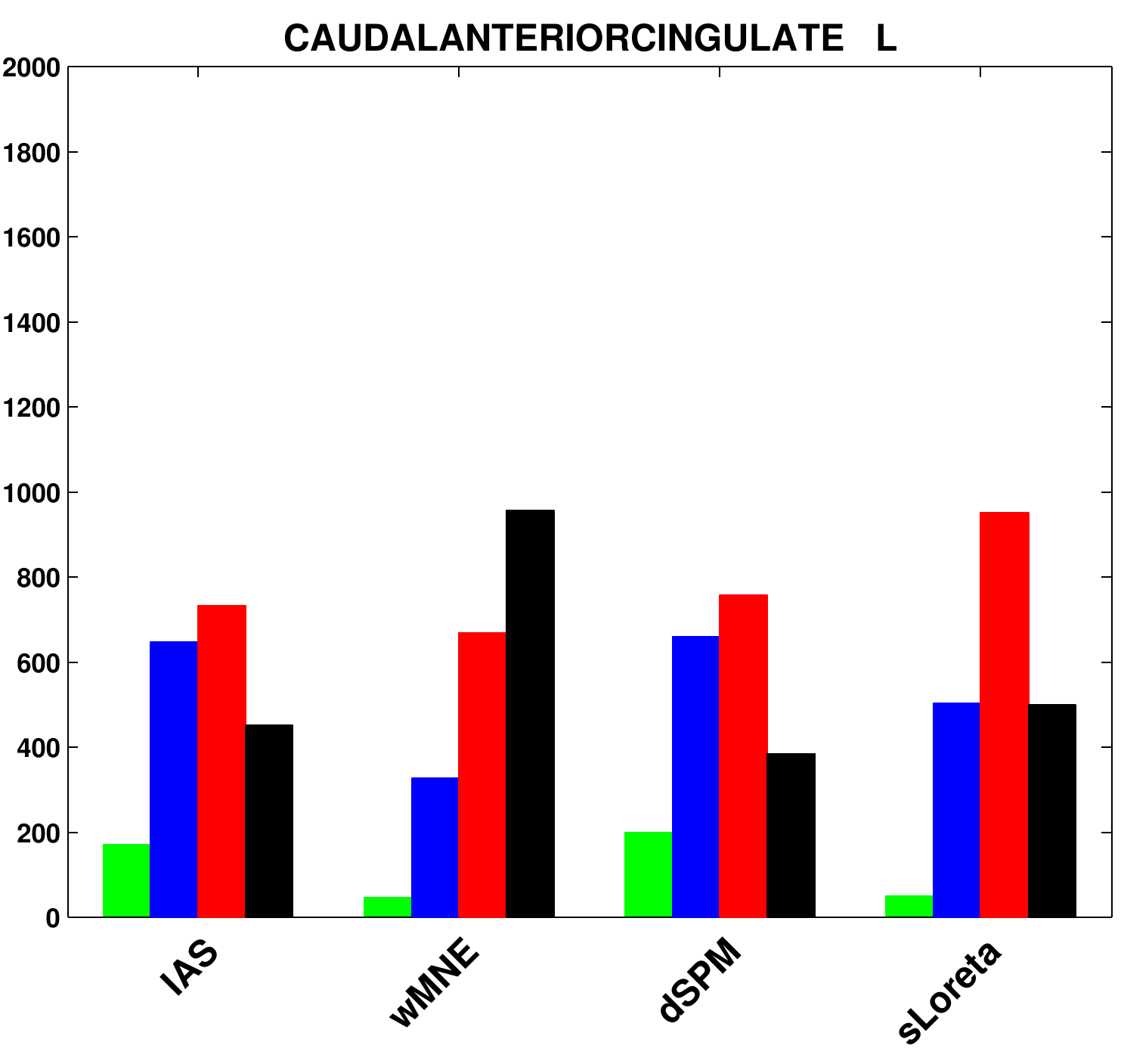}
}
\centerline{
\includegraphics[width=5cm]{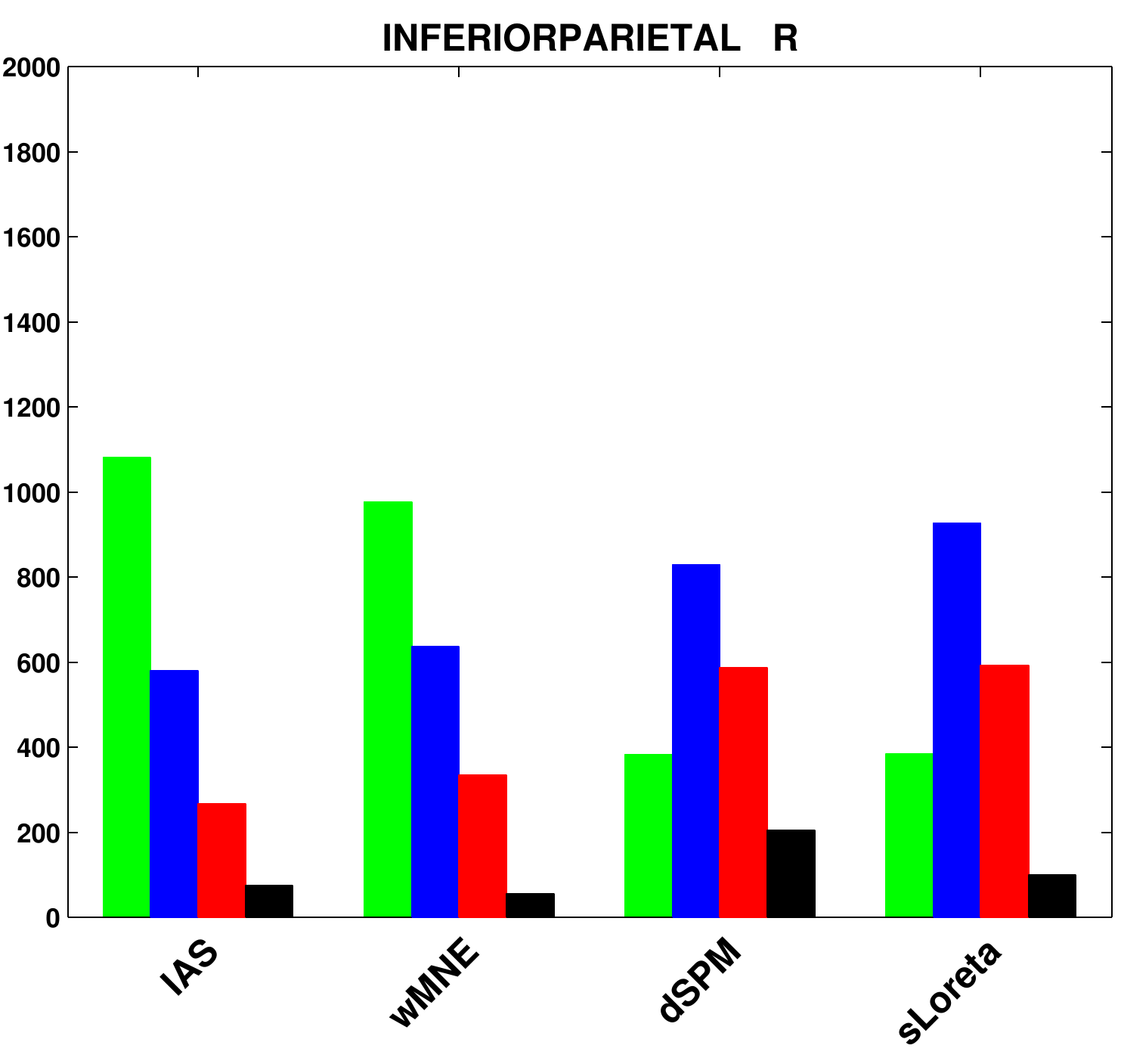}\hspace{1cm}
\includegraphics[width=5cm]{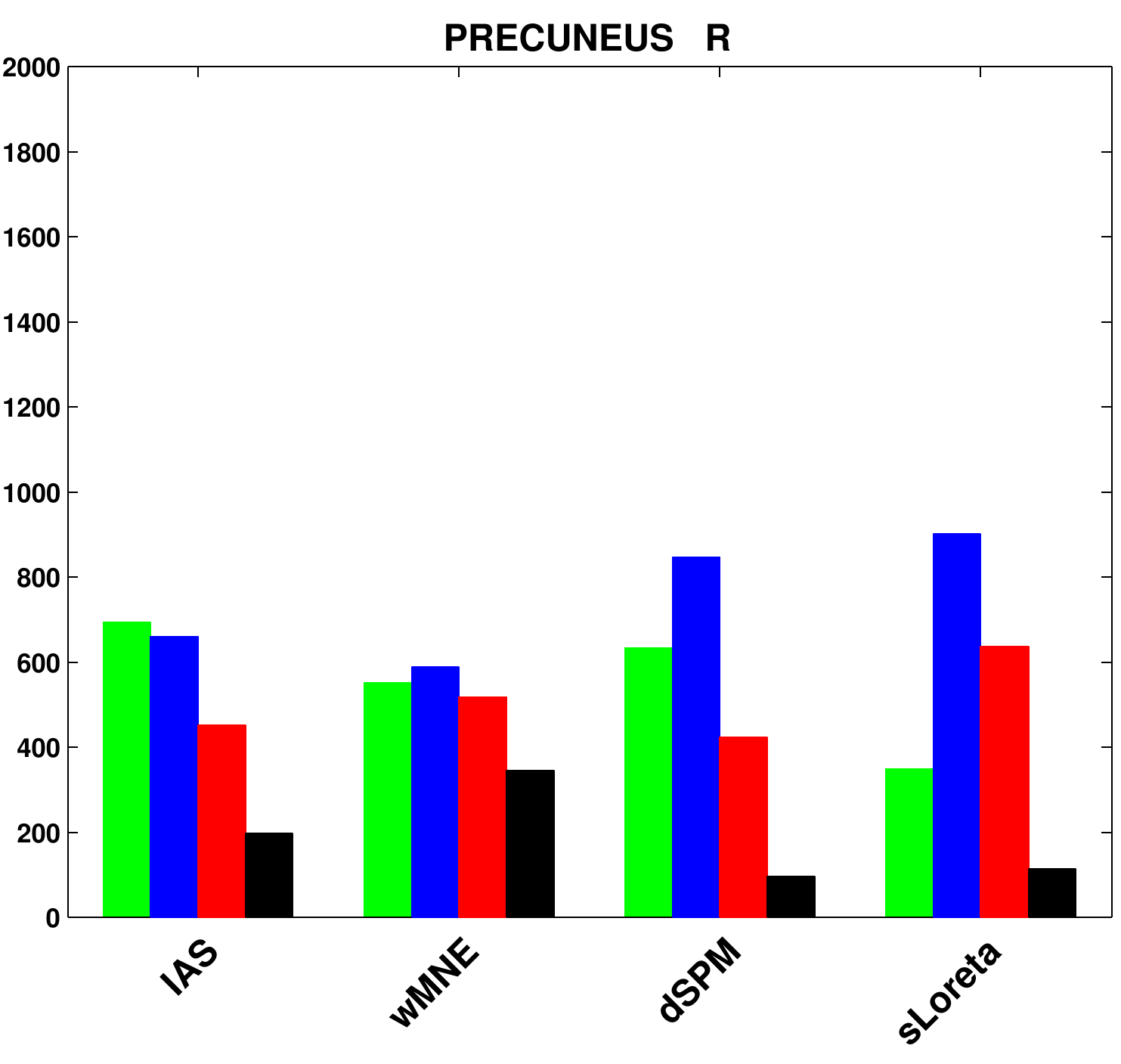}\hspace{1cm}
\includegraphics[width=5cm]{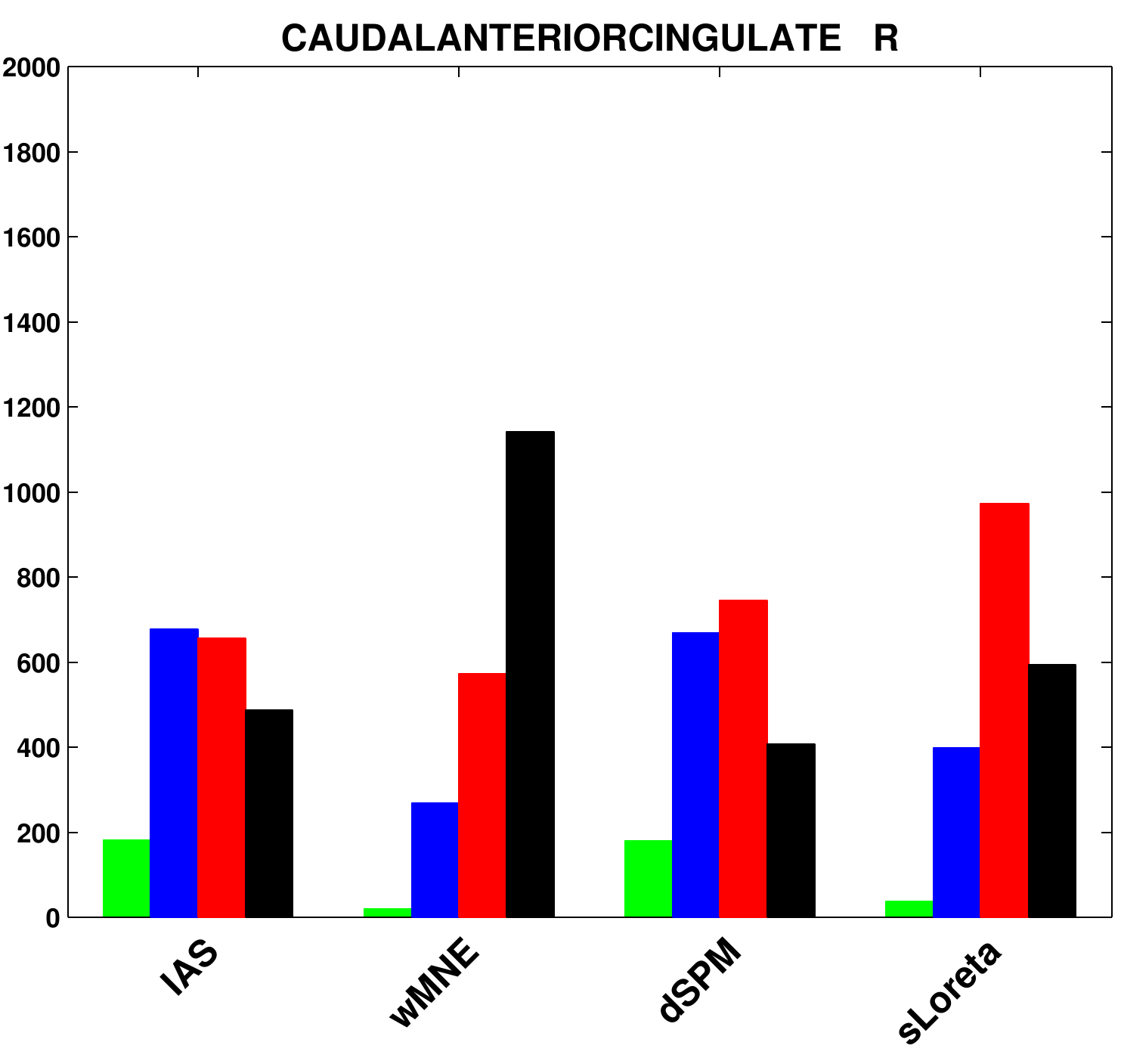}
}
\caption{\label{fig:BF_DMN} Bayes factors for the case where there are active patches in all six  BRs comprising the Default Mode Network: inferior parietal gyrus, precuneus and caudal anterior cingulate gyrus in the left (top) and right (bottom) hemispheres to test the specificity of the IAS inverse solver and the three reference inverse solvers.}
\end{figure}

The histograms of the Bayes factors for the four inverse solvers when there are three active patches located in the left hippocampus, precentral gyrus and right thalamus are displayed in Figure~\ref{fig:BF_Thal_Hip_PC}. Finally, the histograms for the Bayes factors corresponding to the different solvers in the case where there are active patches in the six different BRs in the DMN are displayed in Figure~\ref{fig:BF_DMN}.

\subsection{Reconstruction of the brain activity from real data}

\begin{figure}[tbh]
\centerline{  
\subfloat[\textbf{IAS}]{
\includegraphics[width=9.5cm]{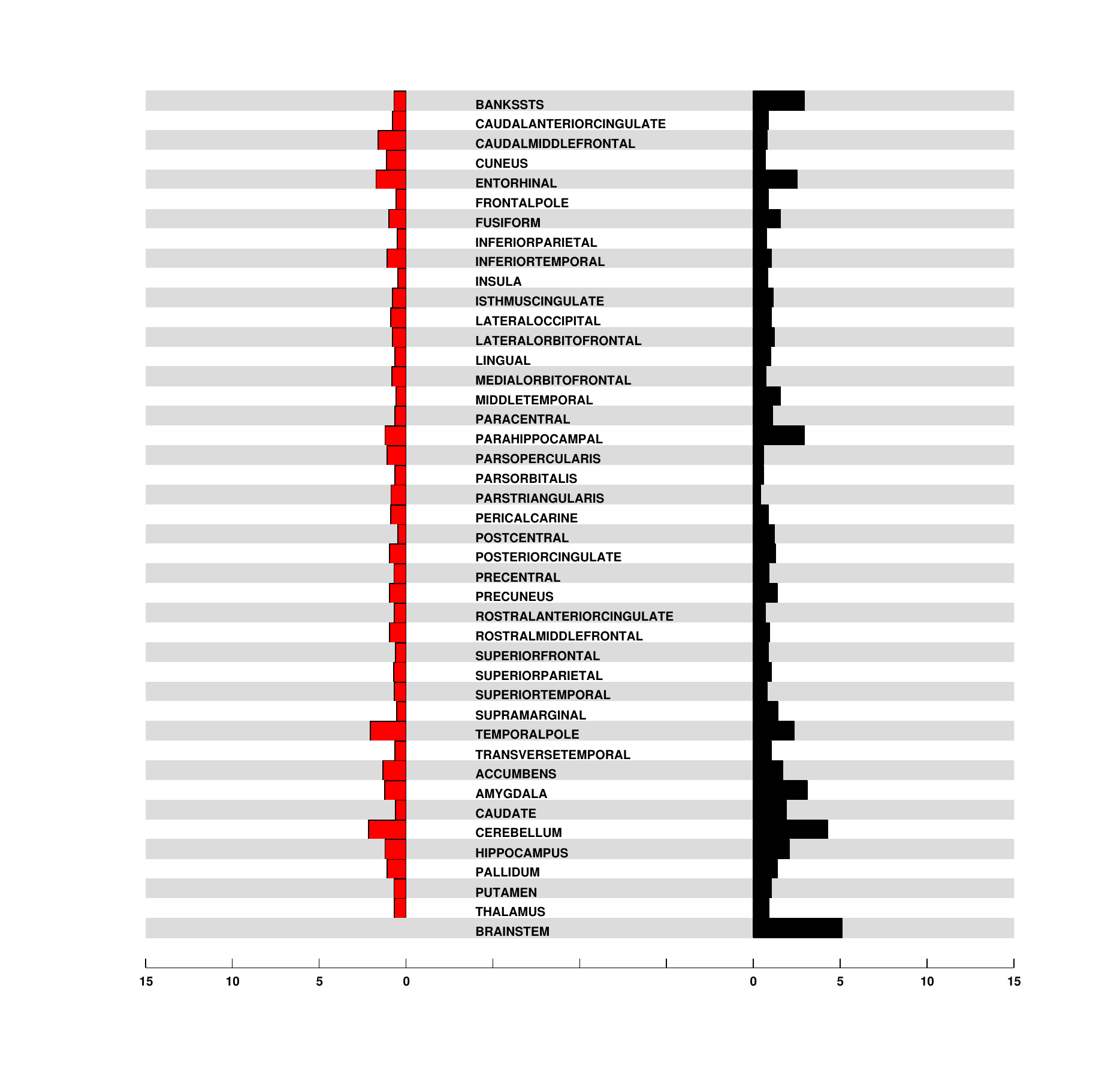}}
\subfloat[\textbf{wMNE}]{
\includegraphics[width=9.5cm]{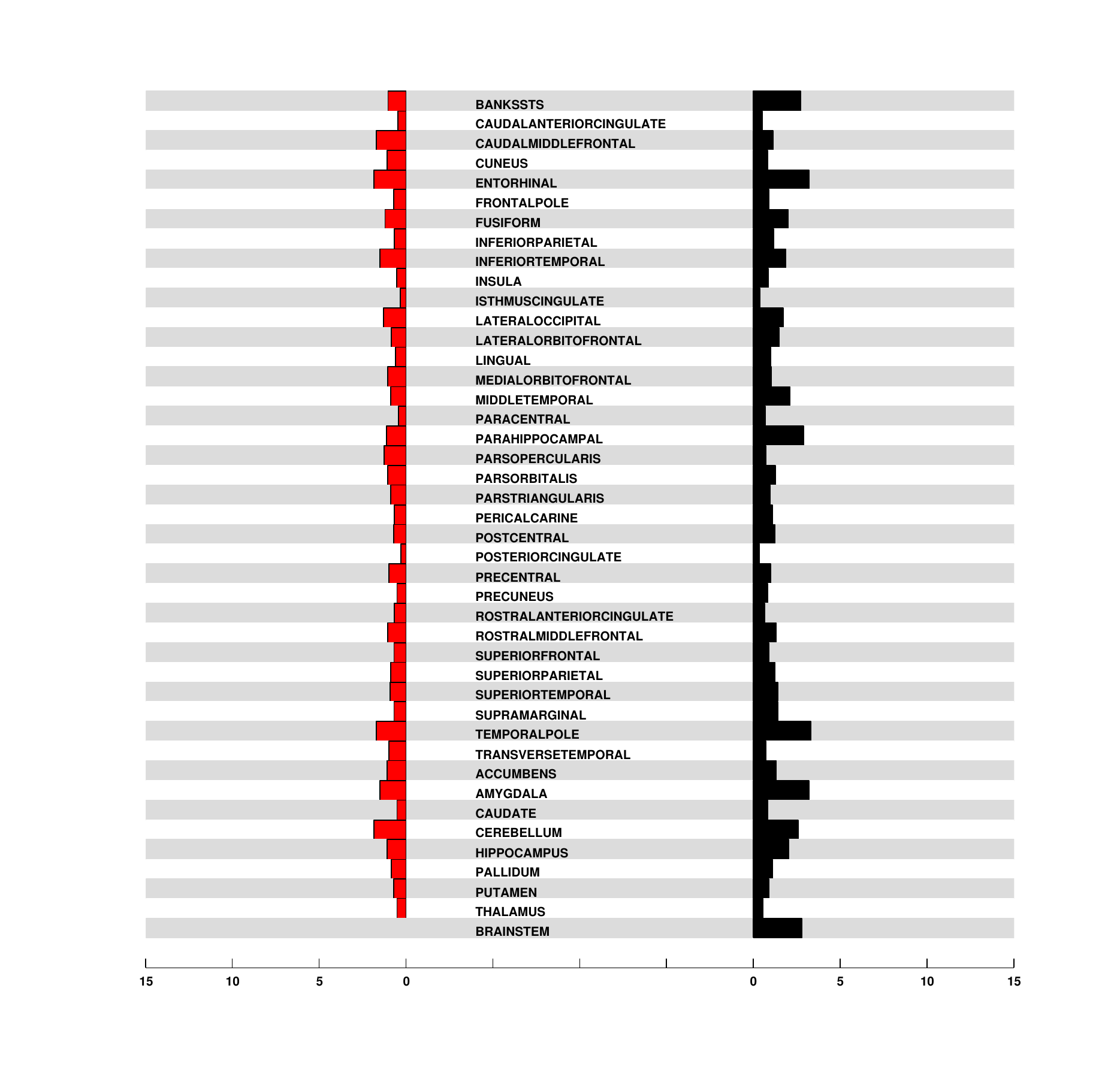}}}
\centerline{ 
\subfloat[\textbf{dSPM}]{
\includegraphics[width=9.5cm]{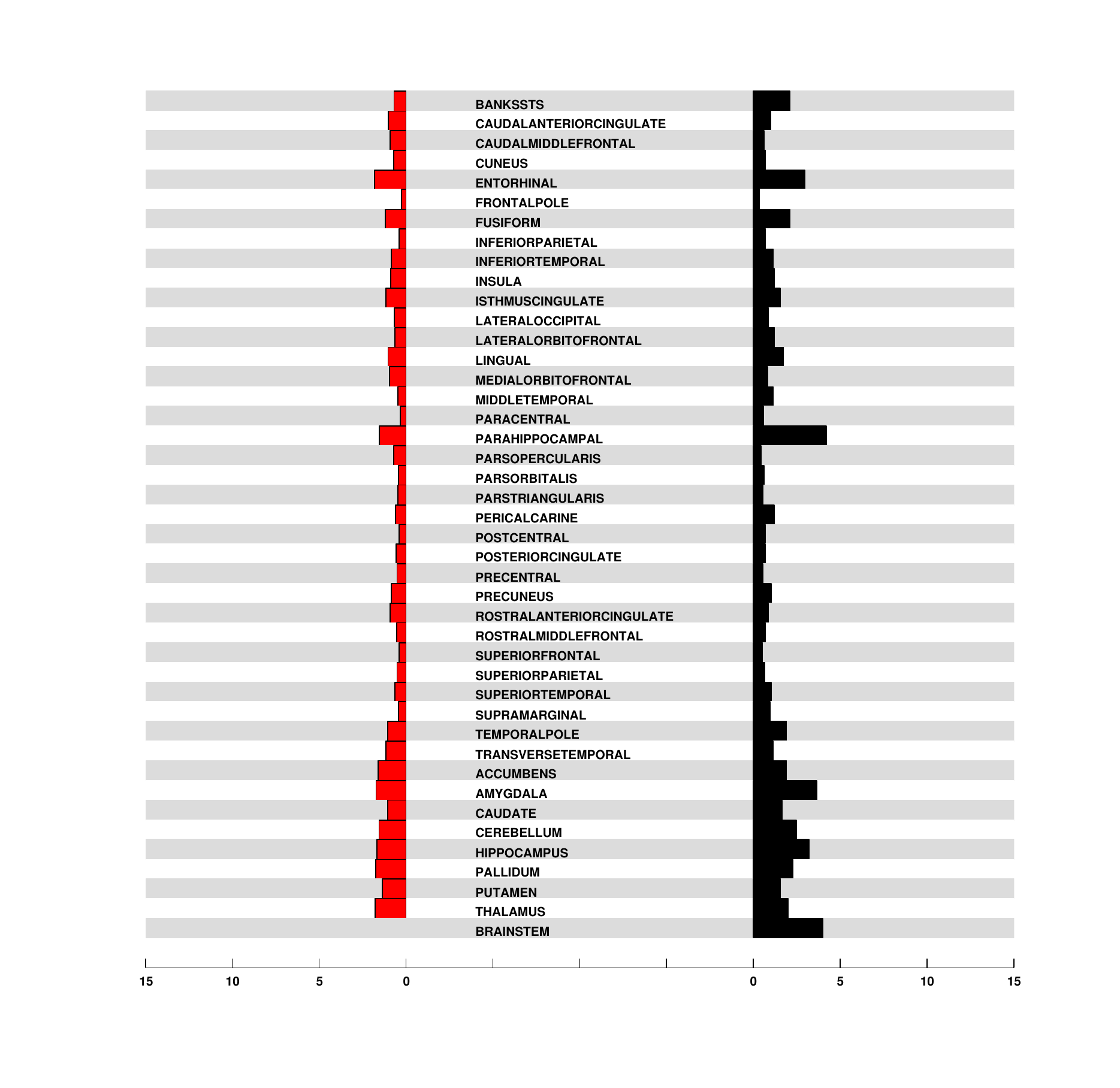}}
\subfloat[\textbf{sLORETA}]{
\includegraphics[width=9.5cm]{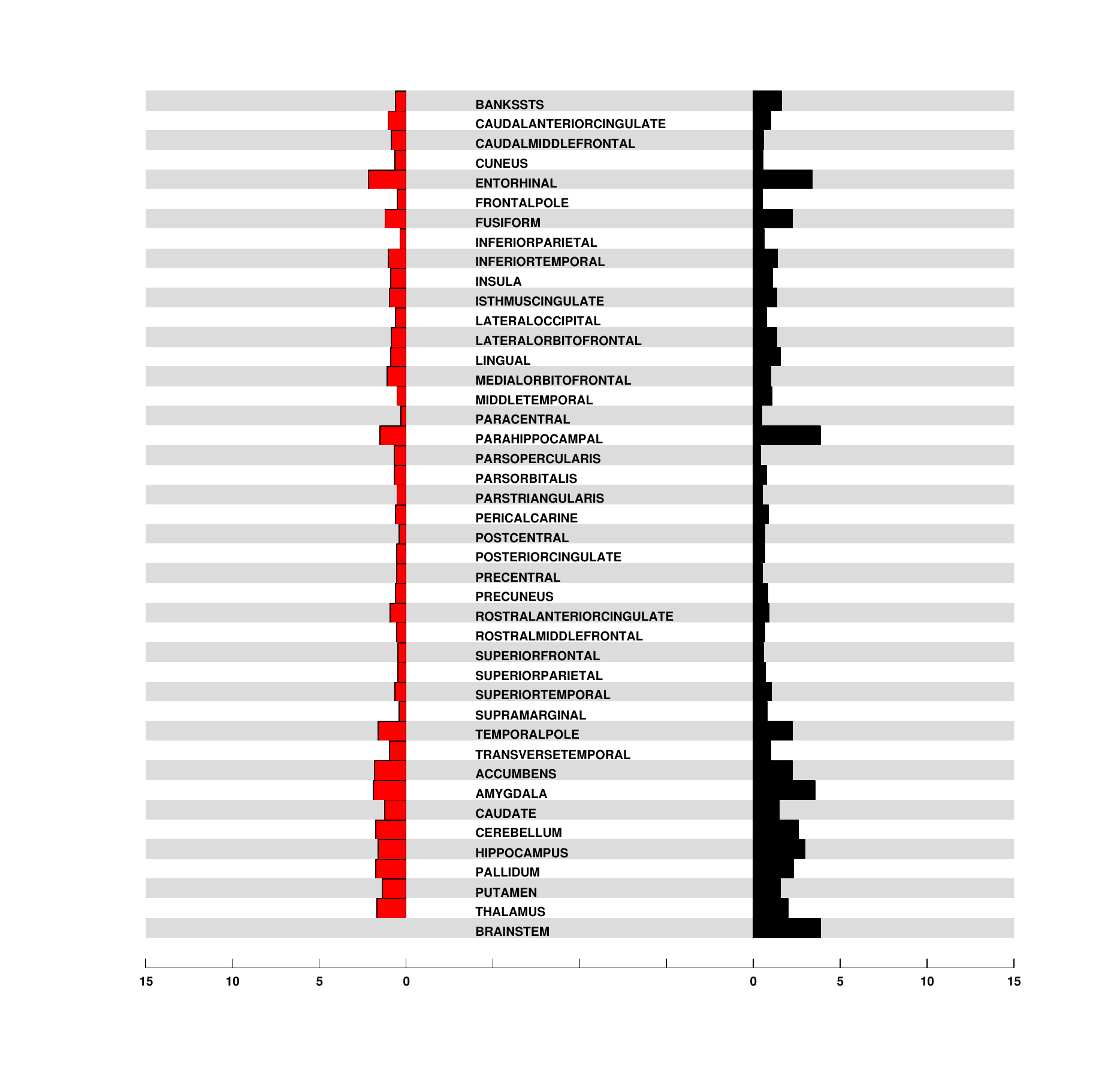}}
}
\caption{\label{fig:real_data_Baffo_5ms}Mapping of the brain activity to 85 different BRs  for the time
$T=5$ ms with real/realistic data example reconstructed with, respectively,  IAS (a), wMNE (b),  dSPM (c) and sLORETA (d). The histograms bin the average activity in each BR: in red the BRs of the left hemisphere and in black the ones of the right hemisphere.}
\end{figure}

\begin{figure}[tbh]
\centerline{
\subfloat[\textbf{IAS}]{
\includegraphics[width=9.5cm]{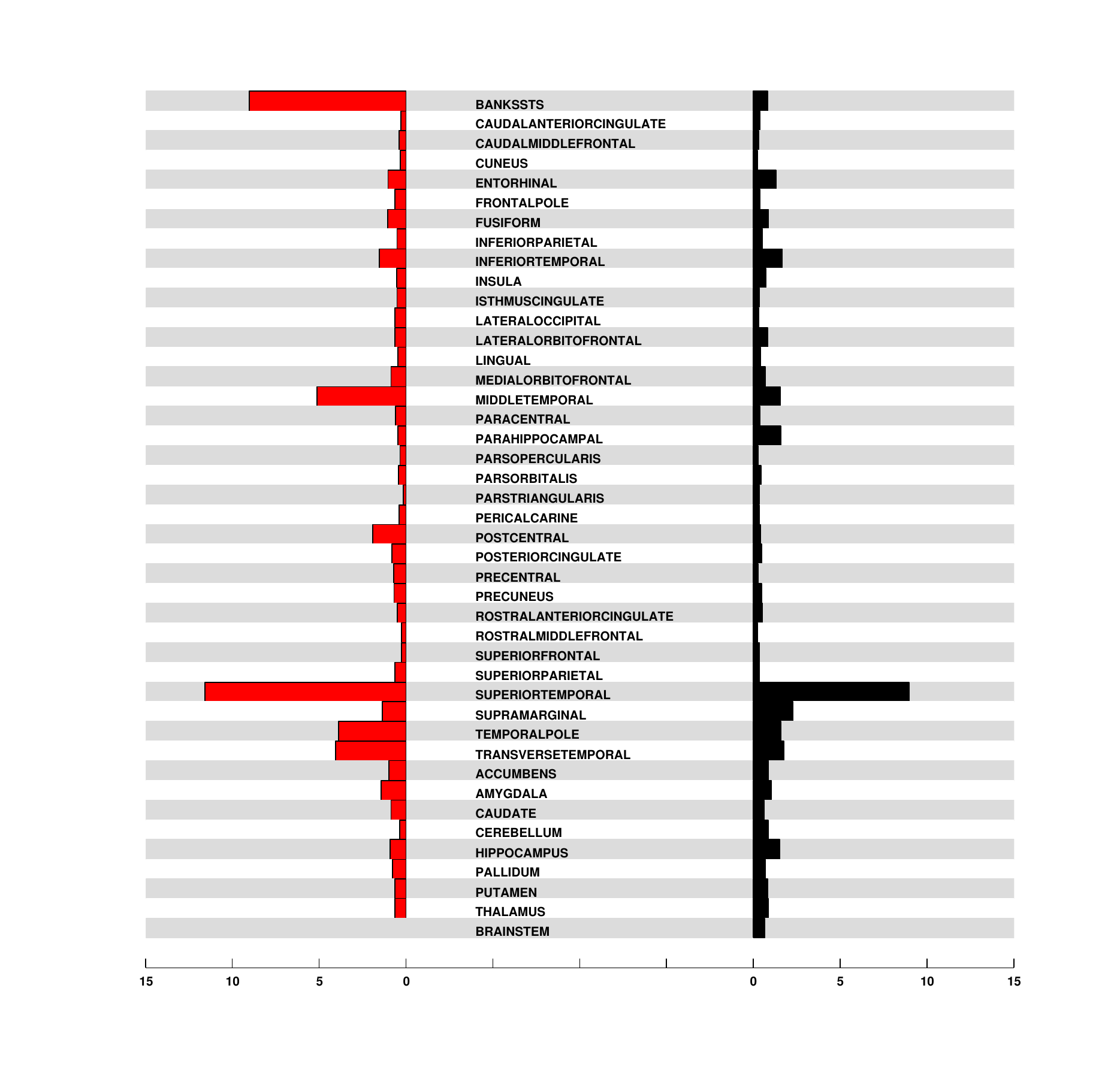}}
\subfloat[\textbf{wMNE}]{
\includegraphics[width=9.5cm]{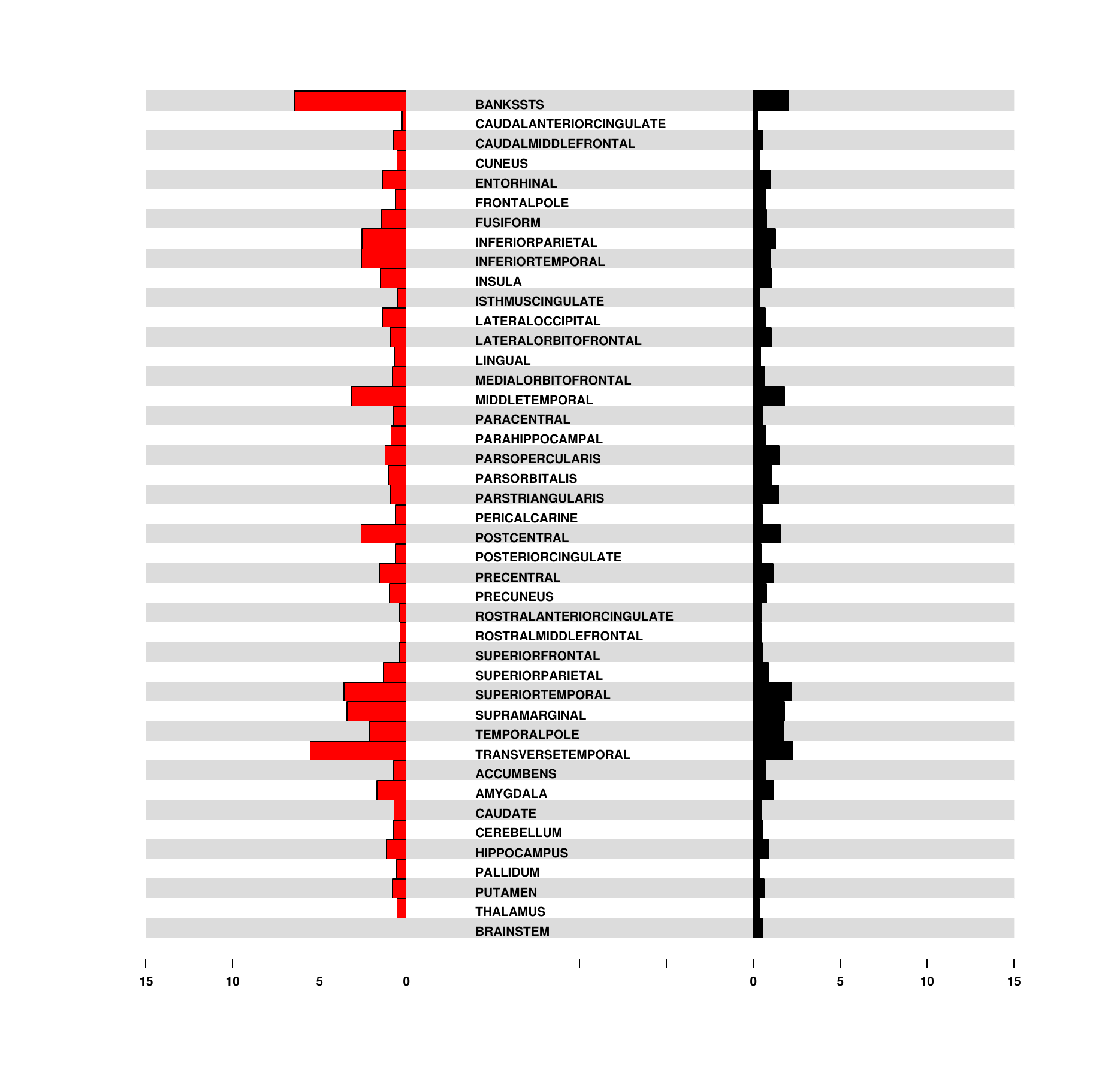}}}
\centerline{
\subfloat[\textbf{dSPM}]{
\includegraphics[width=9.5cm]{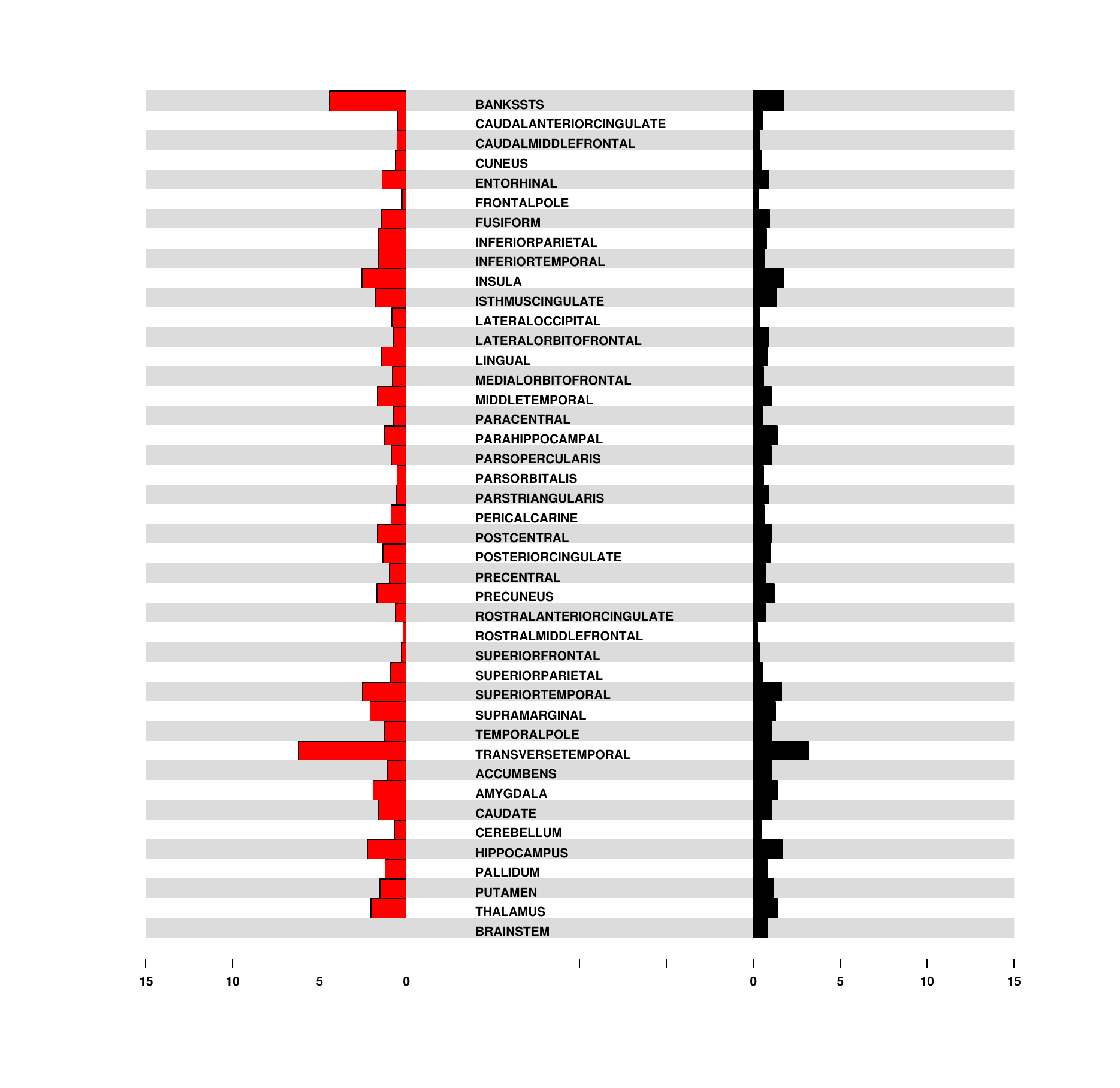}}
\subfloat[\textbf{sLORETA}]{
\includegraphics[width=9.5cm]{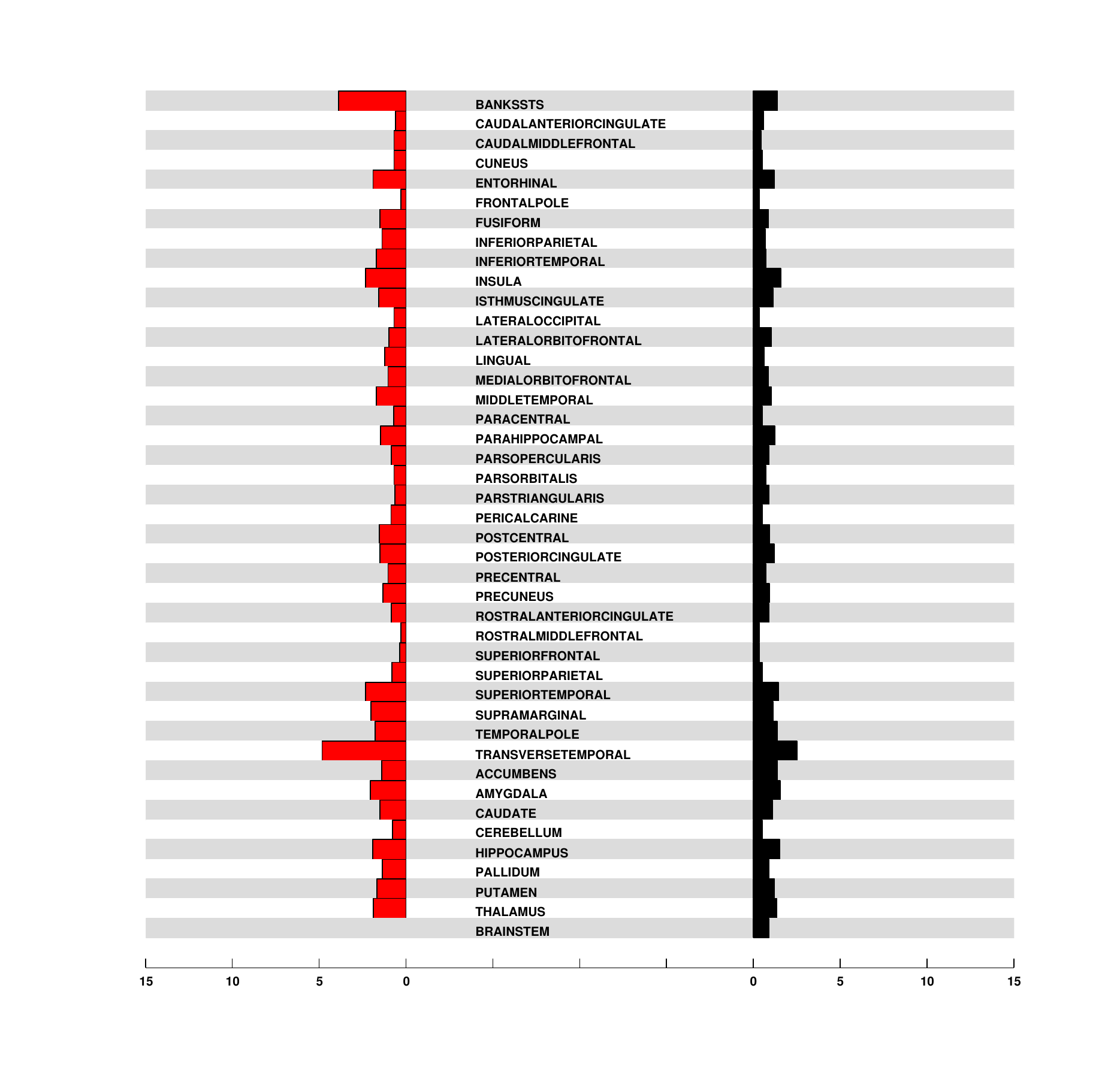}}
}
\caption{\label{fig:real_data_Baffo} Mapping of the brain activity to 85 different BRs  for the
time $T=100$ ms with real/realistic data example reconstructed with,
respectively,  IAS (a), wMNE (b),  dSPM (c) and sLORETA (d). The histograms bin
the average activity in each BR: in red the BRs of the left hemisphere and in
black the ones of the right hemisphere.}
\end{figure}

\begin{figure}[tbh]
\centerline{
\includegraphics[width=9.5cm]{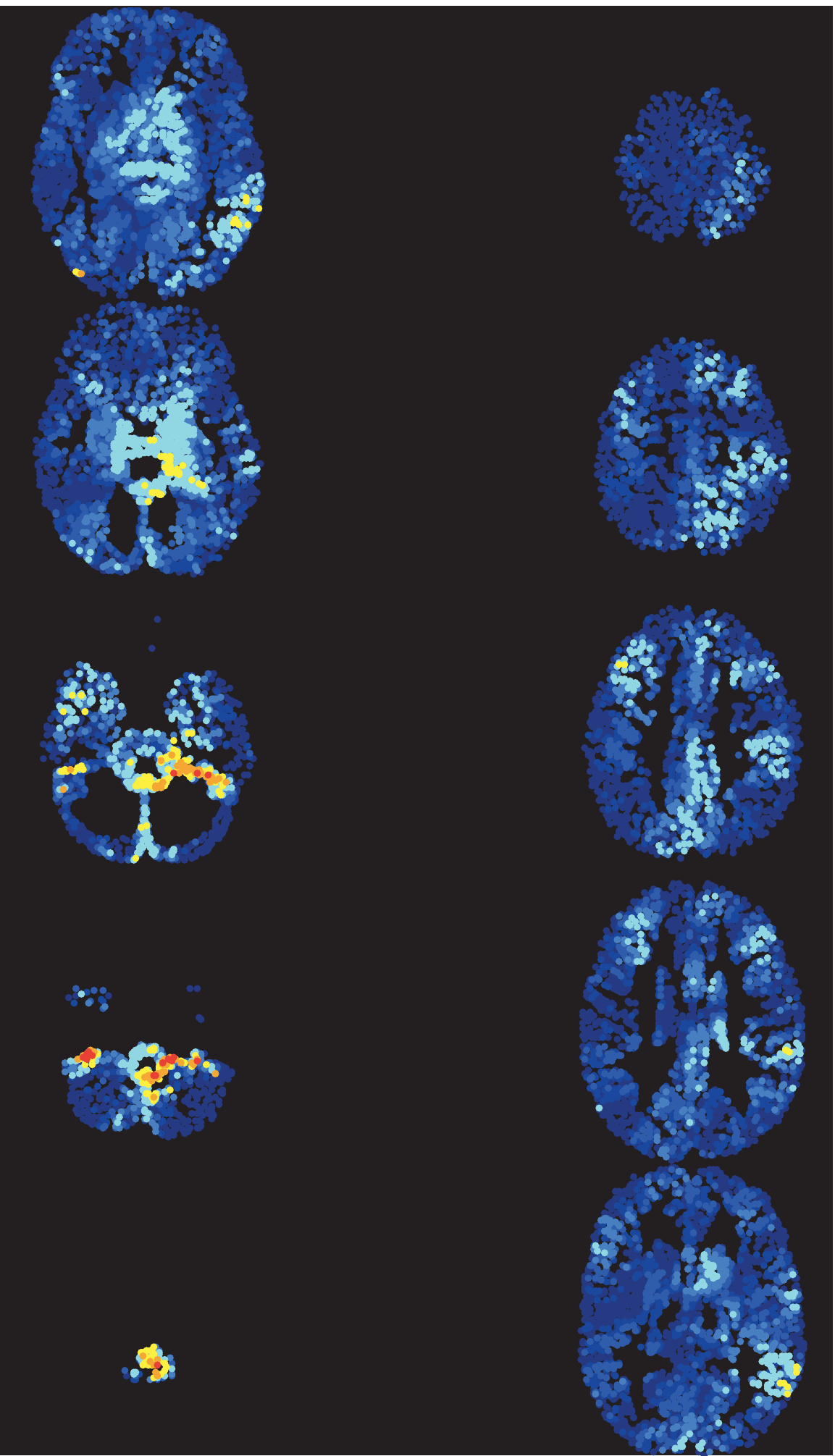}\hspace{0.5cm}
\includegraphics[width=9.5cm]{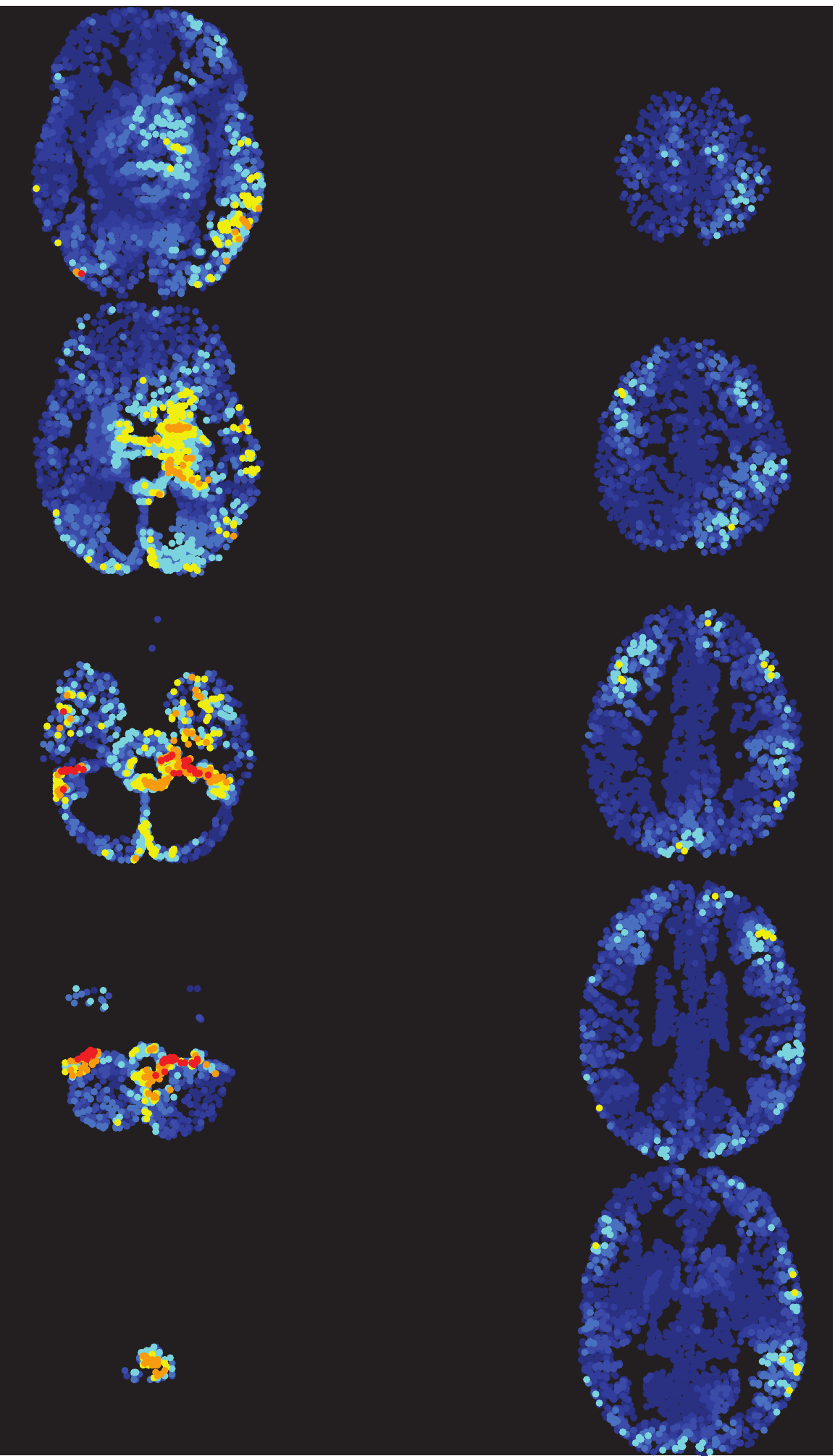}}
\caption{\label{fig:slice_IAS_wMNE_5}Visualization of the axial view of the brain activity at $T=5$
ms as reconstructed by IAS (left) and wMNE (right).}
\end{figure}

\begin{figure}[tbh]
\centerline{
\includegraphics[width=9.5cm]{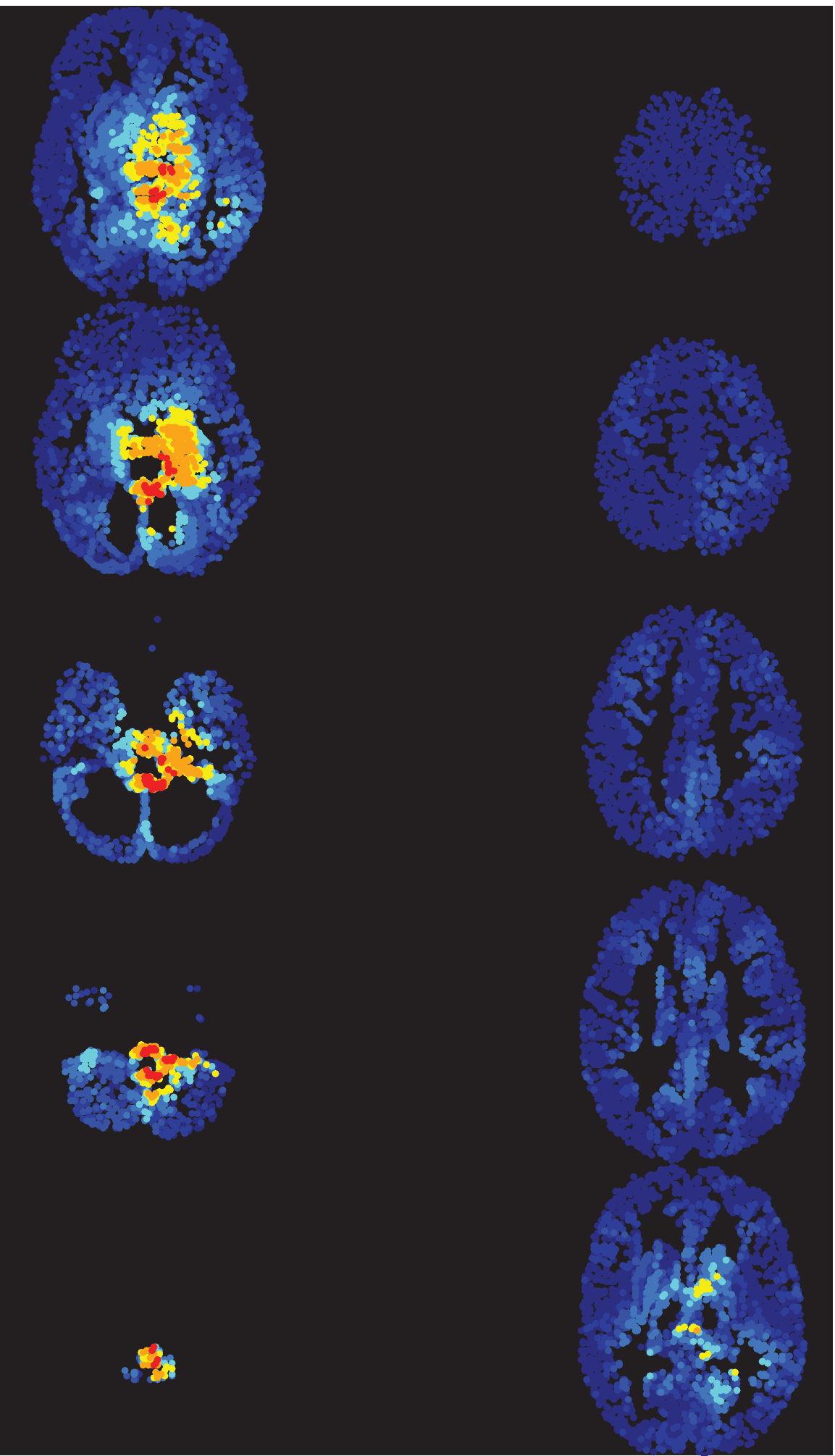}\hspace{0.5cm}
\includegraphics[width=9.5cm]{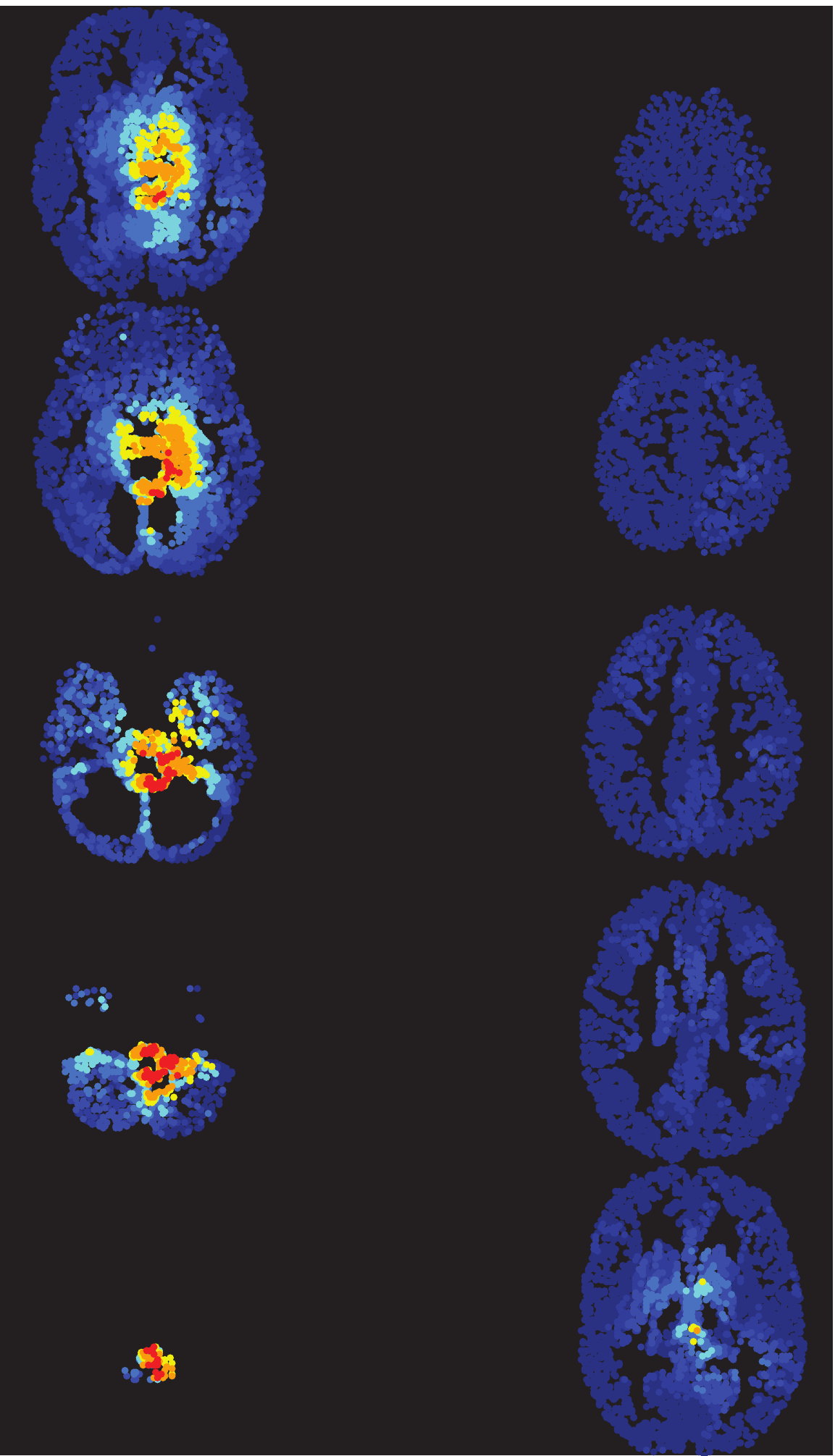}
}
\caption{\label{fig:slice_dSPM_sloreta_5} Visualization of the axial view of the brain activity at $T=5$
ms as reconstructed by dSPM (left) and sLORETA (right).}
\end{figure}

\begin{figure}[tbh]
\centerline{
\includegraphics[width=15cm]{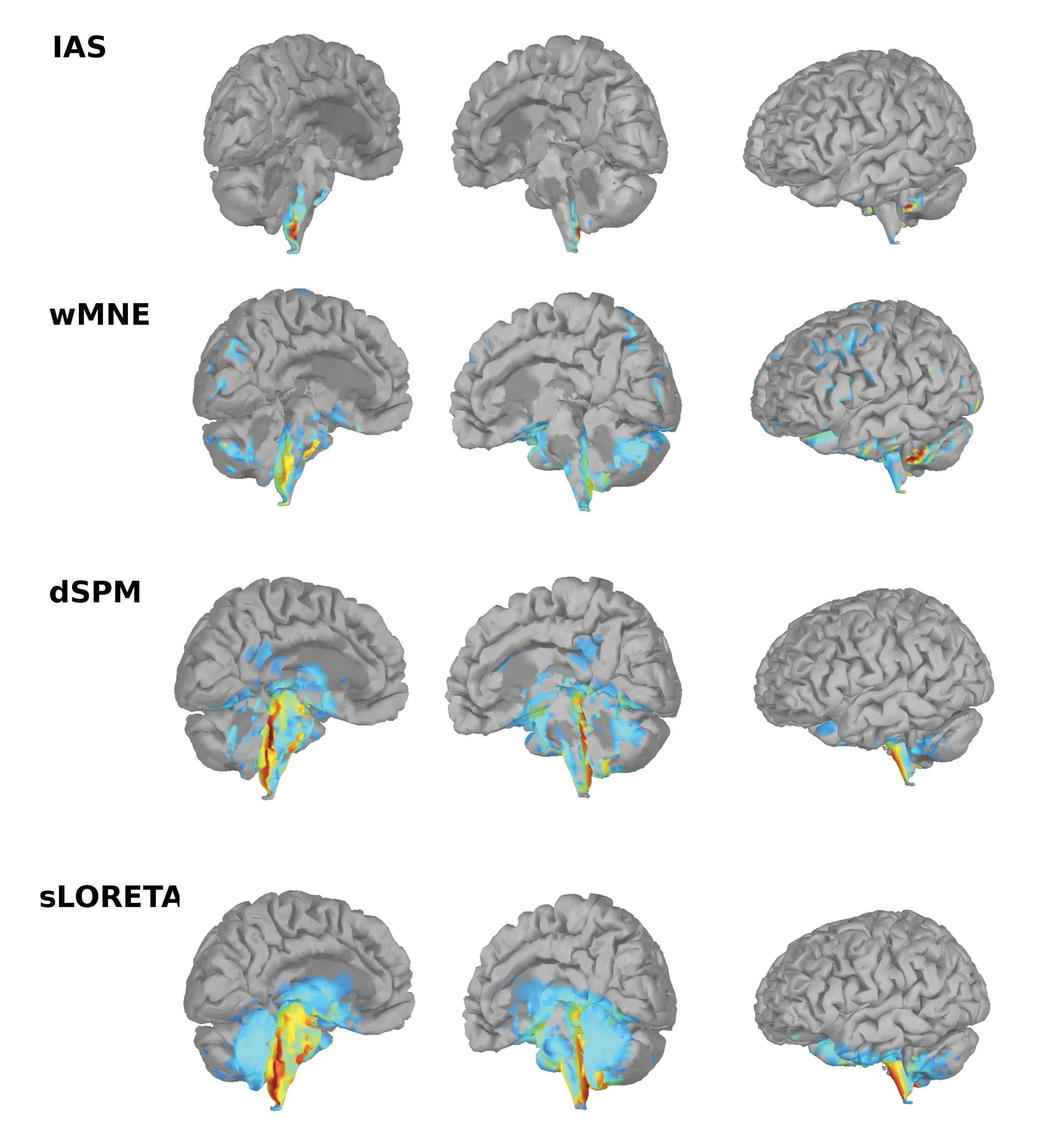}}
\caption{\label{fig:patch_brainstem_M100} Reconstructions obtained from real data at time $T=5$ ms with, 
top to bottom, IAS, wMNE, dSPM and sLORETA, respectively. At this time instant,
most of the signal is generated by the patch activity in the brainstem shown in
Figure~\ref{fig:patch_brainstem_activity}.}
\end{figure}

\begin{figure}[tbh]
\centerline{
\includegraphics[width=9.5cm]{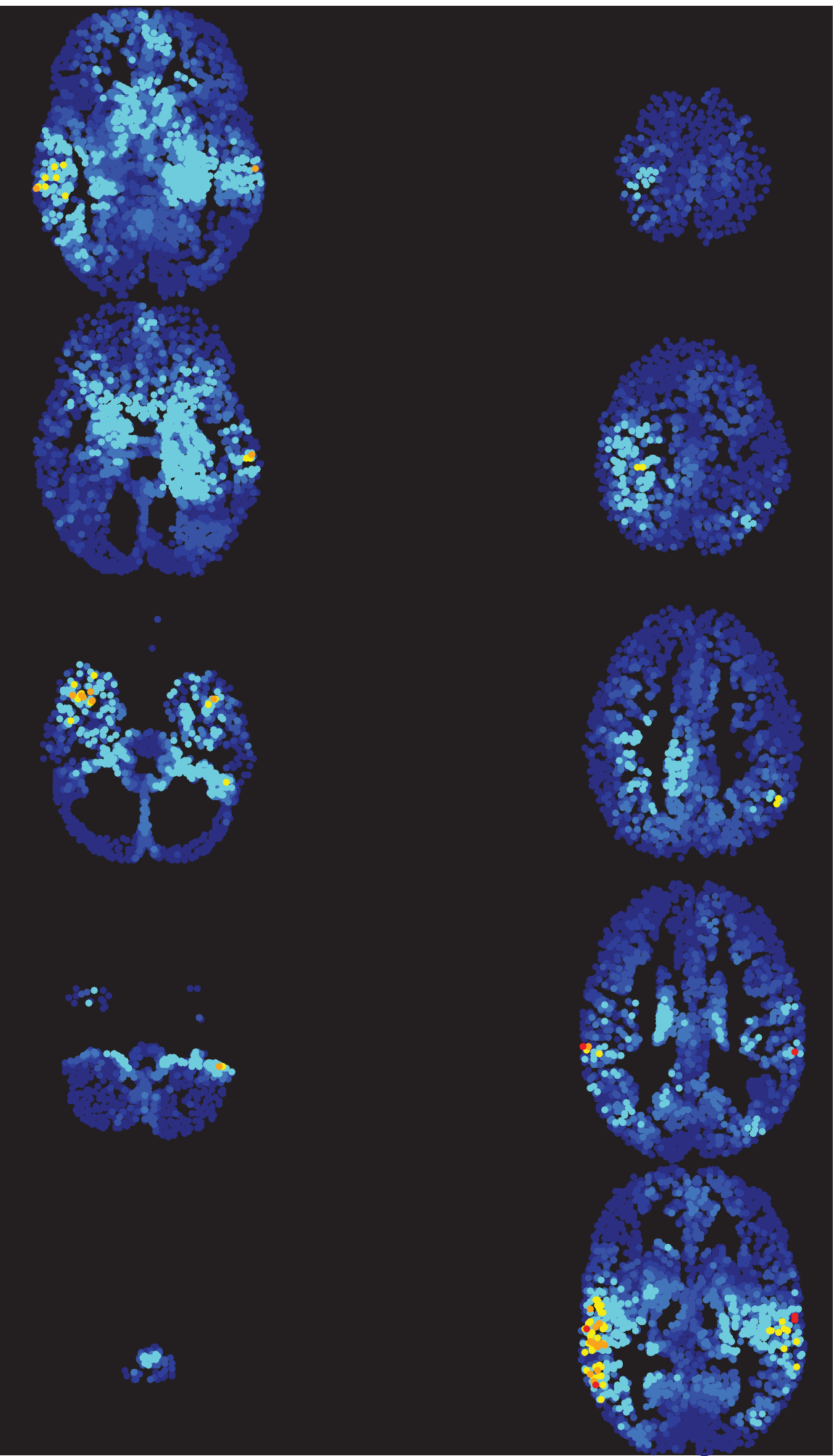}\hspace{0.5cm}
\includegraphics[width=9.5cm]{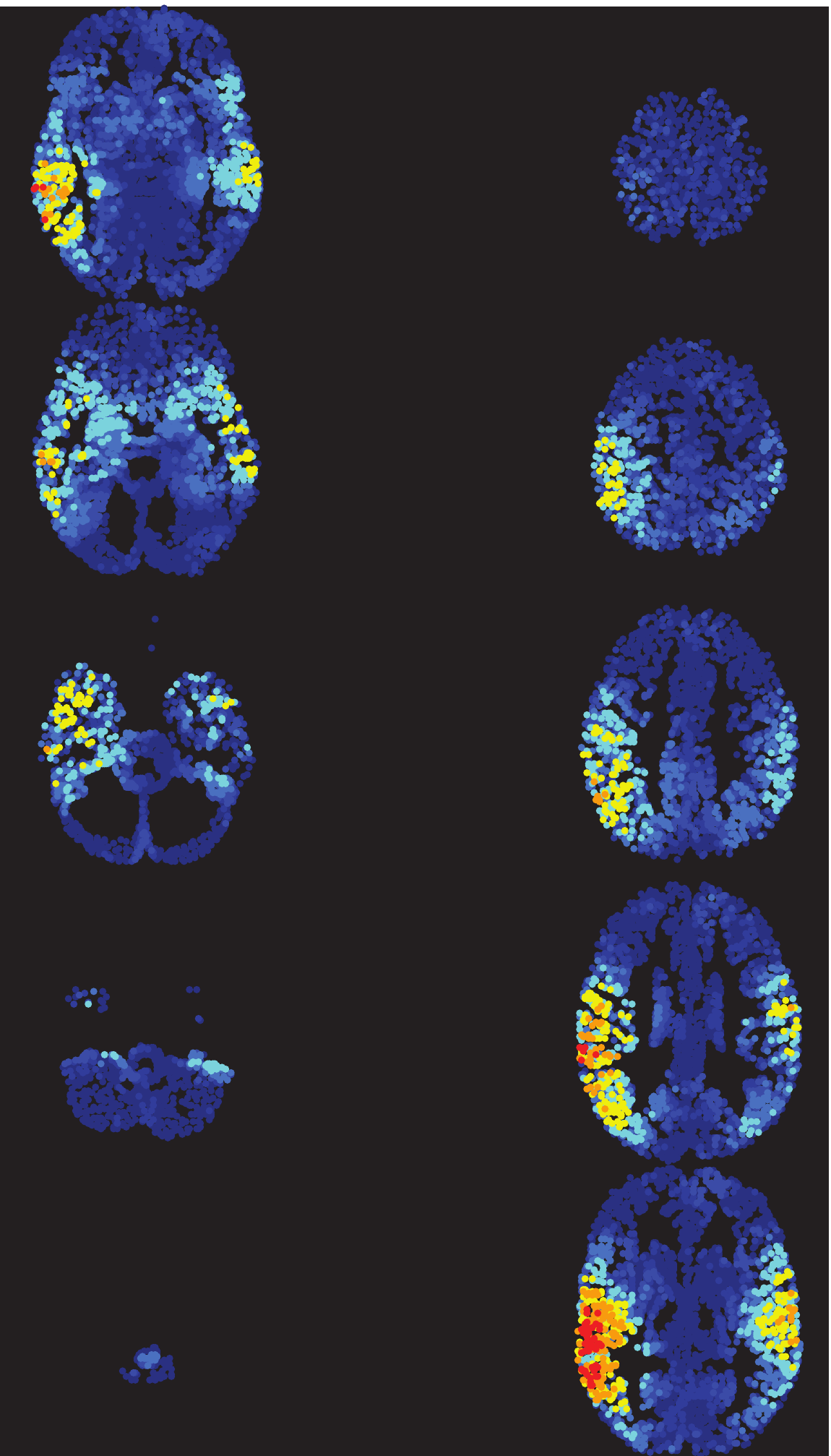}
}
\caption{\label{fig:slice_IAS_wMNE} Visualization of the axial view of the brain activity at 
$T=100$ ms as reconstructed by IAS (left) and wMNE (right).}
\end{figure}

\begin{figure}[tbh]
\centerline{
\includegraphics[width=9.5cm]{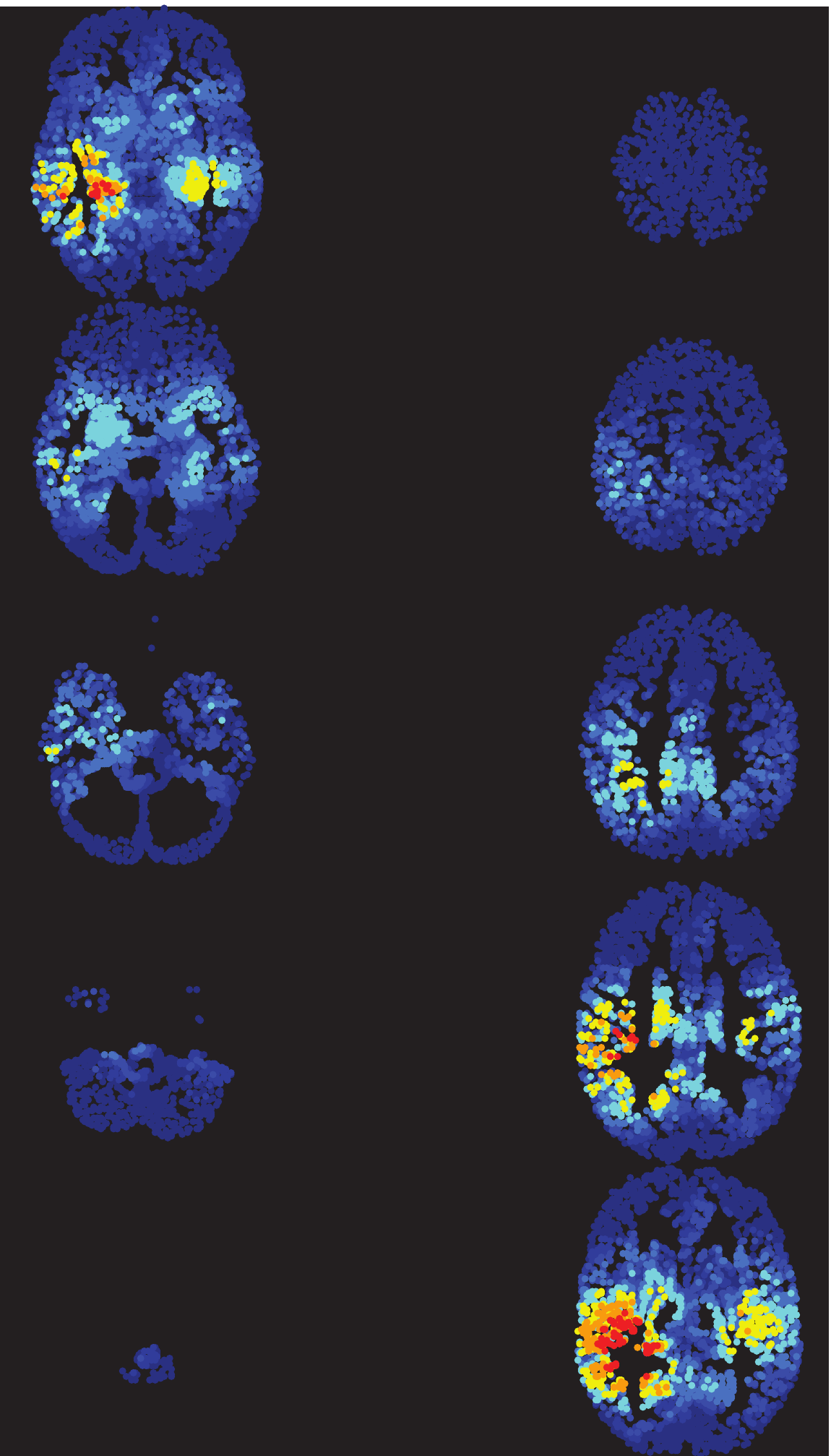}\hspace{0.5cm}
\includegraphics[width=9.5cm]{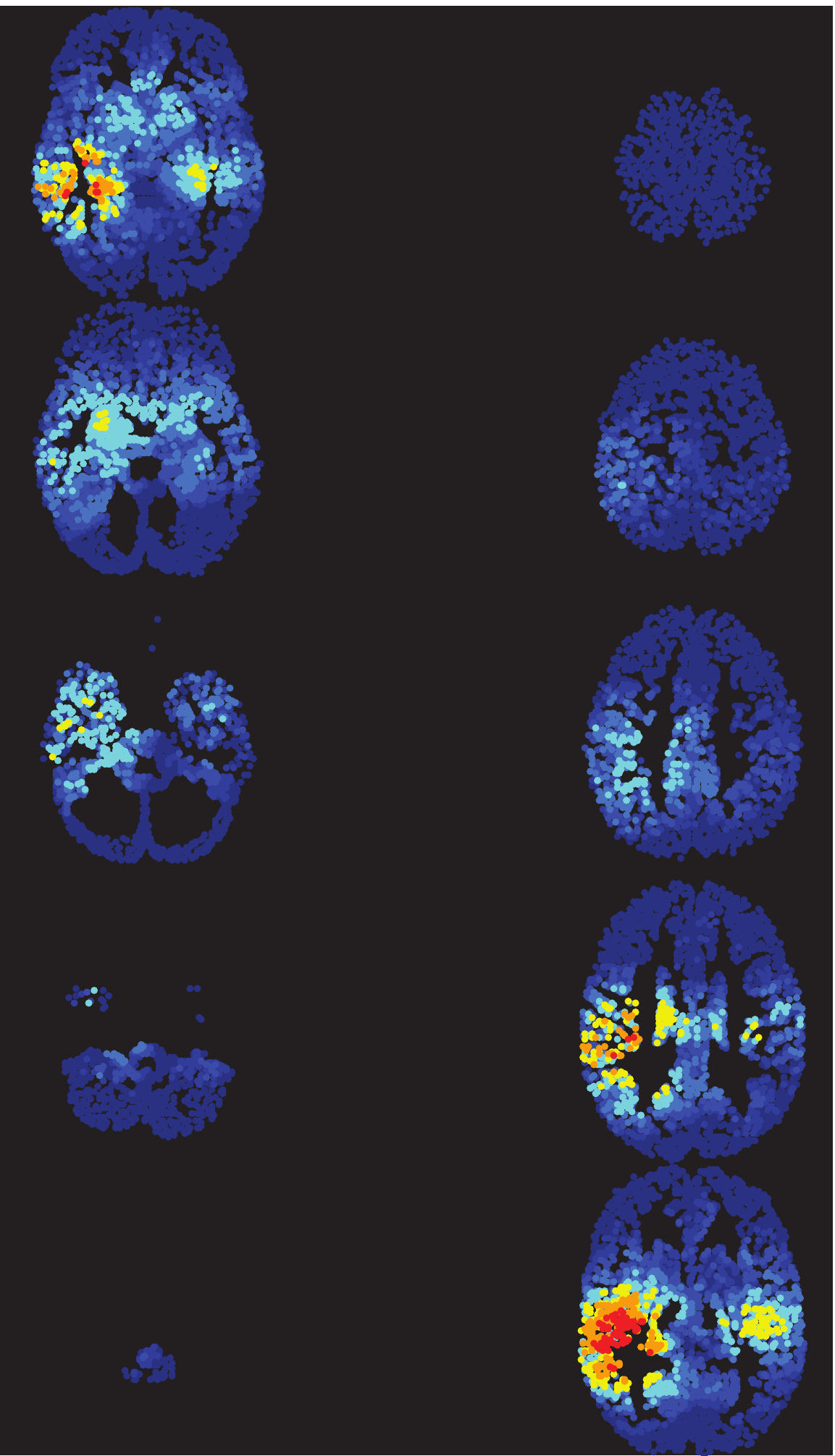}
}
\caption{\label{fig:slice_dSPM_sloreta}Visualization of the axial view of the brain activity  at
$T=100$ ms as reconstructed by dSPM (left) and sLORETA (right). }
\end{figure}

\begin{figure}[tbh]
\centerline{
\includegraphics[width=12cm]{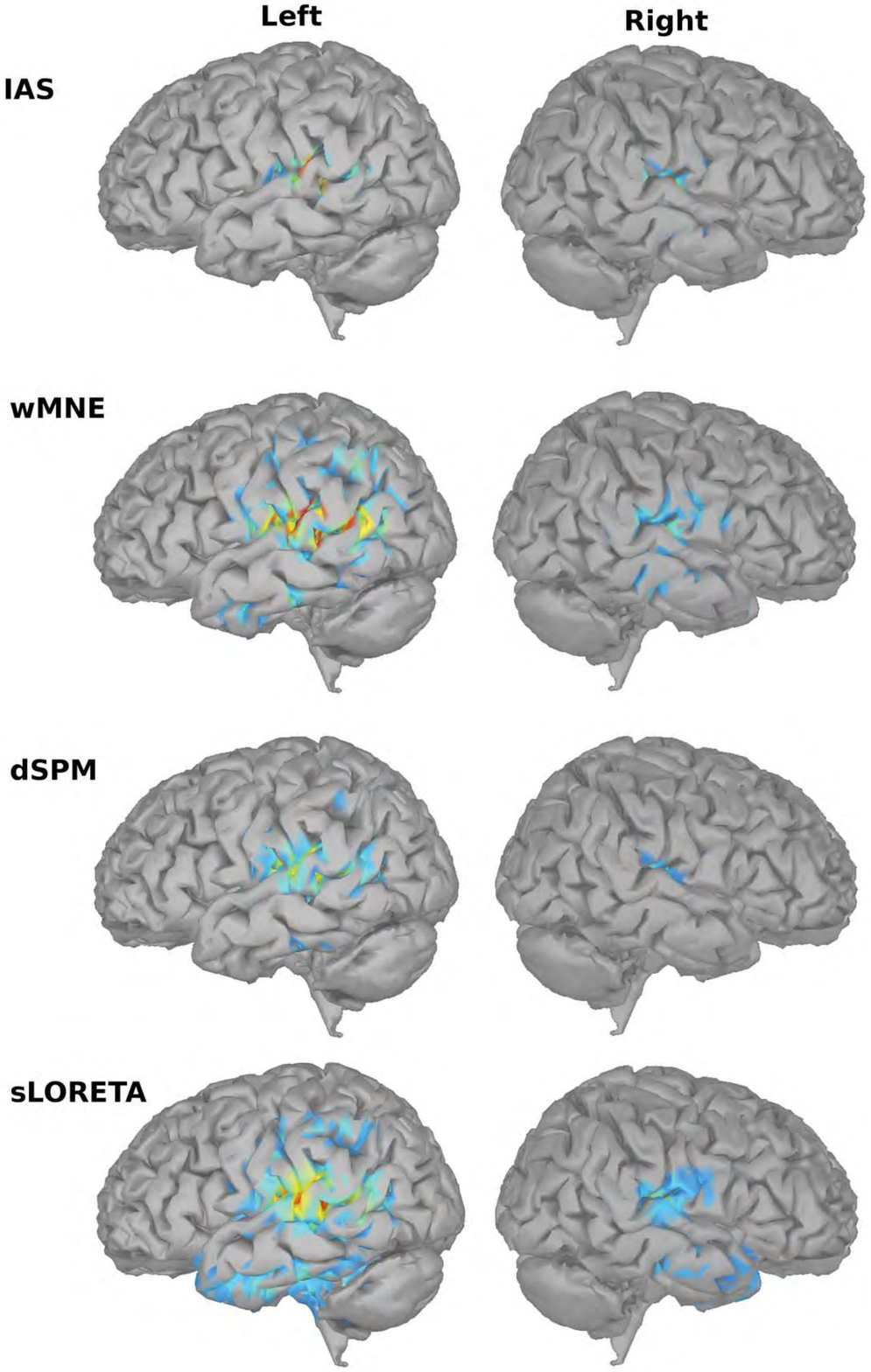}}
\caption{\label{fig:patch_M100}Reconstructions obtained from real data at time $T=100$ ms
with, top to bottom, IAS, wMNE, dSPM and sLORETA, respectively.}
\end{figure}

Finally, we apply the different inverse solvers to the real data, augmented with realistic activity in the brainstem as explained in Section~\ref{sec:real data}. Figures~\ref{fig:real_data_Baffo_5ms} and \ref{fig:real_data_Baffo} show
respectively the average BR-ALI vectors at times $T=5$ ms and $T=100$ ms for the four inverse solvers.
In Figures~\ref{fig:slice_IAS_wMNE_5} and \ref{fig:slice_IAS_wMNE} we show ten
axial slices of the two reconstructions of the brain activity at times $T=5$ ms
and $T=100$ ms obtained using the IAS (left) and the wMNE (right) inverse solvers. In
Figure~\ref{fig:slice_dSPM_sloreta_5} and ~\ref{fig:slice_dSPM_sloreta} we show
the corresponding reconstructions obtained by dSPM (left) and sLORETA (right). 
In Figures~\ref{fig:patch_brainstem_M100}, ~\ref{fig:patch_M100} we show the 3D
reconstructions obtained at $T=5$ ms and $T=100$ ms by the four inverse solvers.

\captionsetup[subfigure]{position=bottom}
\captionsetup[subfloat]{captionskip=-20pt}

\captionsetup[subfigure]{position=bottom}
\captionsetup[subfloat]{captionskip=-20pt} 

\section{Discussion}

\subsection{Sensitivity: single patch activity}

The sparsity of the  BR-ALI map is a good indicator for the sensitivity and specificity of an inverse solver: The more concentrated the histograms are on the active areas, the more likely it is that the activity is correctly identified without too much of confounding. In general, we expect a higher sensitivity and specificity for BRs closer to the sensors than in BRs deep in the brain, as confirmed by the computed experiments.

\subsubsection{Single patch in cortical BRs}

It emerges clearly from the panels in the top row of Figure~\ref{fig:LatOc_Baffo} that the IAS and wMNE inverse solvers have high sensitivity to activation restricted to the left lateral occipital cortex, indicated by the outstanding histogram bar, and that the confounding is confined mostly to anatomically proximal regions, most notably the nearby pericalcarine fissure in the same hemisphere. Neither algorithm suggest any significant activity in the right hemisphere, or in the deep brain structures. In the BR-ALI vector computed by the dSPM or sLORETA algorithms, on the other hand, the left lateral occipital BR is not as clearly identifiable, and the likely locus of activity is attributed to several nearby BRs in the same hemisphere, e.g., the nearby pericalcarine fissure, lingual, and cuneous BRs. Some of the brain activity is suggested also in the right hemisphere and in the deep brain structures. Overall, the sensitivity performance of the IAS algorithm appears to be significantly more similar to that of wMNE than to dSPM or sLORETA, the latter ones producing more confounding activity.

\begin{figure}[tbh]
\centerline{
\subfloat[\textbf{IAS}]{
\includegraphics[width=9.5cm,trim = 0.5in 0.1in 0.5in 0.3in, clip,
scale=0.5]{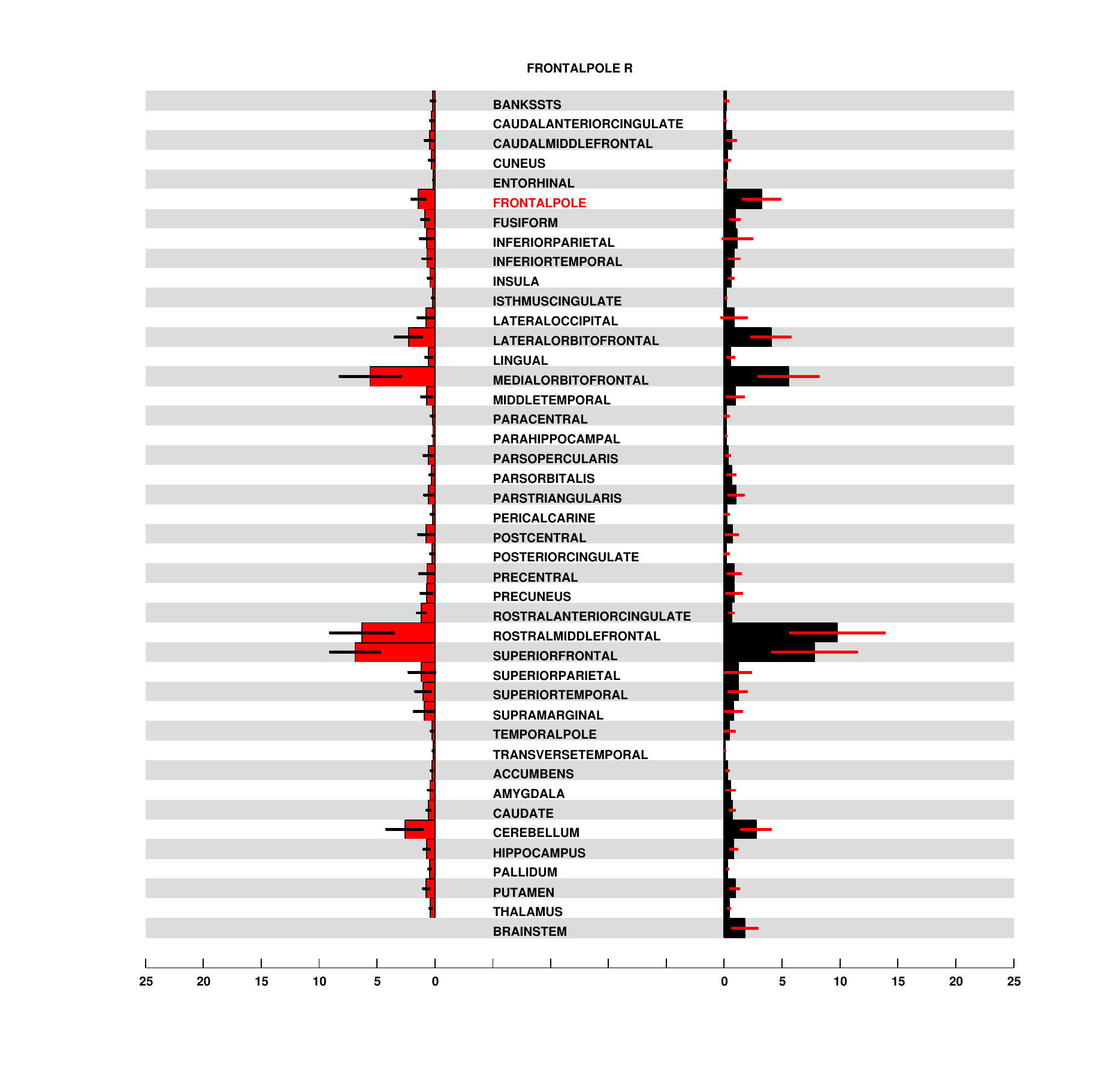}}
\subfloat[\textbf{wMNE}]{
\includegraphics[width=9.5cm,trim = 0.5in 0.1in 0.5in 0.3in, clip,
scale=0.5]{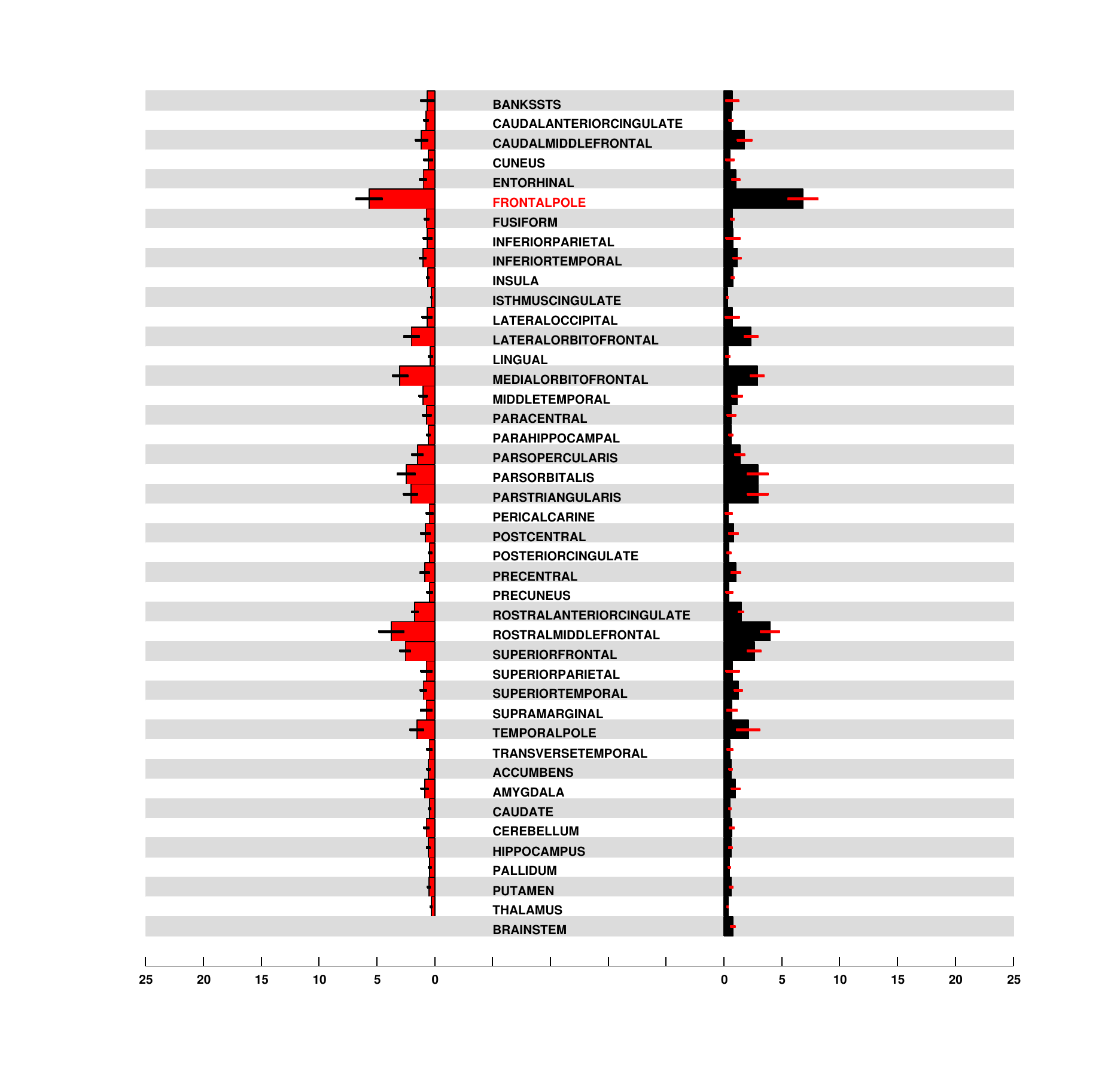}} }
\centerline{
\subfloat[\textbf{dSPM}]{
\includegraphics[width=9.5cm,trim = 0.5in 0.1in 0.5in 0.3in, clip,
scale=0.5]{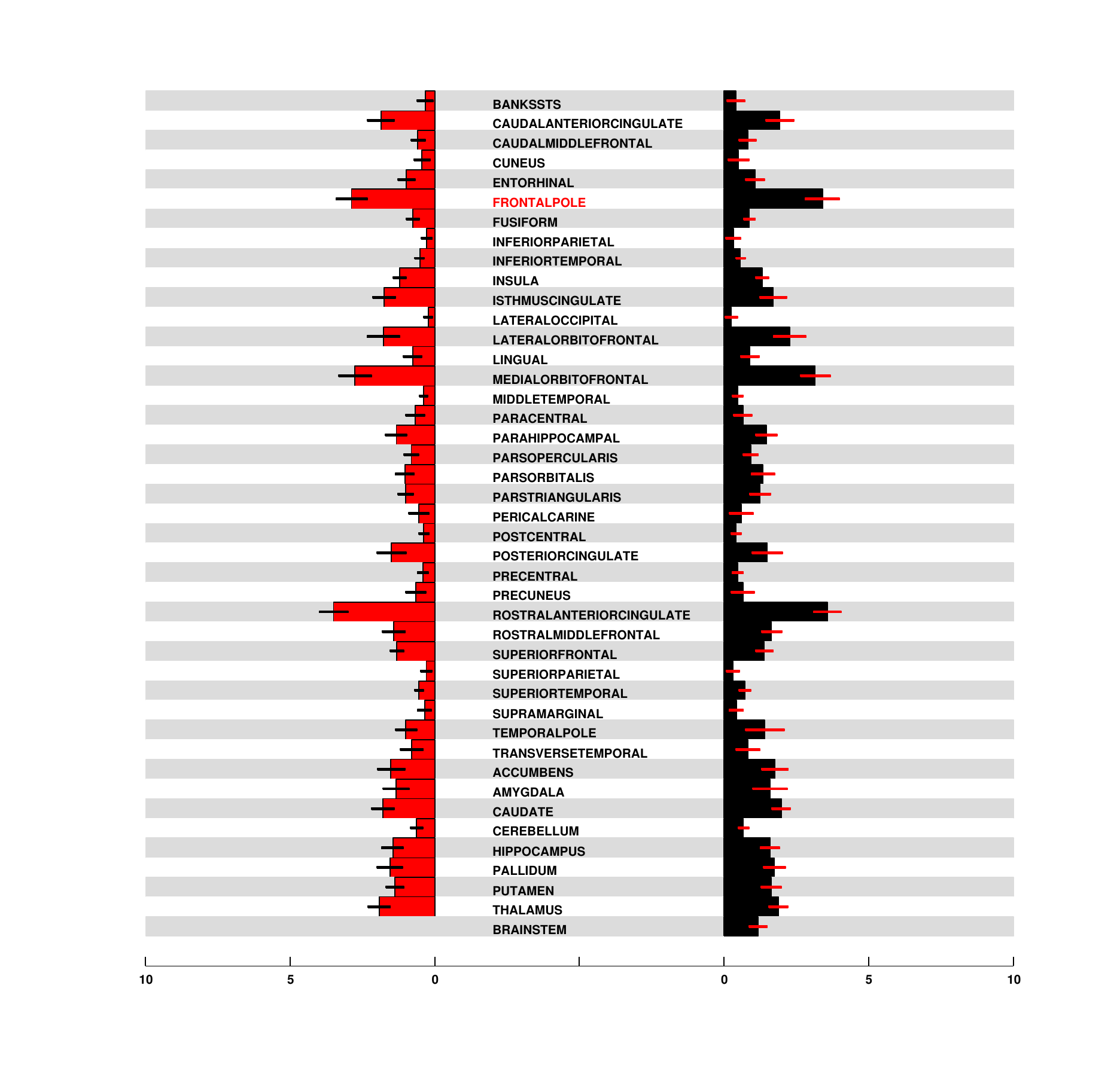}}
\subfloat[\textbf{sLORETA}]{
\includegraphics[width=9.5cm,trim = 0.5in 0.1in 0.5in 0.3in, clip,
scale=0.5]{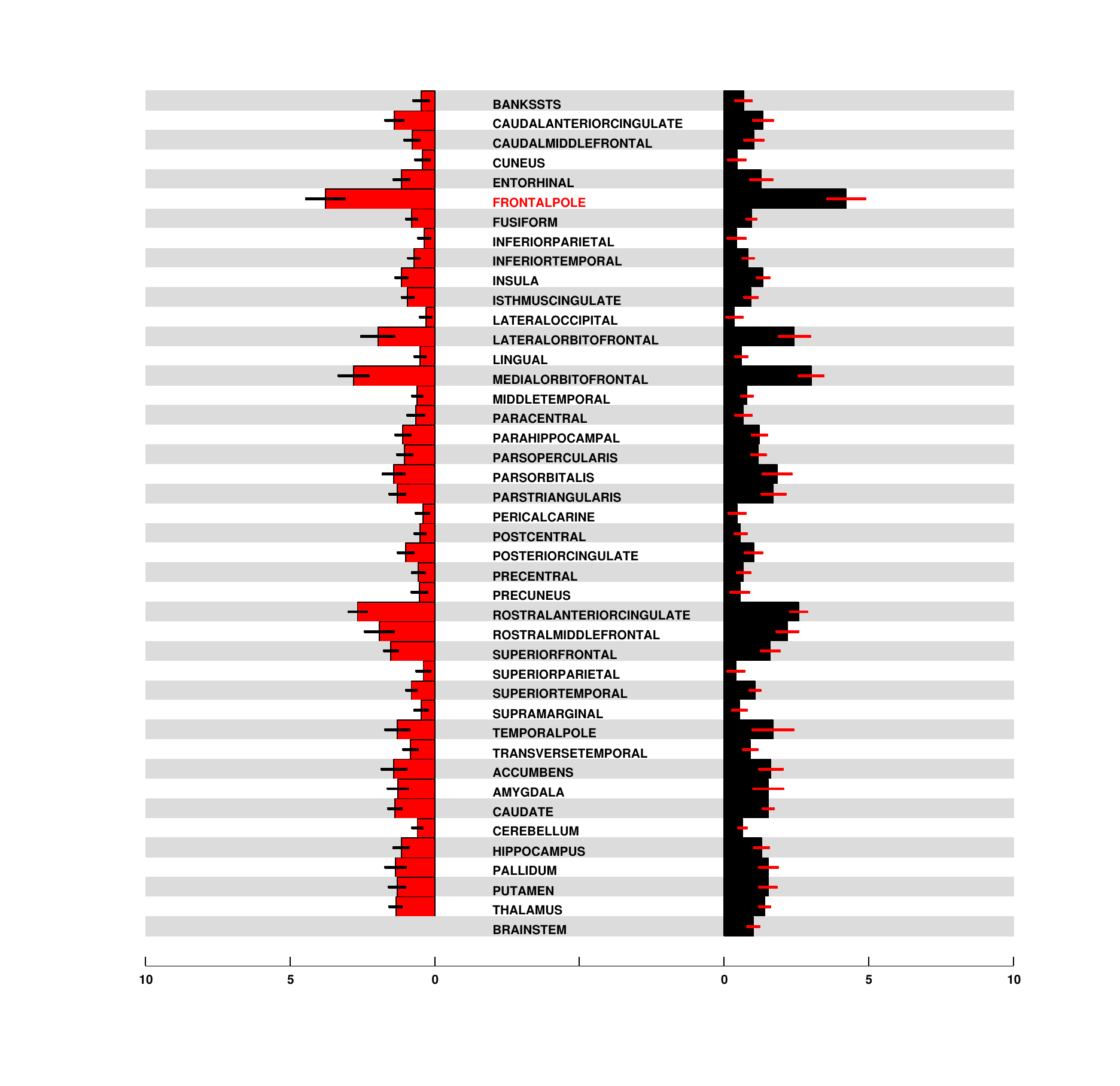}} }
\caption{\label{fig:FP_Baffo}Mapping of the brain activity to 85 different BRs  over 100
simulations using synthetic data corresponding to randomly generated activity patches in the {\em right frontal pole}, indicated in red in the list of the BRs reconstructed with, respectively,  IAS (a), wMNE (b),  dSPM (c) and sLORETA (d). The histograms bin the average activity in each BR: in red the BRs of the left hemisphere and in black the ones of the right hemisphere.}
\end{figure}

The suite of simulations with the activity confined to the right frontal pole region confirms the conclusions of the relative sensitivity of the four different approaches. The corresponding BR-ALI vectors, visualized in the form of histograms in Figure~\ref {fig:FP_Baffo}, show that both the IAS and wMNE algorithms reconstruct a substantial portion of the dipole activity in the right frontal pole, while finding some activity also in the nearby left frontal pole and, to a lesser extend, to the medial orbito frontal cortex, also anatomically close. In neither case, any significant leakage of the reconstructed activity to BRs in the deep brain occurs. The localization of the active BR is much weaker for the dSPM and sLORETA methods, whose BR-ALI vectors indicate a smearing of the reconstructed activity over several BRs, including subcortical ones. In this test sLORETA appeared to recognize the BR where the signal generating the data came from better than dSPM which attributes the highest average activity  to the rostral anterior cingulate cortex.

The first two tests are representative for the four methods when the data correspond to activity in a restricted patch located in a cortical BR, with IAS and wMNE performing quite satisfactorily when it comes to identifying the active BR, while dSPM and sLORETA have the tendency to favor activity in deeper regions of the brain.

\subsubsection{Single patch in subcortical BRs}

The sensitivities of the four solvers when the active patch is confined to the right cerebellum are summarized in Figure~\ref{fig:Cereb_Baffo}. As in the case of active patches in cortical regions, the attribution of the brain activity to the correct BR is less confounded when the inverse problem is solved using the IAS or wMNE algorithms. The BR-ALI vectors corresponding to these two algorithms point more clearly to the right cerebellum, with a little activity smeared to the left cerebellum and brainstem. The BR-ALI vectors for the dSPM and sLORETA inversion methods, on the other hand, tend to distribute the activity in the deeper regions of the brain, without the cerebellum standing out clearly among them. Moreover, a substantial fraction of the activity is mapped to cortical regions, e.g., lingual, parahippocampal and pericalcarine gyrus, suggesting potential problems when it comes to specificity.

\begin{figure}[tbh]
\centerline{
\subfloat[\textbf{IAS}]{
\includegraphics[width=9.5cm,trim = 0.5in 0.1in 0.5in 0.3in, clip,
scale=0.5]{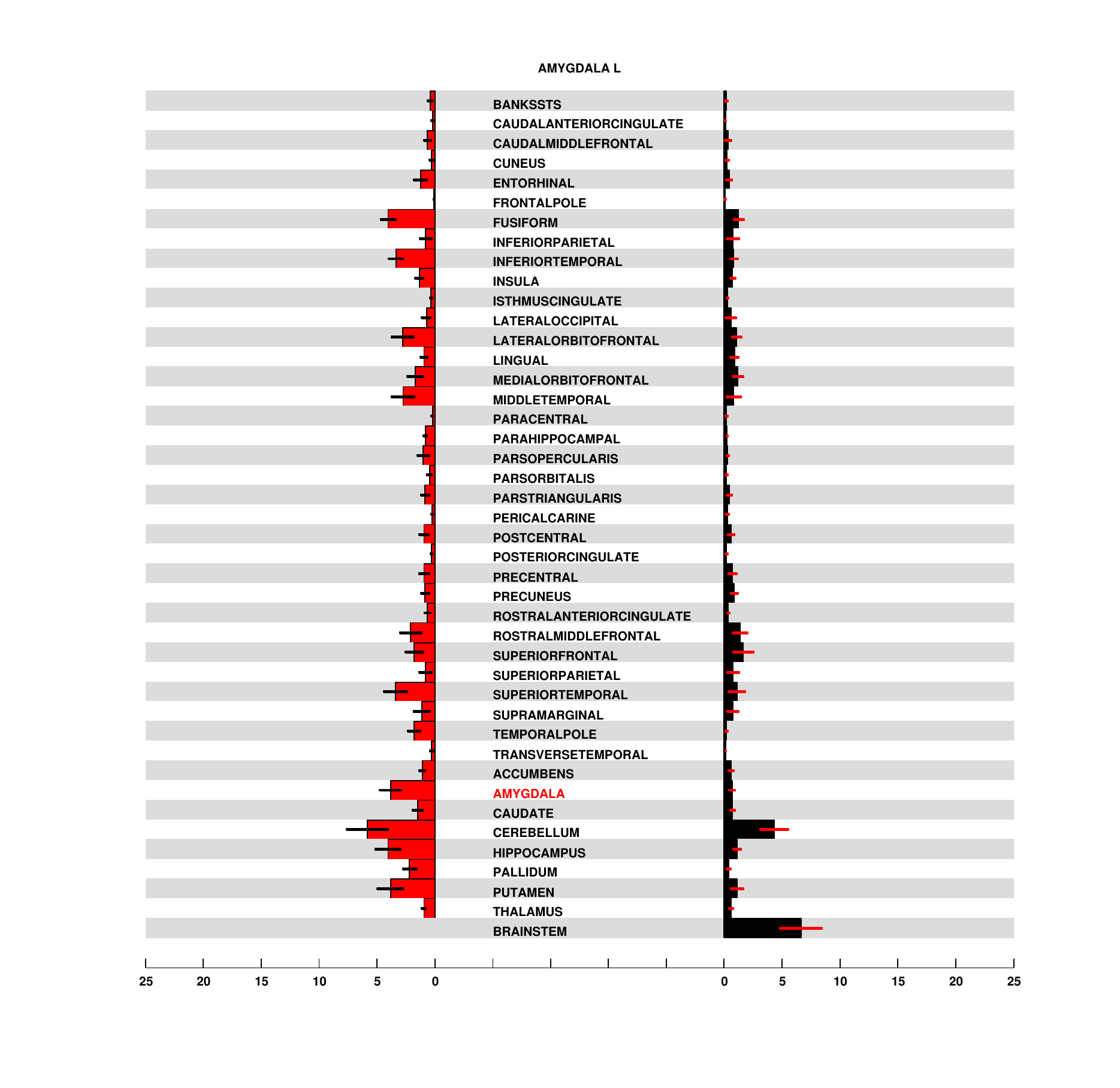}}
\subfloat[\textbf{wMNE}]{
\includegraphics[width=9.5cm,trim = 0.5in 0.1in 0.5in 0.3in, clip,
scale=0.5]{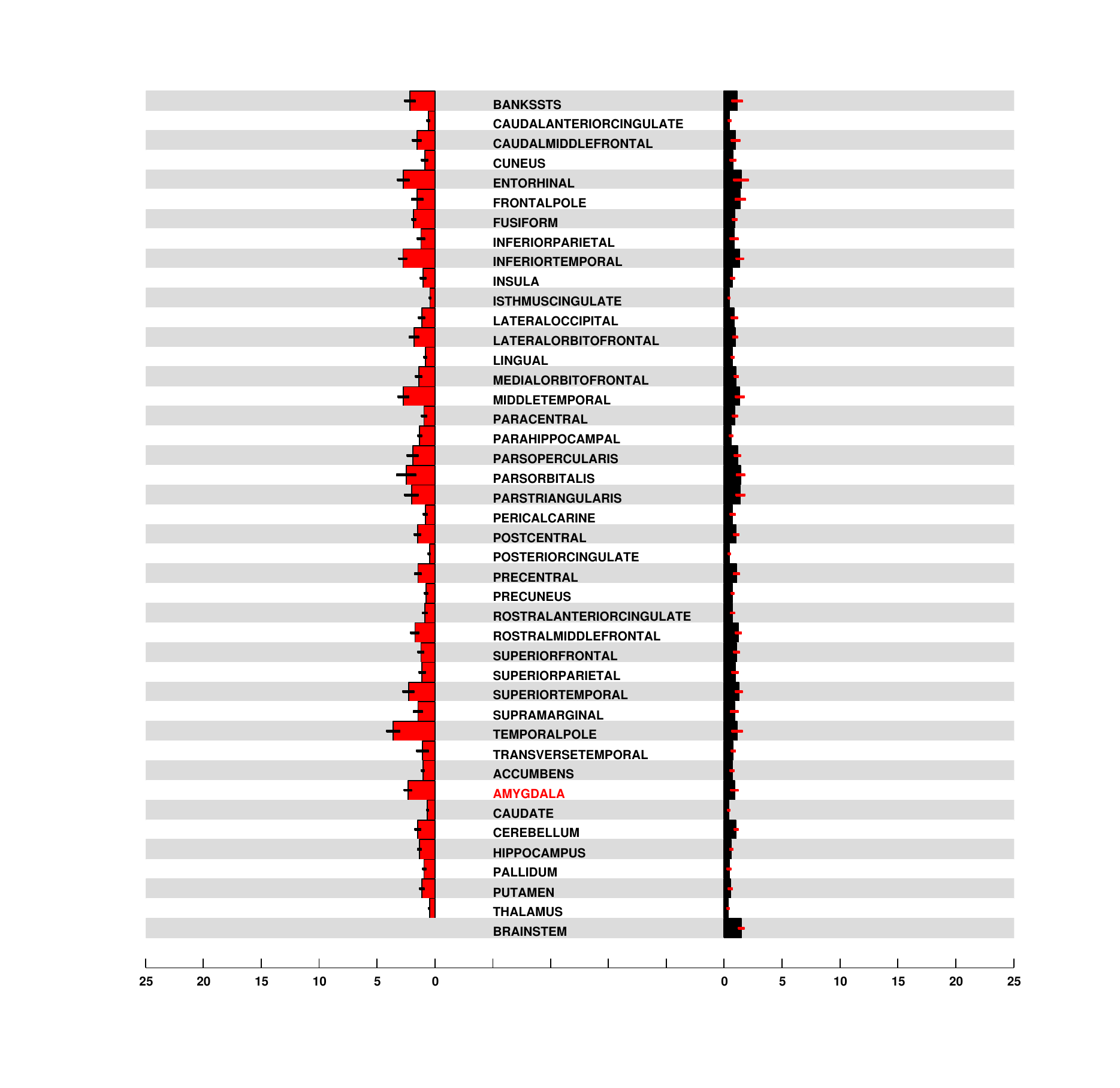}} }
\centerline{
\subfloat[\textbf{dSPM}]{
\includegraphics[width=9.5cm,trim = 0.5in 0.1in 0.5in 0.3in, clip,
scale=0.5]{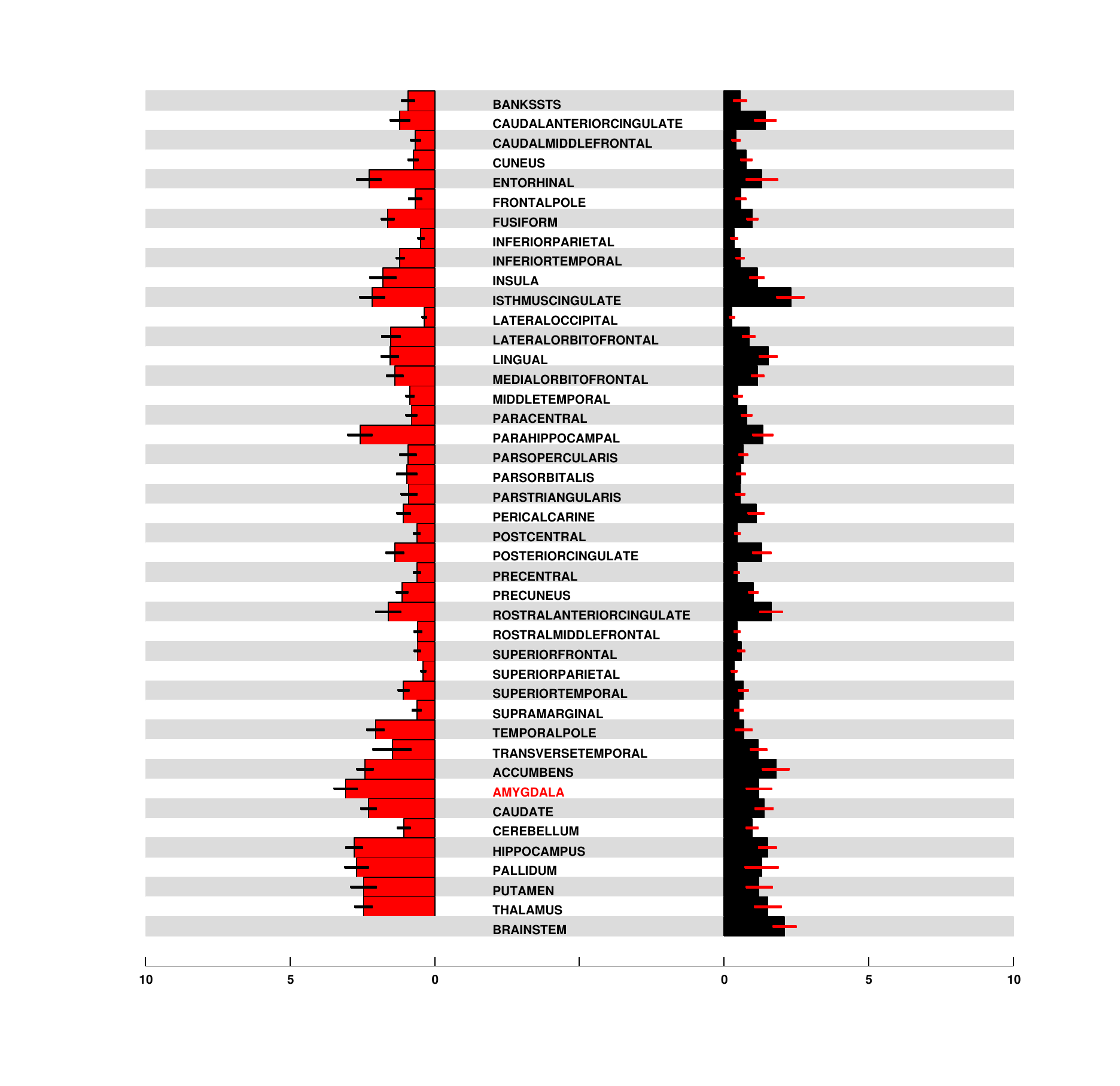}}
\subfloat[\textbf{sLORETA}]{
\includegraphics[width=9.5cm,trim = 0.5in 0.1in 0.5in 0.3in, clip,
scale=0.5]{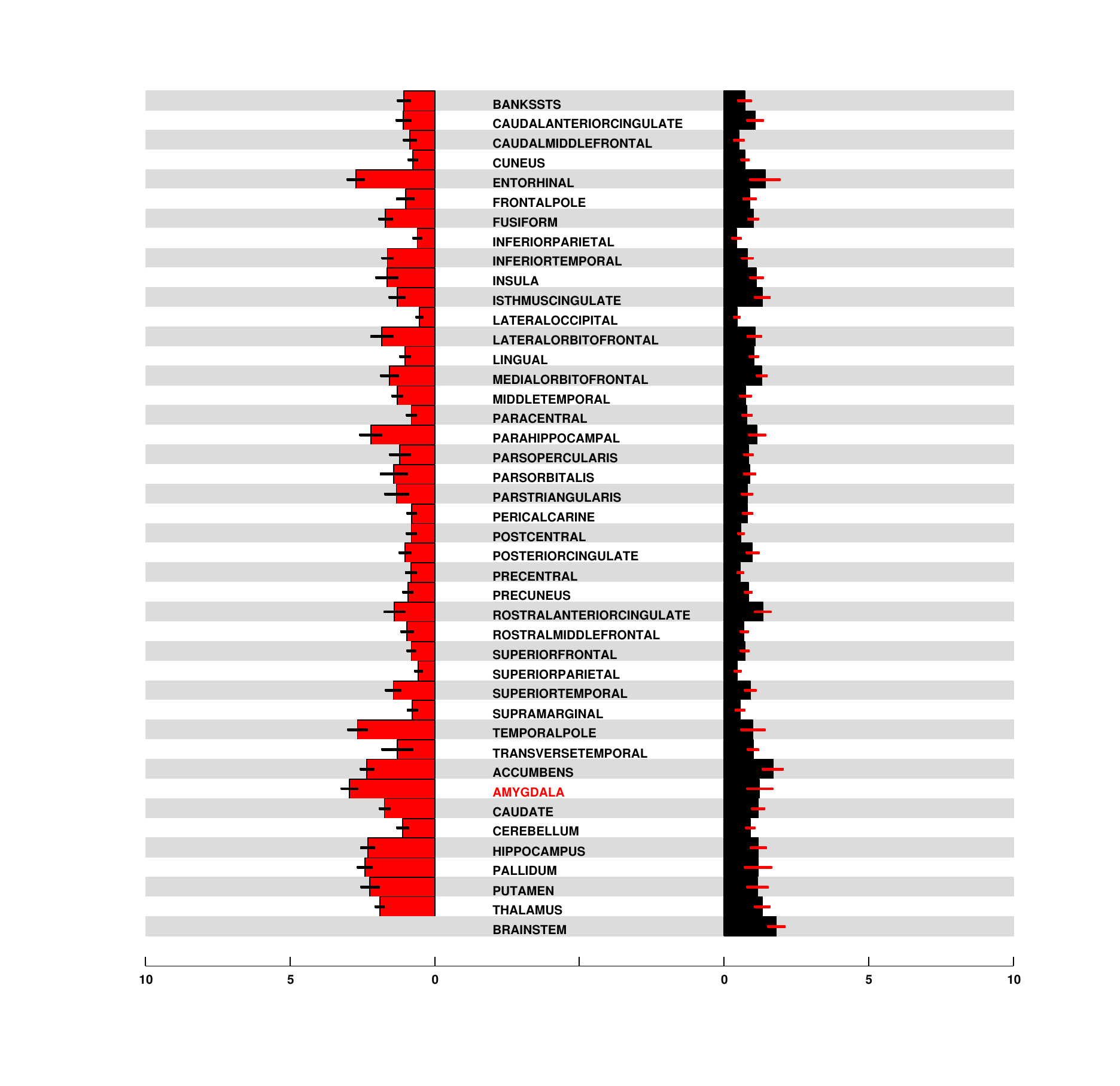}} }
\caption{\label{fig:Amyg_L_Baffo}Mapping of the brain activity to 85 different BRs  over 100
simulations using synthetic data corresponding to randomly generated activity patches in the {\em left amygdala}, indicated in red in the list of the BRs reconstructed with, respectively,  IAS (a), wMNE (b),  dSPM (c) and sLORETA (d). The histograms bin the average activity in each BR: in red the BRs of the left hemisphere and in black the ones of the right hemisphere. }
\end{figure}

The same patterns are observed when the active patch is in the left amygdala. As shown in Figure~\ref{fig:Amyg_L_Baffo}, the IAS inverse solver is quite effective at identifying the presence of activity in the left amygdala, with some leakage to the anatomically adjacent left temporal pole and left entorhynal cortex. The wMNE BR-ALI histogram does not show a localization to the left amygdala of the activity; instead, the activity is distributed over the left hemisphere, with a slight preference for the BRs closer to the left amygdala.  The BR-ALI vectors relative to dSPM  and sLORETA, on the other hand, map the reconstructed activity without much distinction onto all BRs corresponding to internal structures, and a few cortical BRs in the proximity of the left amygdala, suggesting a bias towards regions of the brain away from the surface.

\subsection{ Specificity analysis via Bayes factors}

The specificity analysis using Bayes factors is based on the counts of how often, and at which level, out of the $2\,000$ tests per brain region, the reconstruction of the brain activity by a given algorithm supports the hypothesis  that the activity is indeed in the correct region. 
Due to our choice of color coding, we can be confident that the activity within a BR when the corresponding histogram has taller green or blue bars can be identified by the inverse solver rather unequivocally, while taller black or red bars are an indication that activity in that BR is likely to be erroneously attributed to activity in another part of the brain.

\subsubsection{Single active patch}

The overwhelming predominance of green in the left panels of
Figure~\ref{fig:BF IAS}, displaying the histograms of the Bayes factors for the cortical BRs, is a strong indicator that the solution of the MEG inverse problem produced by the IAS algorithm is much more likely to find active current dipoles in the correct BR than in other regions. A slightly lesser dominance of the green and blue bars for the insula and isthmus cingulate BR suggests that activity concentrated in either of these two regions may have a slightly larger tendency to be incorrectly attributed elsewhere. In most of the Bayes factors histograms for the subcortical BRs the green and blue bars dominate over the red and black ones, with the exception of the thalamus and, to a lesser extent, the caudate gyrus, where the red and black take over, suggesting that the IAS has difficulties to identify unequivocally activity localized in this BR.

In the histograms of the classification of the Bayes factors for the wMNE
algorithm, shown in the right panels of Figure~\ref{fig:BF IAS}, the green and
blue bars tend to be more prominent than the red and black ones in the top row, although not as markedly as for IAS.  This is an indication that when the measured signal comes from a patch in a cortical BR, the solution computed by wMNE is likely to be concentrated in the correct cortical region, with the exceptions for the caudal anterior cingulate, insula, isthmus cingulate, parahippocampal, posterior cingulate and rostral anterior cingulate regions.

The situation is quite different when it comes to correctly attributing activity generated in a subcortical region, as shown in the two panels in the bottom row of the same figure. Here the taller bars are red or black, with the exception of the cerebellum and, to some extent, the brainstem, pointing to a low specificity of wMNE in subcortical regions.

The Bayes factor analysis for dSPM is summarized by the histograms in
the left panels of Figure~\ref{fig:BF DSPM}.  In the histograms pertaining the
cortical BRs the blue and green bars dominate the red and black ones, although the green presence is not as massive as for the IAS or wMNE inverse solvers, indicating that the dSPM inversion is still capable of identifying active cortical BR, but with lower specificity than IAS and wMNE. In many of the cortical BRs the height of the blue bars varies between 800 and 1000, but there is a strong presence of red bars, with height between 200 and 700, pointing to the fact that in some cases, dSPM has difficulties with recovering the activity in the correct region and attributing it instead, to a different region.  The dominance of blue and red in the histograms for the subcortical BRs, together with a green presence, suggests that the localization of the activity in a patch deeper in the brain by dSPM is more accurate than with wMNE, but not as precise as with IAS.

The conclusions of our specificity study with Bayes factors for sLORETA, whose
histograms are displayed in the right panels of Figure~\ref{fig:BF DSPM} are
very similar to those for dSPM, with a better performance than wMNE, but not as good as IAS,  in subcortical BRs and not as specific when it comes to recovering activity in cortical BRs.

\subsubsection{Multiple activity patches}

The histograms of the Bayes factor for the IAS and the three other reference methods, in case where there are two patches of activity, one in the left amygdala and the other in the frontal pole, displayed in Figure~\ref{fig:BF Amyd_FP}, show that the reconstructions computed with the IAS suggest very strongly the presence of activity in the left amygdala, as shown by the tall green bar in the corresponding histogram as well as that in the cortical frontal pole. In this experiment, sLORETA outperforms both wMNE and dSPM, the latter showing the least precision in mapping the activity in the frontal pole area.

The second test with two patches of activity, in the left cerebellum and in the left precentral region, respectively, confirm that the IAS is the most precise of the four inverse solvers when it comes to finding the location of both active patches, as shown in the histograms in Figure~\ref{fig:BF_Cereb_PC}. The performance of wMNE in this case is very good for both the cortical and subcortical BRs, while in this case the specificity of sLORETA is not as good as that of dSPM.

The Bayes factors for the case where there are three active patches,  located in right precentral gyrus, left hippocampus and right thalamus, are summarized in Figure~\ref{fig:BF_Thal_Hip_PC}. The IAS algorithm identifies with least confounding the activity in the precentral region and hippocampus, but is not as accurate when it comes to the thalamic patch. Both interior patches  turn out to be very challenging for wMNE, which has no problems with accurately recovering the patch in the precental region. In this case, dSPM is best at localizing the activity in the thalamus and in the hippocampus.

In the last protocol with Bayes factors, the MEG signal came from six active patches located in the six BRs involved in the DMN. The histograms for the six different BRs with the four inverse solvers, displayed in Figure~\ref{fig:BF_DMN}, indicate that the localization of the activity in the caudal anterior cingulate region in either hemisphere is challenging for all methods, and particularly so for wMNE whose histograms show the black bars to be tallest. The localization of the activity in the inferior parietal BRs is resolved very well by IAS and wMNE, and satisfactorily by dSPM and sLORETA.

\subsection{ Real data: IAS favors sparse solution}

Figures \ref{fig:real_data_Baffo_5ms}  and \ref{fig:real_data_Baffo} show the BR-ALI vectors for the four inverse methods at times T=5 ms and T=100 ms, respectively. At T=5 ms all inverse methods find activity also but not only in the brainstem region: the estimated activities using sLORETA and dSPM are mutually very similar, suggesting activation also in the amygdala and in the parahippocampal cortex; the deep activity is dominating over cortical activity in the reconstructions. The reconstructed activity by wMNE, compared to IAS, is more smooth, while IAS finds activity also in cerebellum. At T=100 ms, IAS suggests activity in the temporal region, mainly in the upper bank of the superior temporal cortex and in the transverse temporal region, that are part of the primary auditory cortex. The IAS MEG solution shows activity in the superior temporal regions of both hemispheres, while when using other methods such bilaterality is not evident. In general, the dSPM and sLORETA algorithms seem to be in favor of activity deeper in brain, and the spatial smoothing characteristics of the algorithms is clearly visible. 

\subsection{Methodological issues and future work}

As shown in Section 2.5, the choice of the parameters in the IAS algorithm can be driven by their physiological meaning. Several tests we performed show that
the reconstructed activity map is not very sensitive to the values of the parameters $\delta$, $\theta_{\rm max}$ and $\xi$. The values
we used in the tests are reasonable for most applications in neuroscience. Thus, the parameter that the user has really to select is $\eta=\beta-\frac 52$. 
Since $\eta$ controls the sparsity of the reconstruction, its choice is related to the a priori knowledge we have on the protocol of the neuroscience 
experiment under study. 
Values of $\eta$ in the order of 0.1 favor more spread reconstruction and for these values the IAS algorithm produces activity maps very similar to 
the maps obtained by the wMNE algorithm.
On the other hand, values in the order of $0.001$ favor focal reconstructions,
which is the more interesting case in neuroscience studies, especially when deep brain sources are involved.

In a future work we want to analyze in more details the sensitivity of the IAS algorithm on the localization of deep brain activity.
To this end, we need a more accurate model for describing deep brain sources, e.g., the model proposed in \cite{attal,attal2013}
or the mixed source space model available in the MNE software\footnote{\tt http://martinos.org/mne}. 
In this case, different values of the parameter $\eta$ for cortical and subcortical regions can be used
in order to increase the sensitivity of the algorithm to deep sources.
Finally, to take into account the imprecision due the use of an averaged physiological atlas that cannot reproduce exactly the individual anatomy, 
one can use some mathematical methods related to fuzzy logic (see, e.g., \cite{algorri2004,ciofolo2009}).

The IAS algorithm described in this paper is designed for a single-time analysis: 
the algorithm is re-initialized each time and does not retain any previous information.
Actually, one of the main advantage of the MEG devices is in that they can measure the neuromagnetic field with a high temporal resolution.
In order to deal with MEG time series some preliminary results show that when the IAS algorithm starts from a $\theta^*$ that is related to 
the values of $\theta_j$ at the previous time step, the rate of convergence of the algorithm is increased. 
Then, for a further analysis we can model $\theta$ as a Markov process, so the estimation of $\theta$ at time 
$t$ depends just on the  value of $\theta$ at time $t-\delta t$.

\section{Conclusions}

Finding a robust metric for assessing the performance of an inverse solver in MEG is not a simple task. Algorithms that are based on the goal of localizing single dipoles may be judged according to the precision of the localization, but such metric may not be a judicious one for methods that estimate distributed sources. Vice versa, single dipole methods may have a limited success when distributed activities are to be estimated. In this paper we propose a metric for the algorithm assessment based on the reliability of an algorithm to identify active brain regions, superficial and deep, regardless of whether a single or several active regions occur. The precision of finding a single dipole is not of concern here, although the focality of the reconstructions help discerning between brain regions that are anatomically close. To avoid the pitfall of anecdotal successes, the methodology is based on extensive independent Monte Carlo sampling, and the results are processed into a form of first and second order statistics, and Bayes factor analysis. The long-term goal of this work is to build reliable uncertainty quantification tools to analyze extensive data sets with non-event based brain data, e.g., various resting states or states of consciousness, brain connectivity, or fingerprinting of diseased or abnormal states. Assessment of success rates of algorithms as presented here, as opposed to precision case studies with activity localization, is in line with the current algorithm testing paradigm in data mining and big data analysis. The proposed methodology was tested with four inverse solvers, one of which is the recently developed IAS algorithm, and three standard methods. Our conclusion is that the IAS method seems to perform relatively consistently in the tasks that it is originally designed for, that is to identify both active cortical and deep brain regions without a significant confounding beyond the inevitable leakage of the estimated activity to anatomically close regions, which is due to the inherent ill-posedness of the problem.

\section*{Appendix: Interpretation of hyperparameters}

In \cite{IAS}, the interpretation of the hyperparameters $\beta\in\R$ and $\theta^*\in\R^N$ was discussed: It was shown that $\beta$ allows the user to control the sparsity of the IAS solution, while the empirical Bayesian approach provided a way to relate $\theta^*$ to the sensitivity scaling. We summarize here the analysis on hyperparameters, developing the discussion of $\theta^*$ further, so that the parameter tuning can be done easily and semi-automatically.

\subsection*{Parameter $\beta$ and control of sparsity}

In \cite{IAS}, it was proved (Theorem 2.1) that the sequential minimization that constitutes the core of the IAS algorithm can be interpreted as a fixed  point iteration to find a minimizer $\widehat Q = [\widehat{\vec q}_1,\widehat{\vec q}_2,\ldots,\widehat{\vec q}_n]^\mT\in\R^{3n}$ of the energy functional (\ref{energy}),
\[
 \widehat Q = {\rm argmin}\big\{ {\mathscr E}(Q,S(Q))\big\}, \quad \widehat \Theta = S(\widehat Q),
\]
where $\widehat\Theta = [\widehat\theta_1;\widehat\theta_2;\ldots,\widehat\theta_n]\in\R^n$, and $S:\R^{3n} \to \R^n$ is defined componentwise as
\[
 \theta_j  =  S_j(\vec q_j) = \theta_j^*\left(\frac{\eta_j}2 + \sqrt{ \frac{\eta_j^2}{4} + \frac{\|\vec q_j\|_{{\mathsf C}_j}^2}{2\theta_j^*}}
 \right),\quad \eta_j = \beta_j -\frac52, \quad 1\leq j\leq n.
\]
Furthermore, it was shown that if we write $\beta_j = 5/2 + \eta$, $1\leq j\leq n$, then, as $\eta \to 0^+$, we have the asymptotic expression
\begin{equation}\label{asymptotics}
  {\mathscr E}( Q, S(Q))
=  \frac 12 \|b -\sum_{j=1}^n \mM_j\vec q_j\|^2_{\mathsf \Sigma} +
  {\sqrt{2}} \sum_{j=1}^n \frac{ \|\vec q_j\|_{{\mathsf C}_j}}{\sqrt{\theta_j^*}} + {\mathscr O}(\eta).
 \end{equation}
 In particular, the penalty term in the above expression is a weighted
 $\ell^1$-norm for the dipole amplitudes that are measured in the metric
 defined by the anatomical prior matrices ${\mathsf C}_j$. This argument
 demonstrates that at the limit, the IAS algorithm provides an effective
 algorithm for finding a weighted Minimum Current Estimate (MCE), with the
 modification given by the anatomical prior \cite{Uutela}. As a conclusion, we
 see that the role of the hyperparameter $\beta$ is to control the sparsity of
 the IAS estimate. In \cite{IAS}, this effect was demonstrated using numerical
 simulations.

\subsection*{Parameter $\theta^*$ and sensitivity}

The asymptotic expression (\ref{asymptotics}) is indicative also from the point of view of the interpretation of $\theta_j^*$. It is well-known that MEG algorithms based on penalized minimization of the fidelity to data tend to favor superficial sources, and to compensate this effect, a sensitivity weight is often introduced, see, e.g. \cite{Lin}. From the Bayesian point of view, sensitivity weighting is a problematic practice, since traditionally the prior should be independent of the observation model, a condition that the sensitivity weight does not satisfy. However, it is possible to find a satisfactory Bayesian interpretation for $\theta_j^*$ so that it effectively works as a sensitivity weight. The connection between sensitivity and hypermodels is built through the analysis of the signal-to-noise ratio as follows. Consider the linear forward model
\[
 b = \mM Q + \varepsilon = \sum_{j=1}^n \mM_j\vec q_j + \varepsilon = b_0 +\varepsilon,\quad \varepsilon\sim{\mathcal N}(0,{\mathsf \Sigma}).
\]
The signal-to-noise ratio (SNR) is defined as
\[
 {\rm SNR} = \frac{\mE\{\|b_0\|^2\}}{\mE\{\|\varepsilon\|^2\}},
\]
where
\[
 {\mE\{\|\varepsilon\|^2\}} = {\rm trace}({\mathsf\Sigma}),
\]
while from the prior model, conditional on $\Theta\in\R^n$, we have
\[
 \mE\{\|b_0\|^2\mid\Theta\} = \sum_{j=1}^n \theta_j{\rm trace}\big(\mM_j {\mathsf C}_j \mM_j^{\mT}\big) = \sum_{j=1}^n \theta_j \|\mM_j{\mathsf C}_j^{1/2}\|^2_{\rm F},
\]
where the subscript refers to the Frobenius norm of the matrix. Furthermore, by using the Gamma hyperprior model $\theta_j \sim\Gamma(\beta,\theta_j^*)$ for the vector $\Theta$, we arrive at
\[
\mE\{\|b_0\|^2\} = \sum_{j=1}^n \mE\{\theta_j\} \|\mM_j{\mathsf C}_j^{1/2}\|^2_{\rm F} = \sum_{j=1}^n \beta \theta_j^* \|\mM_j{\mathsf C}_j^{1/2}\|^2_{\rm F}.
\]
The choice of the hyperparameters $\theta_j^*$ must therefore be compatible of what we {\rm a priori} assume about the distribution of the activity and the resulting SNR.  To begin with, assume that we have a reason to believe that only one source is active, but we do not know which one. Then, the active source $j_1$ must satisfy
\[
 \beta \theta_{j_1}^*  \|\mM_{j_1}{\mathsf C}_{j_1}^{1/2}\|^2_{\rm F} = {\rm SNR}\times{\rm trace}({\mathsf\Sigma}),\quad \mbox{ or }\quad   \theta_{j_1} ^*= \frac {{\rm SNR}\times{\rm trace}({\mathsf\Sigma})}{\beta \|\mM_{j_1}{\mathsf C}_{j_1}^{1/2}\|^2_{\rm F}}.
\]
This must be true, whichever the active source is, and if each source has equal probability to be active, the exchangeability argument yields the scaling law
\[
\theta_{j} ^*= \frac {{\rm SNR}\times{\rm trace}({\mathsf\Sigma})}{\beta \|\mM_j{\mathsf C}_j^{1/2}\|^2_{\rm F}}, \quad 1\leq j\leq n.
\]
This argument can be generalized to several, and unknown, number of active sources. Assume that we believe that $k$ of the sources are non-zero, but we do not know which ones. Denoting the indices to the active sources by $j_1,j_2,\ldots,j_k$, we must have
\begin{equation}\label{k sources}
 \sum_{\ell=1}^k \beta \theta_{j_\ell}^* \|\mM_{j_\ell}{\mathsf C}_{j_\ell}^{1/2}\|^2_{\rm F} = {\rm SNR}\times{\rm trace}({\mathsf\Sigma}).
\end{equation}
For the exchangeability argument, let us denote by $\gamma\in\R^n$ the vector with entries $\gamma_j = \beta \theta_{j}^* \|\mM_{j}{\mathsf C}_{j}^{1/2}\|^2_{\rm F}$, and by $\mP_k\in\R^{n_k\times n}$ the matrix such that the $p$th row contains exactly $k$ entries equal to one, other entries being zero, and each permutation appears only once in the matrix. Therefore,  the number of rows in $\mP_k$ is $n_k = n!/(k!(n-k)!)$. Since we assume that (\ref{k sources}) is valid regardless of the selection of the active sources, the vector $\gamma$ must satisfy the linear system
\[
 \mP_k\gamma = {\rm SNR}\times{\rm trace}({\mathsf\Sigma}) {\mathsf 1}_{n_k},
\]
where ${\mathsf 1}_{n_k}\in\R^{n_k}$ denotes a vector with unit entries. The only possible solution of this system is
 \[
  \gamma_j = \frac{{\rm SNR}\times{\rm trace}({\mathsf\Sigma})}{k},
 \]
 and therefore, we arrive at the scaling law
 \[
\theta_{j} ^*= \frac 1k \frac {{\rm SNR}\times{\rm trace}({\mathsf\Sigma})}{\beta \|\mM_j{\mathsf C}_j^{1/2}\|^2_{\rm F}}, \quad 1\leq j\leq n.
\]

Finally, assume that we only have a prior idea of how many non-zero sources we might have, and we express this belief in a form of a probability density
\[
 \mP\{\mbox{$\#$ of active sources = $k$}\} =p_k, 1\leq k\leq n,
\]
where $\sum_k p_k = 1$. If we expect that of the $n$ dipoles, it is reasonable that $\overline k = sn$ are active, $0<s<1$, we may use binomial distribution for $p_k$,  $k_k\sim{\rm Binom}(n,s)$, which in practice can be approximated by a Poisson distribution,  $p_k\sim{\rm Poisson}(\overline k)$ with mean $\overline k$ provided by the user.
Using the previous result, conditioned on $k$, we arrive at the scaling law
 \[
\theta_{j} ^*= \frac{C}{ \|\mM_j{\mathsf C}_j^{1/2}\|^2_{\rm F}}, \quad 1\leq j\leq n,
\]
with
\[
C = {\rm SNR}\times{\rm trace}({\mathsf\Sigma}) \, \sum_{k=1}^n\frac {p_k}k.
\] 
This argument confirms that, in order to match the model with the SNR, the parameters $\theta_j^*$  should indeed be chosen to be inversely proportional to the sensitivity. 

\section*{Appendix: Construction of the activity patch}

To select the vertices in the activity patch ${\mathcal{P}}$, we first pick randomly a seed vertex from the BR of interest, then grow the patch by adding iteratively the neighboring vertices, at each step pruning off those that fall outsize the pertinent BR, and stopping the process as soon as the desired number $N_{\mathcal{P}}$ of vertices have been included. The selected nodes along with the edges of the triangular mesh form a local graph.
To generate the activity in the patch, we start by computing a positive graph Laplacian of the patch, which is the matrix 
$\mL  \in\R^{N_{\mathcal{P}}\times N_{\mathcal{P}}}$ with entries
\[
L_{i,j} = \left \{\begin{array}{ll}
          -{\rm deg}(v_i) & \hbox{if} \ i=j, \\
          1 & \hbox{if} \ i\ne j \ \hbox{and} \ v_i \ \hbox{is adjacent to} \ v_j, \\
          0 &   \hbox{otherwise},
					\end{array} \right.
\]
where ${\rm deg}(v_i)$ is the number of the edges that terminate at the vertex $v_i$.

After defining a correlation length $\lambda$, given in units of the number of steps, we draw a random amplitude vector by setting
\[
 Q = (\mL + \lambda^2 \mI_{N_{\mathcal{P}}})^{-1}W, 
\]
where $\mI_{N_{\mathcal{P}}}\in\R^{N_{\mathcal{P}}\times N_{\mathcal{P}}}$ is the unit matrix and $W\in\R^{N_{\mathcal{P}}}$ is a standard normal Gaussian random vector, that is,  $W\sim {\mathcal N}(0,\mI_{N_{\mathcal{P}}})$. The amplitudes are scaled so that the amplitude of the dipole at the seed vertex is one. Finally, we draw the dipole moment directions from the anatomical prior, making sure that adjacent dipoles are not pointing in the opposite sides of the cortex patch.

\section*{Acknowledgements} This work was completed during the visit of DC and ES at University of Rome ``La Sapienza'' (Visiting Researcher/Professor Grant 2015). The hospitality of the host university is kindly acknowledged. The work of ES was partly supported by NSF, Grant DMS-1312424. The work of DC was partially supported by grants from the Simons Foundation (\#305322  and \# 246665) and by NSF, Grant DMS-1522334.

\bibliography{BiblioMEGSIM}

\captionsetup[subfloat]{captionskip=-20pt}

\captionsetup[subfloat]{captionskip=-20pt}

\end{document}